\documentclass[aps,rmp,showpacs,twocolumn,nofootinbib,superscriptaddress,
reprint,amsmath,amssymb]{revtex4-1}
\usepackage{graphicx}  
\usepackage[normalem]{ulem}
\usepackage[colorlinks,linkcolor=blue, citecolor=blue]{hyperref}

\begin{document}
\title{\boldmath Colloquium: Hadron Production in Open-charm Meson Pair at $e^+e^-$ Collider}
\author{Xiongfei Wang}
\email[Corresponding author:]{wangxiongfei@lzu.edu.cn}
\author{Xiang Liu}
\email[Corresponding author:]{xiangliu@lzu.edu.cn}
\affiliation{School of Physical Science and Technology, Lanzhou University, Lanzhou 730000, China}
\affiliation{Lanzhou Center for Theoretical Physics, Key Laboratory of Theoretical Physics of Gansu Province,
and Key Laboratory for Quantum Theory and Applications of MoE, 
Gansu Provincial Research Center for Basic Disciplines of Quantum Physics,
Lanzhou University, Lanzhou 730000, China}
\author{Yuanning Gao}
\affiliation{School of Physics and Center of High Energy Physics, Peking University, Beijing 100871, China}

\date{\today}

\begin{abstract}
\bf The standard model of particle physics is a well-established theoretical framework, yet several unresolved issues remain that warrant further experimental and theoretical exploration. In the realm of quark physics, these issues include understanding the nature of quark confinement and elucidating the mechanism linking quarks and gluons to strongly interacting particles within the standard model theory, which may offer insights into the underlying physics mechanisms. These issues inquiries can be addressed through the study of hadrons produced at $e^+e^-$ collisions and decaying to open-charm meson pairs
utilizing the capabilities of {\it BABAR}, Belle, BESIII, and CLEO-c experiments, which have yielded valuable insights into nonstandard hadrons in recent decades. This Colloquium examines the contributions of $e^+e^-$ colliders from the {\it BABAR}, Belle, BESIII, and CLEO-c experiments to such studies in the past two decades and discusses future prospects for $e^+e^-$ collider experiments.
\end{abstract}

\maketitle
\tableofcontents

\section{Introduction}
\begin{figure*}[!htbp]
\includegraphics[width=1.0\textwidth]{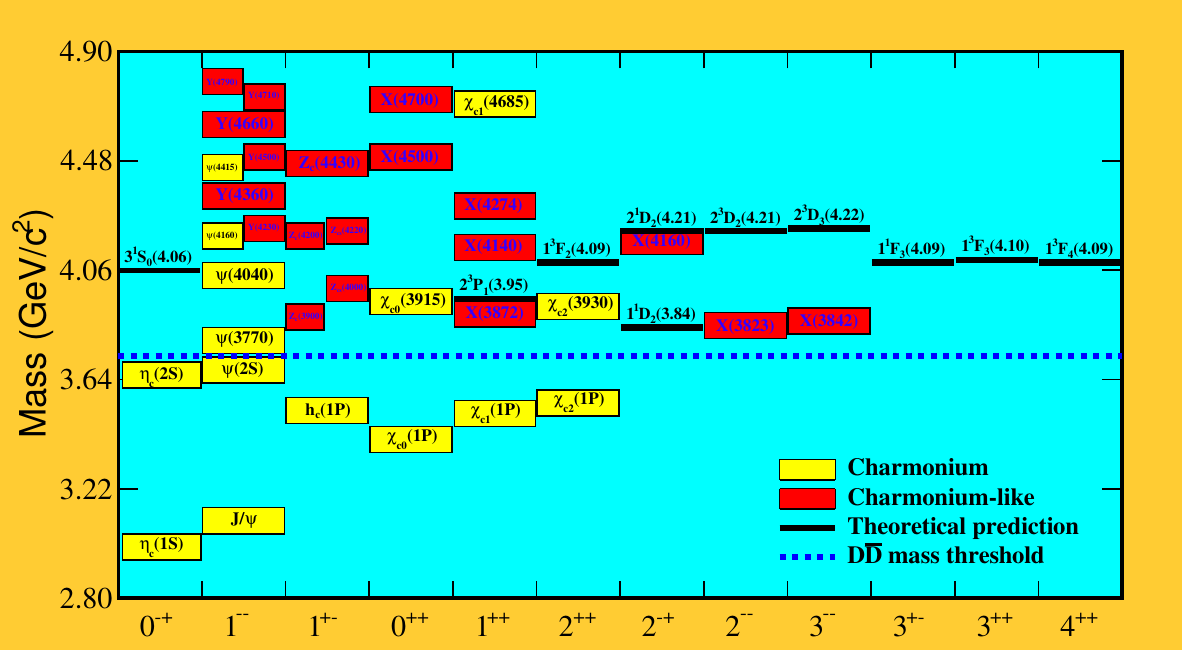}
\caption{\label{Mot:Charmonium}Charmonium(like) family.}
\end{figure*}
The 50th anniversary of the discovery of the $J/\psi$ particle, a landmark achievement in the field of particle physics, was marked in 2024.
Prior to this discovery, the charm quark ($c$) was proposed through the Glashow-Iliopoulos-Maiani mechanism~\cite{Bjorken:1964gz,Cabibbo:1963yz, Glashow:1970gm,Archilli:2017xmu}  to explain the puzzling decay behavior of $K_L\to \mu^+\mu^-$. 
In 1974 two experimental groups led by Ting and Richter independently discovered the $J/\psi$ particle~\cite{E598:1974sol,SLAC-SP-017:1974ind}, an event celebrated as the “November revolution” in the field. This breakthrough not only confirmed the $c$ quark as a new member of the quark family but also unveiled the charmonium meson family, significantly enriching hadron spectroscopy. 
Charmonium is well described by the potential model as a nonrelativistic $c\bar{c}$ bound state~\cite{Eichten:1974af, Eichten:1978tg, Eichten:1979ms}. It provides an ideal platform for probing the nonperturbative dynamics of the strong interaction.
The strong interaction is one of the fundamental forces in nature that govern the behavior of quarks and gluons within hadrons. However, understanding its nonperturbative aspects remains a significant challenge. This area of research represents a frontier that still awaits precise quantitative exploration~\cite{Liu:2013waa,Hosaka:2016pey,Chen:2016qju,Richard:2016eis,Esposito:2016noz,
Ali:2017jda, Olsen:2017bmm,Guo:2017jvc,Liu:2019zoy,Yuan:2019zfo,Brambilla:2019esw,Chen:2022asf,Meng:2022ozq}.

Over the past half century, an increasing number of charmonium(like) states, including the $XYZ$ particles, have been discovered. To date, the Particle Data Group (PDG) has cataloged more than a dozen charmonium(like) states~\cite{ParticleDataGroup:2024cfk}. These states can not only serve as candidates for conventional charmonium but also open up new avenues for exploring exotic hadron states including multiquark configurations, which has promoted the ``Particle Zoo 2.0" initiative.
By studying charmonium(like) states or exotic hadron states, we hope to unlock valuable insights into the underlying mechanisms of the strong interaction, such as the nature of quark confinement and the role of gluon fields in binding quarks together. 
This could potentially lead to significant advancements in our understanding of the structure and behavior of matter at the subatomic scale. 
In addition to its theoretical importance, the study of charmonium(like) states related to the nonperturbative dynamics of the strong interaction also has implications for various other areas of physics, including the understanding of the early Universe and the development of more accurate theoretical models.

\begin{figure}[!htbp]
\begin{center}
\includegraphics[width=0.48\textwidth]{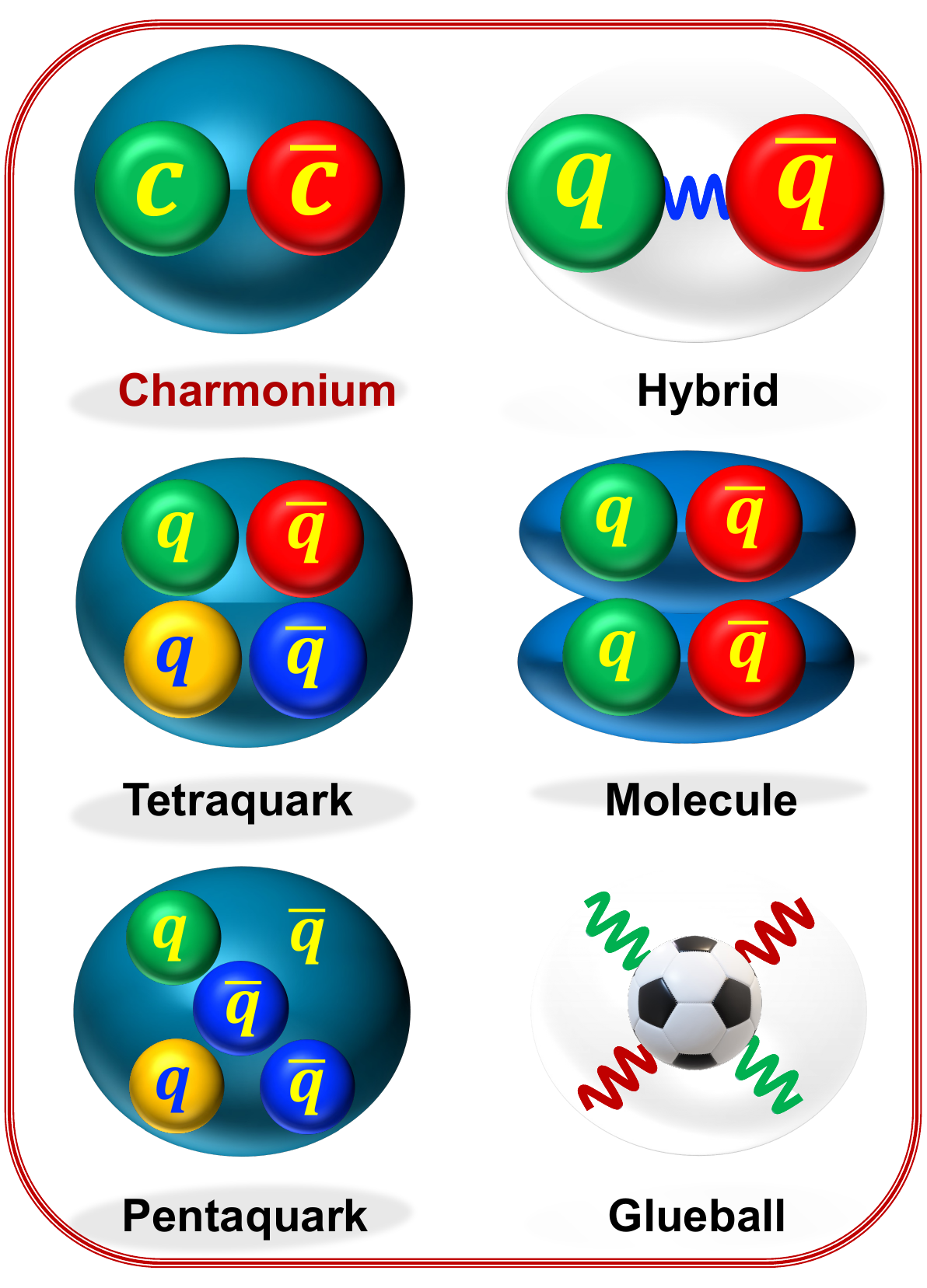}
\caption{\label{Mot:hadron} 
Charmonium and some exotic hadronic states. 
Conventional charmonia are composed of one charm quark and one charm antiquark forming a $c\bar{c}$ bound state. 
Exotic hadronic states include quark-antiquark-gluon hybrids, tetraquark, hadroquarkonia, hadron-hadron molecules, pentaquark, multigluon glueballs, etc..
}
\end{center}
\end{figure}

In Fig.~\ref{Mot:Charmonium} we summarize the current status of observed charmonia and selected charmoniumlike $XYZ$ states. 
Charmoniumlike states or sometimes so-called $XYZ$ particles refer to a series of new hadronic states with hidden charm that have been discovered over the past two decades. As their nature was initially unknown, they were temporarily assigned the labels $X$, $Y$, and $Z$ (the last letters
of the Latin alphabet), a convention that has persisted. Generally, the $Y$ states are vector particles with quantum numbers $J^{PC} = 1^{--}$.
Most $Z$ states are charged, though some neutral ones also exist.
While the $X$ category serves as a temporary designation for the remaining states with more ambiguous properties. Note that these names are provisional, and the naming conventions have not been strictly standardized. As mentioned, the observation of charmoniumlike $XYZ$ states has prompted their extensive discussion as candidates for exotic hadronic states~\cite{Liu:2013waa,Hosaka:2016pey,Chen:2016qju,Richard:2016eis,Esposito:2016noz,Ali:2017jda,Olsen:2017bmm,Guo:2017jvc,Liu:2019zoy,Yuan:2019zfo,Brambilla:2019esw,Chen:2022asf,Meng:2022ozq}. 
To illustrate these candidates, Fig.~\ref{Mot:hadron} presents various proposed configurations of such exotic states. In fact, charmoniumlike states can be classified into different groups according to their production mechanisms. Among them, the states labeled with $Y$ are produced primarily through $e^+e^-$ annihilation and constitute the majority of all observed charmoniumlike states.
It has long been observed that most $Y$ states are discovered in hidden-charm final states from $e^+e^-$ annihilation, for example, $J/\psi\pi^+\pi^-$, $\psi(2S)\pi^+\pi^-$, $h_c\pi^+\pi^-$, and $J/\psi K\bar{K}$~\cite{ParticleDataGroup:2024cfk}. In contrast, theoretical studies have
suggested that charmoniumlike states having masses above 3.9 GeV should predominantly decay
into open-charm final states, which are expected to be the primary contributors to their total decay widths. Consequently, it is particularly important to investigate the processes in which $e^+e^-$ annihilation results in open-charm final states. This is the main objective covered by this Colloquium. Key experiments in this field include the concluded CLEO-c, {\it BABAR}, and Belle experiments, along with the ongoing BESIII and Belle II collaborations, all of which are briefly introduced in this Colloquium.

Since open-charm decay modes of charmonium(like) states are typically allowed by the Okubo-Zweig-Iizuka mechanism ~\cite{Okubo:1963fa,Zweig:1964jf,Iizuka:1966fk},
the study of $e^+e^-$ annihilation into various open-charm meson pairs has drawn significant interest.
In 1980 Eichten {\it et al.}~\cite{Eichten:1974af, Eichten:1978tg, Eichten:1979ms} first attempted a theoretical calculation for the charm cross-section in $e^+e^-$ annihilations based on a coupled-channel potential model. Note that the traditional Breit-Wigner (BW) parametrizations are adequate to describe the narrow, isolated resonances. However, they are ineffective for overlapping states or those close to thresholds for decays to some final states. In such situations, maintaining fundamental amplitude properties of unitarity derived from probability conservation and analyticity derived 
from causality necessitates
more elaborate coupled-channel mode, which is an indispensable framework for understanding the complex and multifaceted nature of hadronic interactions~\cite{Oller:2025leg}.
In this model Eichten {\it et al.} performed a prediction of $\Delta R$ ($\Delta R = \sum_{i}R_{i}$, where $R_{i}$ stands for the ratio of individual hadron cross-section to muon cross-section in $e^+e^-$ collisions, $i$ runs over two-body channels)
with the $D^{(\ast)}\bar{D}^{(\ast)}$ final states. According to this prediction, many large enhancements can be seen in the cross-section at 3.90, 4.04, 4.16, 4.22, and 4.42 GeV, etc., dominated by the final states of open-charm meson pairs. Meanwhile, similar 
broad
resonance peaks above the open-charm region can be seen in the energy dependence of the 
cross-section ratio $R=\sigma(e^+e^-\to Hadrons)/\sigma(e^+e^-\to\mu^+\mu^-)$~\cite{ParticleDataGroup:2024cfk}, which means that the production of open-charm meson pair contributes greatly to the $R$ 
distribution.
In other words, the $c\bar{c}$ mesons above the open-charm threshold
in the $e^+e^-$ annihilations could strongly decay with an additional produced $q\bar{q}$ pair
that forms the system of $c\bar{q}$-$\bar{c}q$ and subsequently
dissociate into two charmed mesons, $D^{(\ast)}\bar{D}^{(\ast)}$.
These processes provide critical insights into the properties of high-mass charmonium and charmoniumlike states, a research focus central to both hadron spectroscopy and the exploration of the nonperturbative area of strong interactions.

In this Colloquium we provide an overview of current experimental progress in studying hadrons produced at $e^+e^-$ collisions and decaying to open-charm meson pairs, with a focus on results from BaBar, Belle, CLEO-c and BESIII. Notably, as BESIII continues to accumulate high-precision data, it has emerged as a leading facility for measuring hadron production in charm-meson pair systems. We therefore highlight its recent advances in this field.

\section{Experimental apparatus}
Since 2003 $e^+e^-$ collider experiments such as those of {\it BABAR}, Belle, BESIII, and CLEO-c have played a crucial role in observing new hadronic states. As typical $B$-meson factories, the {\it BABAR} and Belle experiments have collected numerous $B$-meson data samples, providing an effective way to discover new hadronic states, including charmoniumlike $XYZ$ states. Among them, $X(3872)$ was the first observed $XYZ$ particle~\cite{Belle:2003nnu}. In fact, with the development of the initial state radiation (ISR) technique, direct production of charmoniumlike $XYZ$ states became possible in the {\it BABAR} and Belle experiments. $Y(4260)$~\cite{BaBar:2005hhc}, discovered through this process, is a prominent example. 
Note that a precise cross-section measurement of $e^+e^-\to\pi^+\pi^-J/\psi$ at $\sqrt{s}$ = 3.77-4.60 GeV by the BESIII experiment indicates that the $Y(4260)$ resonance structure consists of two resonances, namely $Y(4220)$ and $Y(4320)$~\cite{BESIII:2016bnd}. 
The former is generally referred to as  $Y(4230)$.
Furthermore, more and more $Y$ states have been reported in the past two decades. The question of how to understand these puzzling phenomena forms the "$Y$ problem," as proposed in the BESIII white paper~\cite{BESIII:2020nme}. In this Colloquium hadrons produced at $e^+e^-$ collisions and decaying to open-charm meson pairs are closely related to these novel observations. In the following we describe the details of the aforementioned experimental apparatus.

\subsection{{\it BABAR} experiment}
\begin{figure*}[!htbp]
\begin{center}
\includegraphics[width=1.0\textwidth]{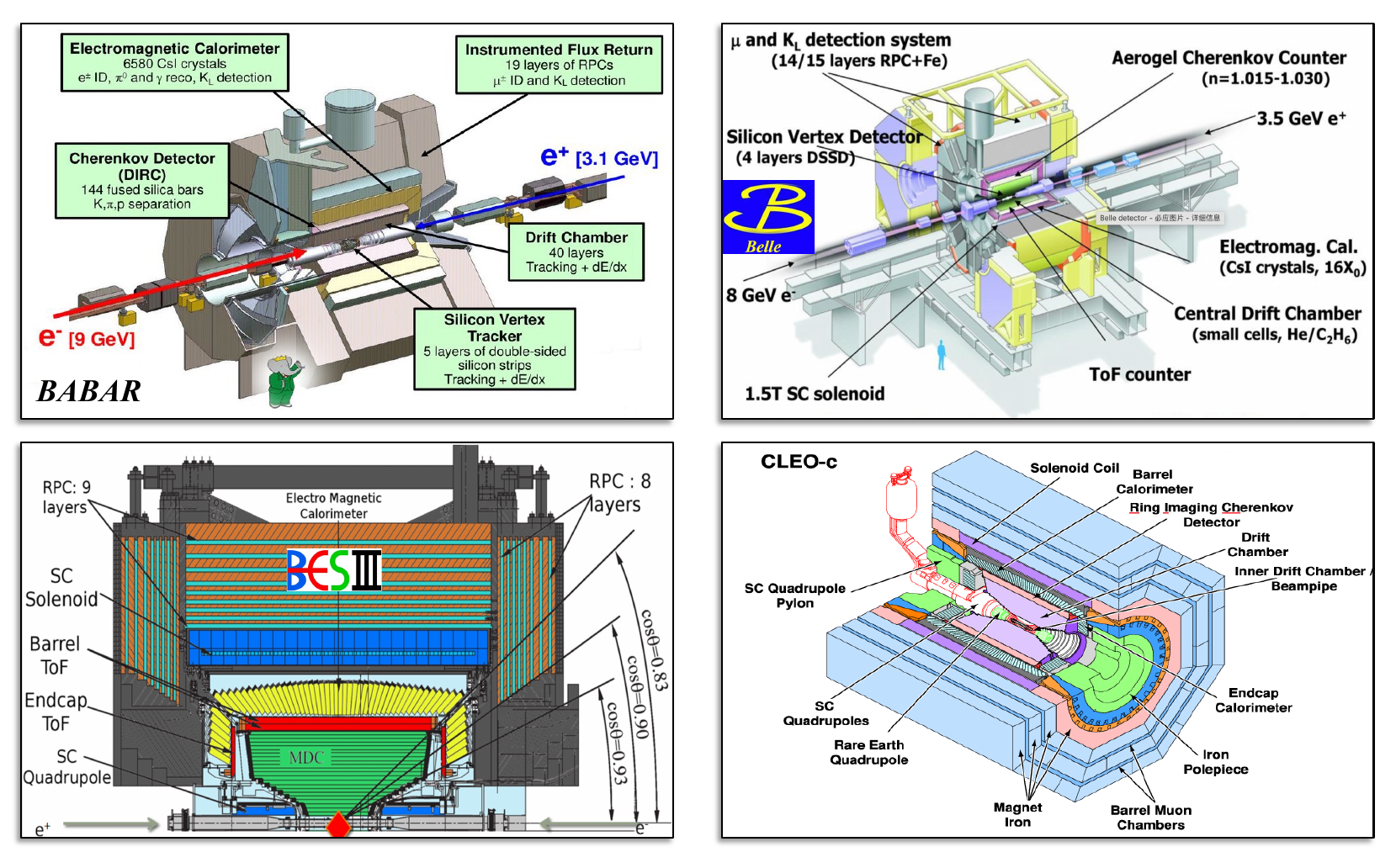}
\caption{\label{SLAC&&BABAR}Overview of detectors for {\it BABAR} (top left panel), Belle (top tight panel), BESIII (bottom left panel) and CLEO-c (bottom right panel).}
\end{center}
\end{figure*}

The {\it BABAR} experiment, located at the Stanford Linear Accelerator Center  in San Francisco 
was
a groundbreaking asymmetric $e^+e^-$ collider with a design luminosity of $3\times 10^{33}\rm cm^{-2}s^{-1}$ that operated
 at the $\Upsilon(4S)$ resonance~\cite{Kozanecki:2000cm, BaBar:2001yhh}. This state-of-the-art facility 
played
a crucial role in making a broad set of measurements capable of confronting the fundamental question regarding what happened to all the antimatter based on the $B$-meson decays.
One of the most interesting aspects of this experiment 
was
its ability to dive into physics issues that occurred less than $10^{-34}$ s after the start of the big bang. The precision and accuracy achieved by {\it BABAR} detector were
truly significant, as they allowed  scientists to study phenomena that were previously inaccessible.
The {\it BABAR} detector, as shown in Fig.~\ref{SLAC&&BABAR} was constructed with the specific purpose of capturing and analyzing high-intensity collisions to study the behavior of subatomic particles. The charged-particle tracks were
meticulously measured in a multilayer silicon vertex tracker that was surrounded by a cylindrical wire drift chamber. This 
allowed for precise tracking and analysis of the paths taken by these particles.
In addition, electromagnetic showers produced by electrons and photons were detected using an array of CsI crystals located just inside the solenoidal coil of a superconducting magnet. This setup enabled researchers to observe and study the interactions between these particles and their surroundings.
Furthermore, muons and neutral hadrons 
were
identified through arrays of resistive plate chambers that 
were
strategically placed within the steel flux return of the magnet. This 
allowed
for the accurate identification and analysis of these types of particles.
Charged hadrons 
were
also identified through $dE/dx$ measurements in the tracking detectors and using a Cherenkov ring-imaging detector surrounding the drift chamber. These methods 
provided
valuable data on the behavior and characteristics of charged hadrons during collision events.
Owing to the outstanding performance of the {\it BABAR} detector, the {\it BABAR} Collaboration discovered $Y(4260)$ in 2005 via the ISR process $e^+e^- \to \gamma_{\text{ISR}} \pi^+\pi^- J/\psi$~\cite{BaBar:2005hhc}. 
The ISR process of $e^+e^-$ annihilation is accompanied by emission of one or several photons from the initial electron or positron~\cite{Druzhinin:2011qd}. 
In the experiment, especially at $B$ factories, there are two approaches to study ISR events, a tagged and an untagged one. In the tagged approach, the ISR photon should be detected.
In the untagged one, the detection of the ISR photon is not required, but all the final hadrons must be detected and fully reconstructed. The ISR events are selected by the requirement that the recoil mass against the hadronic system is close to zero.
This state exhibits strong coupling to hidden-charm final states.
Although this behavior provides important clues for 
understanding the internal structure and underlying dynamics of the $Y(4260)$ state, its properties remain enigmatic until observations of its decay into the final states of open-charm mesons. 
At that time $Y(4260)$ was recalled as $Y(4230)$.
\subsection{Belle and Belle II experiments}
The Belle experiment, located at the KEKB-factory in Tsukuba, Japan, 
was
a groundbreaking project designed to conduct quantitative studies of $B$-meson decay and test $CP$ violation. The experiment 
utilized
 an asymmetric $e^+e^-$ collider with a design luminosity of $10^{34}{\rm cm}^{-2} {\rm s}^{-1}$ that 
operated
at the $\Upsilon(4S)$ resonance~\cite{Belle:2000cnh}.
The Belle detector 
was
a crucial component of the KEKB accelerator facility, which was designed to study the properties of $B$-mesons and other particles produced in high-energy 
collisions, with a composition similar to the
BABAR experiment. 
The detector was carefully configured inside
a 1.5 T superconducting solenoid and iron structure, which effectively surrounded the KEKB beams in the Tsukuba interaction region. This configuration, as shown in Fig.~\ref{SLAC&&BABAR}, 
allowed
precise measurements of $B$-meson decay vertices.
To achieve this level of precision, the Belle detector 
included
several key components. A silicon vertex detector situated just outside of a cylindrical beryllium beam pipe accurately measures $B$-meson decay vertices, while charged-particle tracking 
was
performed by a wire drift chamber [the Central Drift Chamber (CDC)]. Furthermore, particle identification 
was
provided by $dE/dx$ measurements in the CDC, the Cherenkov counter with aerogel threshold, and the time of flight (TOF) counter placed radially outside the CDC.
Furthermore, electromagnetic showers 
were
detected in an array of CsI(Tl) crystals located inside the solenoid coil. Muons and $K_{L}$ mesons 
were
identified by arrays of resistive plate counters interspersed in the iron yoke. The extensive coverage provided by these components 
allowed
complete data collection within a wide $\theta$ region extending from $17^\circ$ to $150^\circ$. Here $\theta$ is the full solid angle.
To ensure thorough detection capabilities across all angles, any part of the uncovered small-angle region 
was
instrumented with a pair of ${\rm Bi_{12}GeO_{2}}$ crystal arrays placed on the surfaces of the quadrupole collision superconducting magnet cryostats in both the forward and backward directions. 
Owing the excellent performance of the Belle detector, the Belle Collaboration discovered the first charmoniumlike state, $X(3872)$, in 2003~\cite{Belle:2003nnu}. This discovery not only offered a novel perspective for understanding the structure of exotic hadron states but also posed a challenge to the traditional quark potential model~\cite{Eichten:1974af, Eichten:1978tg, Eichten:1979ms, Barnes:2005pb,
Radford:2007vd}.
The Belle Collaboration has also been an important contributor to the observation of charmoniumlike $Y$ states such as $Y(4360)$ \cite{Belle:2007umv}, $Y(4630)$~\cite{Belle:2008xmh}, and $Y(4660)$~\cite{Belle:2007umv}.
Currently, the Belle II experiment~\cite{Belle-II:2018jsg} is operational and will continue to collect data relevant to this research area.
Belle II is a substantial upgrade of the Belle detector operating at the SuperKEKB electron-positron collider, which was designed to operate at a peak luminosity of $8\times10^{35}cm^{-2}s^{-1}$ with a target integrated luminosity of 50 $ab^{-1}$, which represents a 40-fold increase compared to its predecessor. Here the luminosity at Belle II is achieved via a novel low-emittance nanobeam approach, which is combined with a new positron damping ring and a positron beam vacuum chamber. 

\subsection{BESIII experiment}
The BEPCII, also known as the second generation of the Beijing Electron Positron Collider, is a symmetric $e^+e^-$ collider that operates in the $\tau$-charm physics region with a design luminosity of $10^{33}{\rm cm}^{-2}{\rm s}^{-1}$ at a beam energy of 1.89 GeV~\cite{Yu:2016cof}. It consists of a 200 m linear accelerator, dual storage rings with a circumference of 240m, and the BESIII detector shown in Fig.~\ref{SLAC&&BABAR}, which is located in the southern end of the BEPCII.
The cylindrical core of the BESIII detector, which encloses 93\% of the $4\pi$ solid angle, incorporates a helium-based multilayer drift chamber, a plastic scintillator TOF system, and a CsI (Tl) electromagnetic calorimeter. All of these components are located within a superconducting solenoidal magnet that generates a 1.0 T magnetic field. An octagonal flux-return yoke, reinforced with steel-interleaved resistive plate chamber (RPC)
muon identifier modules, provides support for the solenoid. This advanced setup achieves a charged-particle momentum resolution of 0.5\% at 1 GeV/$c$ and a $dE/dx$ resolution of 6\% for electrons from Bhabha scattering. The barrel section of the TOF system has a time resolution of 68 ps, and the end cap section has a time resolution of 110 ps. By adopting multigap RPC technology, it now offers a time resolution of 60 ps. 
The BESIII experiment~\cite{BESIII:2009fln} started data collection in 2009 and has thus far accumulated large data samples in center-of-mass (c.m.) energies ($\sqrt{s}$) between 2.0 and 4.9 GeV~\cite{Ablikim:2013ntc,BESIII:2015qfd, BESIII:2017lkp, BESIII:2022dxl,BESIII:2022ulv,BESIII:2024lbn}. 
On April 5, 2016, the BEPCII collider successfully achieved its design luminosity~\cite{Li:2021zrf}.
The BESIII experiment, equipped with advanced instrumentation, provides a unique setup for investigating charmonium and charmoniumlike states, charmed mesons and baryons, light hadron spectroscopy, $\tau$ physics, QCD and Cabibbo-Kobayashi-Maskawa parameters, as well as new physics through the study of rare and forbidden decays~\cite{BESIII:2020nme}.
Thanks to its unique data sample and excellent detector performance, 
the BESIII experiment discovered in 2013 the four-quark state
$Z^{\pm}_c(3900) \to J/\psi \pi^{\pm}$ produced in $e^{+}e^{-} \to Z^{\pm}_c(3900) \pi^{\mp}$ at $\sqrt{s}$ =4.260 GeV~\cite{BESIII:2013ris}. In recent years BESIII has become the leading experiment for measuring hadrons produced at $e^+e^-$ collisions and decaying to open-charm meson pairs.
This has been enabled by its data collection at c.m. energies \(\sqrt{s}>4\) GeV.

\subsection{CLEO-c experiment}
The Cornell Electron Storage Ring (CESR) was also a symmetric $e^+e^-$ collider that 
operated
in the $\tau$-charm physics region with a design luminosity of $2\times 10^{33}{\rm cm}^{-2}{\rm s}^{-1}$~\cite{Fast:1999rf, Richichi:2003md}.
The aim of this project is similar to the BESIII experiment, which uses the CLEO-c detector at the CESR in New York. 
The CLEO-c detector, shown in Fig.~\ref{SLAC&&BABAR}, consists of a new silicon tracker, a new drift chamber, and a ring-Imaging Cherenkov counter, together with the CLEO II/ILV magnet, electromagnetic calorimeter, and muon chambers. The upgraded detector was installed and commissioned during the fall of 1999 and spring of 2000. Subsequently, the operation 
had
been reliable, and a high-quality dataset had
been obtained then. 
Its comprehensive program included a wide range of measurements that contributes to advancing our understanding of fundamental processes within the standard model of particle physics.
These measurements have the potential to shed light on important phenomena such as quark interactions, decay processes, and other aspects of particle behavior. Furthermore, they 
provided
valuable insights into how these processes fit into the broader framework of particle physics theory.
An examination of CLEO-c measurements 
uncovered results regarding open-charm meson pairs. These results are thoroughly reviewed in this Colloquium.
\section{Experimental advances}
\begin{figure*}[!htbp]
\begin{center}
\includegraphics[width=1.0\textwidth]{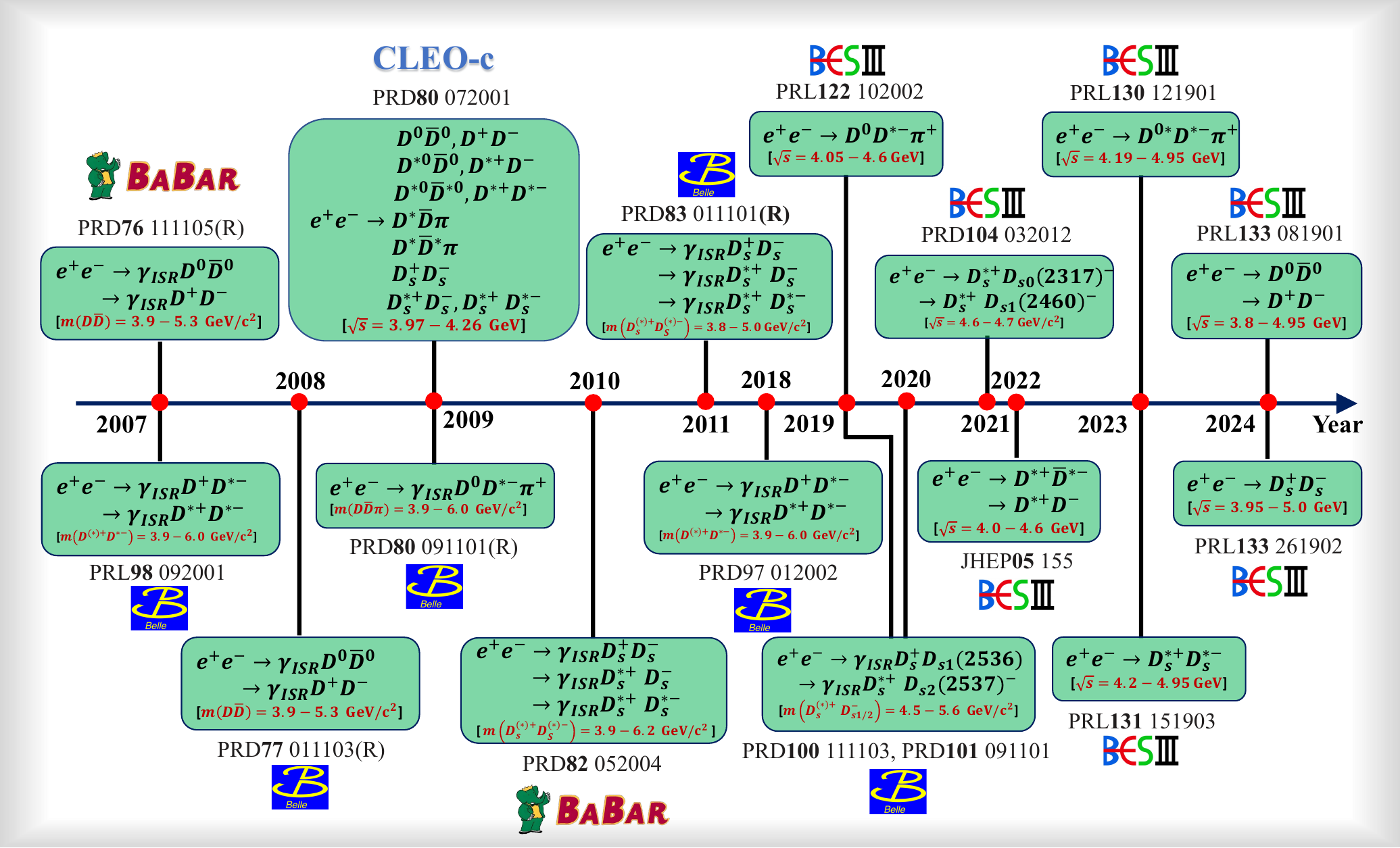}
\caption{\label{DDbar:roadmap}Status of open-charm meson pair production in $e^+e^-$ colliders from the experiments of {\it BABAR}, Belle, BESIII and CLEO-c.}
\end{center}
\end{figure*}
To properly understand the nature of charmoniumlike spectroscopy and to explore the exotic configurations of nonstandard hadrons, the experimental measurement with high precision on electron-positron pairs annihilating into open-charm final states is crucial.
Thanks to the $e^+e^-$ collider experiments, many achievements in the open-charm final states have been obtained by {\it BABAR}, Belle, BESIII and CLEO-c experiments.
These results offer groundbreaking insights into the nature of charmoniumlike states and strong interactions and even shed light on our understanding of nonperturbative QCD. In Fig.~\ref{DDbar:roadmap} we summarize the status of these observations.
Details of the experimental advances are discussed soon.

\subsection{Charmed meson pair}
\subsubsection{$e^+e^-\to D^0\bar{D}^0$ and $D^+D^-$}
\begin{figure*}[!htbp]
\begin{center}
\includegraphics[width=1.02\textwidth]{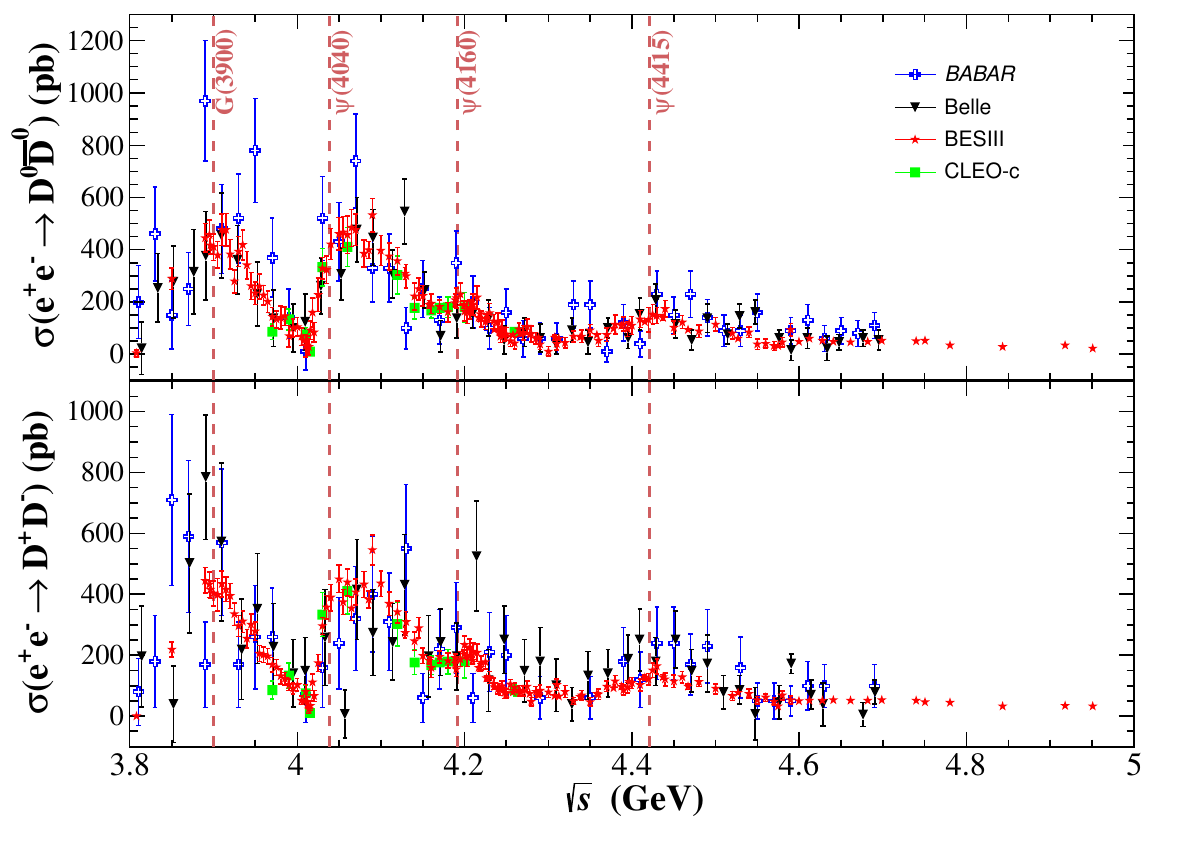}
\caption{ 
Comparisons of cross-sections for the $e^+e^-\to D\bar{D}$ process as a function of c.m. energy from 3.80 to 4.95 GeV from different experiments of {\it BABAR}~\cite{BaBar:2006qlj}, Belle~\cite{Belle:2007qxm}, BESIII~\cite{BESIII:2024ths} and CLEO-c~\cite{CLEO:2008ojp}.
}
\label{BCS:DDbar}
\end{center}
\end{figure*}

The $D^0~(D^-)$ meson is a light-heavy quark structure composed of the charm and the
up quark (down quark).
The studies of exclusive production cross-sections for open-charm final states offer essential insights into the nature of the charmonium(like) states.
Previously, the available cross-sections of the reaction $e^+e^-\to D\bar{D}$ with limited energy points were reported by $B$ factories~\cite{Belle:2007qxm, BaBar:2006qlj} using the ISR process in the experiments of {\it BABAR} and Belle and by direct production in the CLEO-c experiment~\cite{CLEO:2008ojp}.
Recently, the BESIII experiment presented a high-precision measurement of Born cross-sections for reactions of $e^+e^-\to D^0\bar{D}^0$ and $D^+D^-$ at 150 c.m. energies between 3.80 and 4.95 GeV using a data sample corresponding to an integrated luminosity of 10 fb$^{-1}$~\cite{BESIII:2024ths}.
Here a single-tag technique is employed to proceed with the event selection of $e^+e^-\to D\bar{D}$; that is,
only one $D^0~(D^-)$  meson is reconstructed through the $K^-\pi^+\pi^+\pi^-~(K^-\pi^+\pi^-)$ mode, while the accompanying antiparticle is extracted from the recoil side, which has included the charge-conjugate mode.
And the Born cross-section is defined to be a corrected cross-section via the ISR and vacuum polarization (VP) effects as  
\begin{equation}
\sigma^{B} =\frac{\sigma^{\rm obs}}{\delta^{\rm ISR}\times{1/{1-\Pi}^2}} = \frac{\sigma^{\rm dressed}}
{1/{1-\Pi}^2} ,
\end{equation}
where the ISR correction factor $\delta^{\rm ISR}$ is obtained via an iterative procedure, following Refs.~\cite{WorkingGrouponRadiativeCorrections:2010bjp, Sun:2020ehv}, and the VP factor $\frac{1}{{1-\Pi}^2}$ is from Ref.~\cite{Jegerlehner:2011ti}, while $\sigma^{\rm obs}$ denotes the observed cross-section and $\sigma^{\rm dressed}$ represents the dressed cross-section.
Many clear peaks are seen in the line shape of the cross-section of $e^+e^-\to D^0\bar{D}^0$ and 
$D^+D^-$ around the mass ranges of $G(3900)$, $\psi(4040)$, $\psi(4160)$, $Y(4230)$, and $\psi(4415)$, etc., as shown in Fig.~\ref{BCS:DDbar}. 

\begin{figure}[!htbp]
\begin{center}
\includegraphics[width=0.48\textwidth]{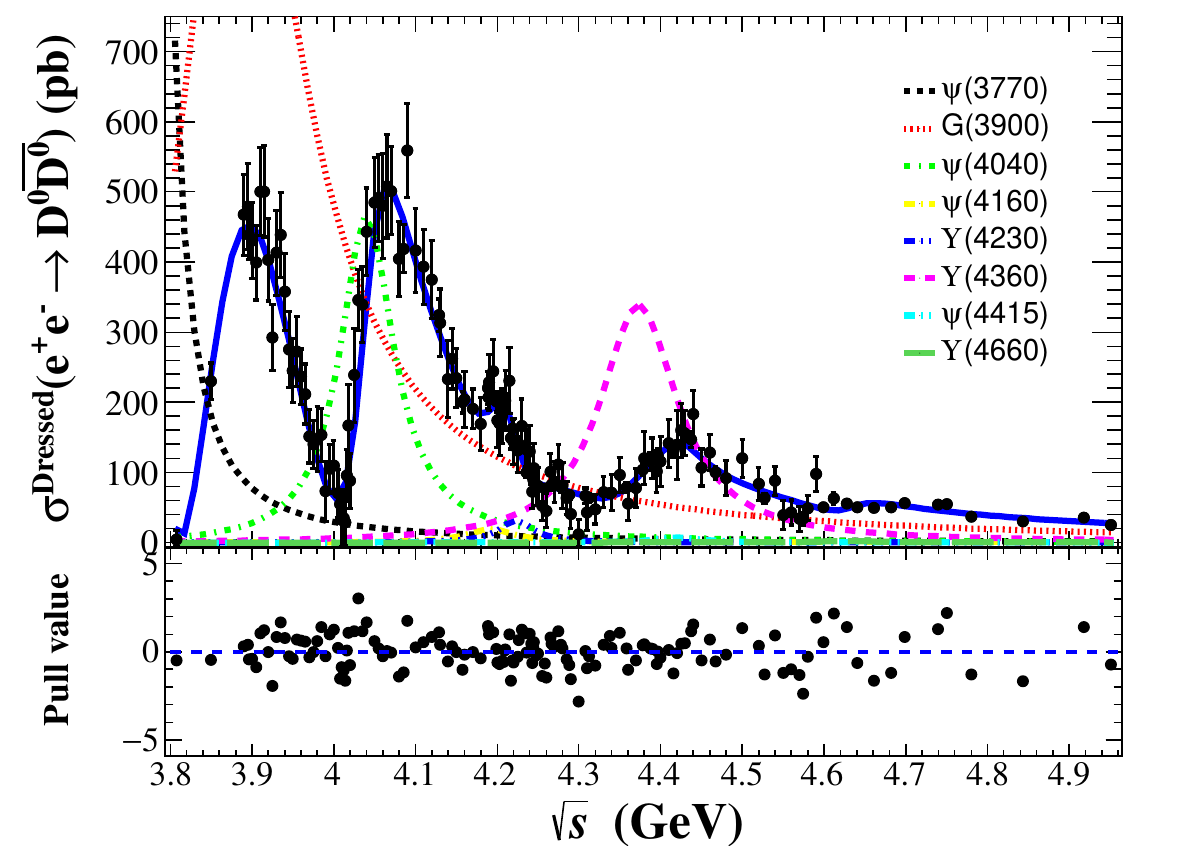}
\includegraphics[width=0.48\textwidth]{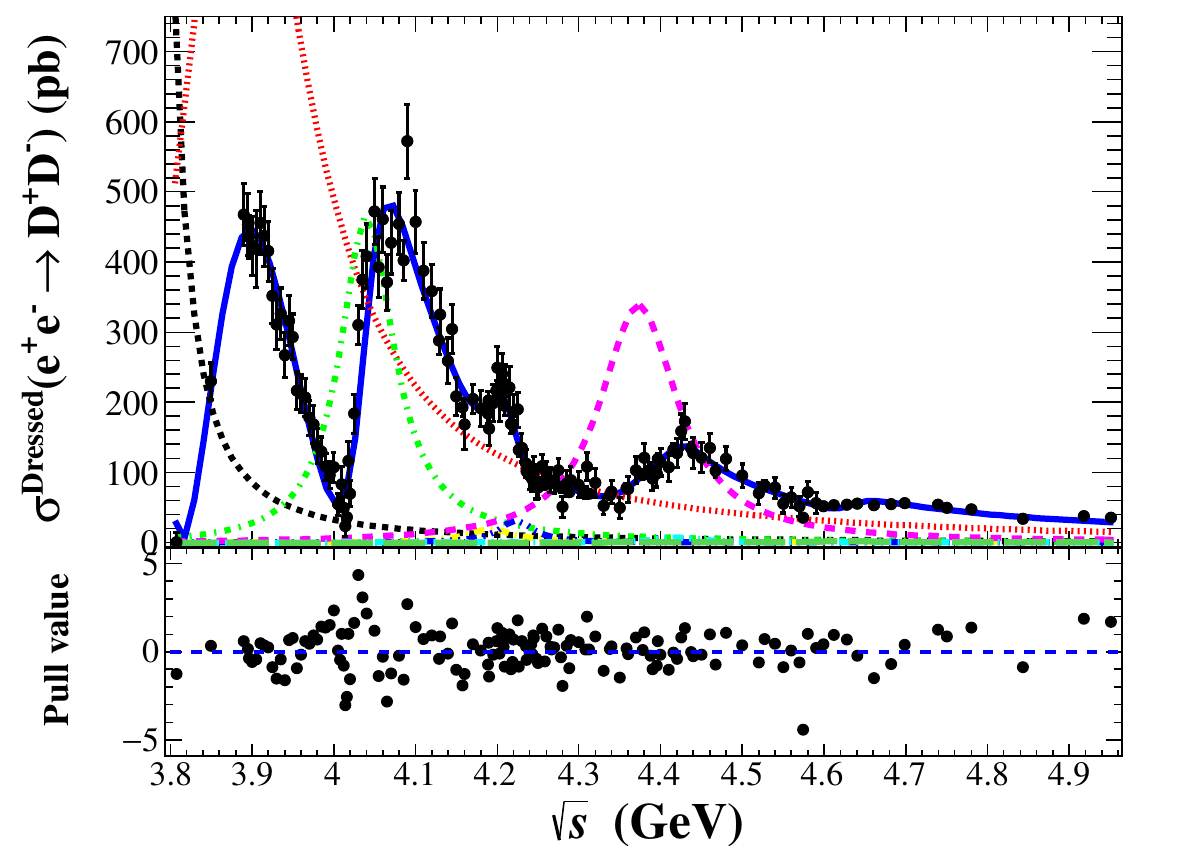}
\caption{Simultaneous fits to the dressed cross-sections for the reactions of $e^+e^-\to D^0\bar{D}^0$ and $D^+D^-$ with the assumption of eight resonances for BESIII data~\cite{BESIII:2024ths}.}
\label{BCS:DDbar:fit}
\end{center}
\end{figure}

\begin{figure}[!htbp]
\begin{center}
\includegraphics[width=0.49\textwidth]{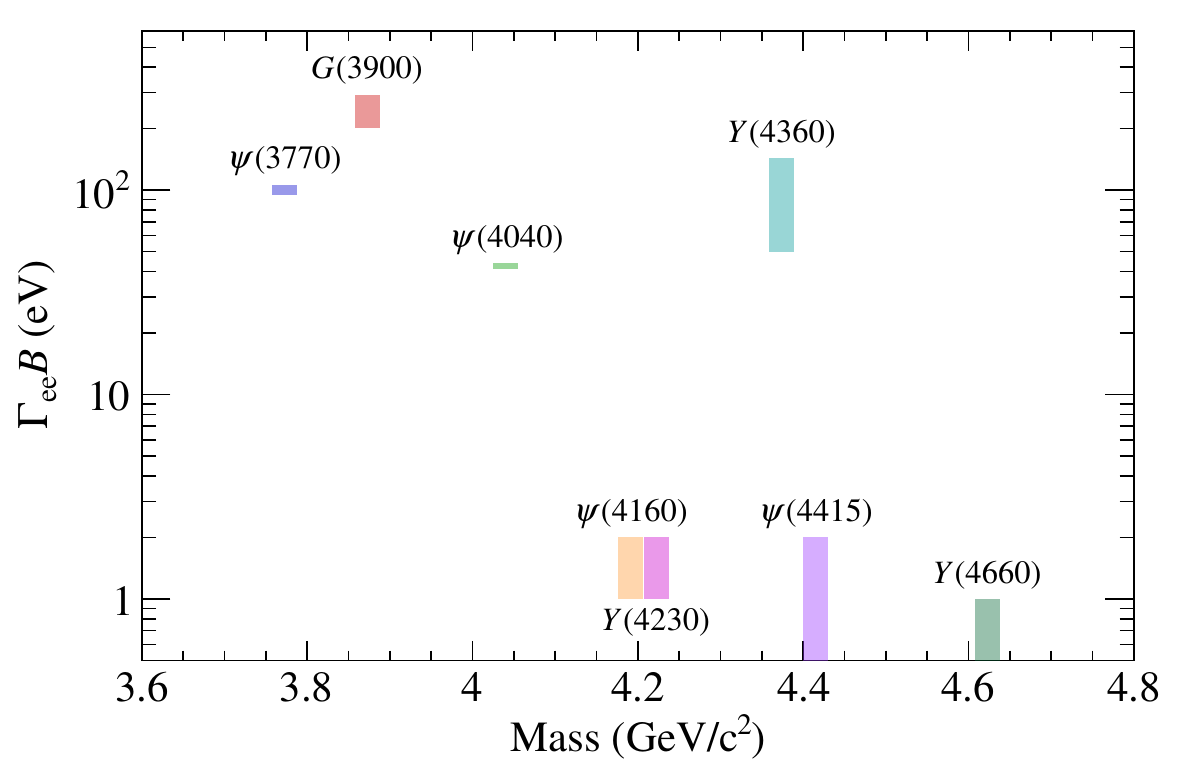}
\caption{Fitted values of $\Gamma_{ee}\cal{B}$
for assumed resonances decaying into the $D\bar{D}$ final states.}
\label{BCS:DDbar02:GammaBr}
\end{center}
\end{figure}

Meanwhile, the BESIII experiment~\cite{BESIII:2024ths} has also provided a model-dependent fit to the dressed cross-section of $e^+e^-\to D\bar{D}$ (as shown in Fig.~\ref{BCS:DDbar:fit}), that is parametrized  as a coherent sum of eight relativistic BW functions. These functions correspond  to seven known resonances
$\psi(3770)$, $\psi(4040)$, $\psi(4160)$, $Y(4230)$, $Y(4360)$, $\psi(4415)$, and $Y(4660)$ plus another structure regarded as $G(3900)$ around 3.9 GeV, i.e.,
\begin{equation}
\sigma^{\rm dressed}(\sqrt{s}) = \left|\sum_{k=1}^{8}e^{i\phi_{k}}{\rm BW}_{k}(\sqrt{s})\sqrt{\frac{P(\sqrt{s})}{P(M)}}\right|^{2},
\end{equation}
with
\begin{equation}
\label{BWABCD}
{\rm BW}(\sqrt{s}) = \frac{\sqrt{12\pi\Gamma_{ee}{\cal{B}}\Gamma}}{s - M^{2} + iM\Gamma}.
\end{equation}
Note that for the name of $G(3900)$, it is sometimes a resonancelike peak and is similar to a kind of particle with quantum number $J^{PC} = 1^{--}$. However, it was not treated as a resonance in previous works~\cite{BaBar:2006qlj,Belle:2007qxm}. 
In Eq.~\ref{BWABCD} the mass
$M$ and total width $\Gamma$ for known resonances are fixed according to individual PDG values~\cite{ParticleDataGroup:2024cfk}, while they are free for $G(3900)$ around 3.9 GeV. Electronic partial widths
($\Gamma_{ee}$) and branching fractions of the decay ($\cal{B}$) are free for all resonances.
The relative phases between different BW functions are denoted by $\phi_k$, and the phases are set to be different in the simultaneous fit. $P(\sqrt{s})$ represents the two-body phase space (PHSP) factor. 
Figure~\ref{BCS:DDbar02:GammaBr} presents the fitted 
values of $\Gamma_{ee}\cal{B}$ for the assumed resonances that decay into the $D\bar{D}$ final states.
Here the fitted mass and width of the $G(3900)$ structure are
$M=(3873 \pm 14 \pm3)$ MeV/$c^2$ and $\Gamma =(180\pm14\pm7)$ MeV, 
which deviate by more than 1$\sigma$ from the corresponding values reported in the {\it BABAR} experiment~\cite{BaBar:2006qlj}.
The BESIII experiment in Ref.~\cite{BESIII:2024ths} has also demonstrated that the parameters for all assumed resonances strongly depend on the chosen fit model, highlighting the need for more comprehensive research, such as coupled-channel $K$-matrix analysis~\cite{Husken:2024hmi}.
Here the $K$-matrix formalism is normally applied to describe two-body scattering processes of the type 
$c_{i}d_{i}\to R\to a_{i}b_{i}$, where $i$ represents each separate channel and $R$ is the  number of resonances that these channels have in common~\cite{Chung:1995dx, Henner:2022ksn}.
In a recent work~\cite{Husken:2024hmi}, the structure around 3.9 GeV was explained as an interference effect between $\psi(3770)$ and $\psi(4040)$ utilizing a coupled-channel $K$-matrix analyses, 
but alternative explanations for such an enhancement have been proposed
~\cite{Zhang:2009gy, Du:2016qcr, Lin:2024qcq, Salnikov:2024wah}.
In this case, a more comprehensive approach based on the $K$-matrix formalism to describe the cross-section of various open-charm final states is expected to test the scenarios of charmonium(like) states  in the future~\cite{Wang:2019mhs,Wang:2020prx, Llanes-Estrada:2005qvr, Li:2009zu, Cao:2020vab} in the future.

\begin{figure}[!htbp]
\begin{center}
\includegraphics[width=0.49\textwidth]{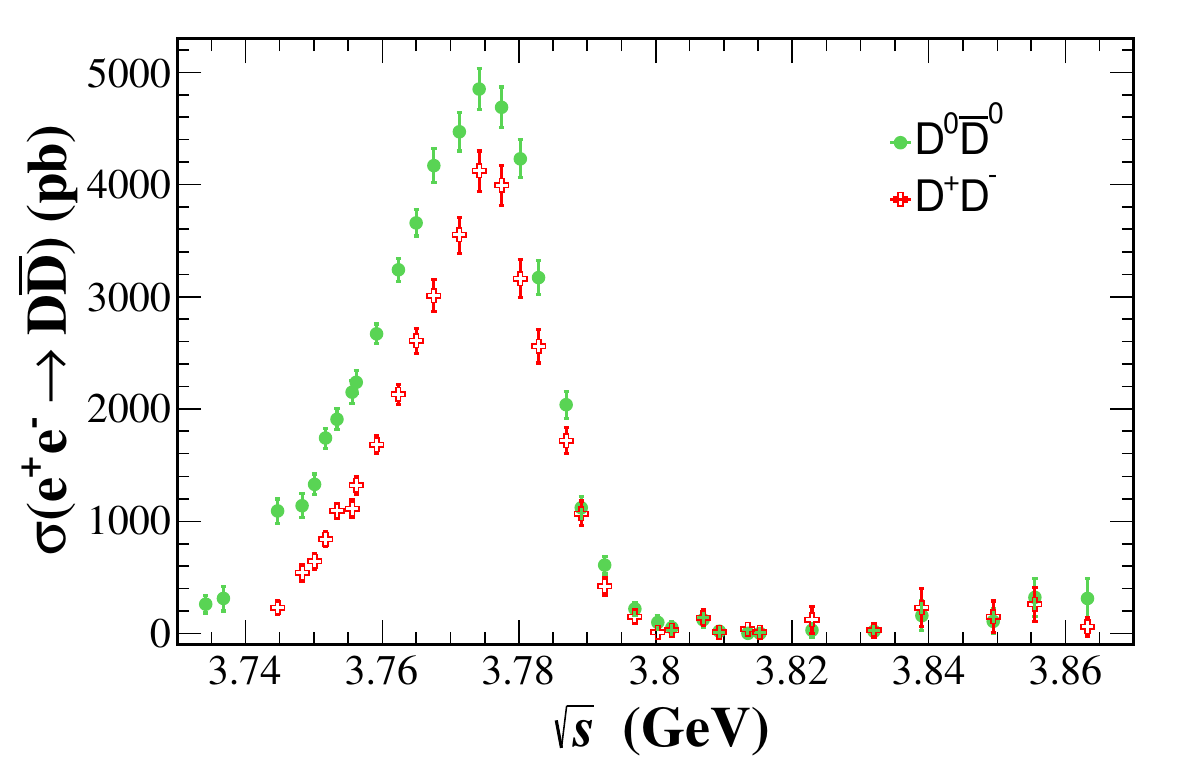}
\caption{Born cross-sections for the reactions of  $e^+e^-\to D^0\bar{D}^0$ and $D^+D^-$ as a function of c.m. energy near $\psi(3770)$ from unpublished BESIII data~\cite{Julin:2017jcl, Husken:2024hmi}.}
\label{BCS:DDbar02:psi3770}
\end{center}
\end{figure}

Furthermore, Refs.~\cite{Julin:2017jcl, Husken:2024hmi} also highlighted the significance of BESIII
measurements with more precision for Born cross-sections of $e^+e^-\to D^0\bar{D}^0$ and $D^+D^-$ near $\psi(3770)$, as shown in Fig.~\ref{BCS:DDbar02:psi3770}. Although this work has not been finished, and does not include a fit to the
cross-section with the $K$-matrix model, one would need to at least account for the effect of interference with
the $\psi(2S)$. In addition, 
the vanishing Born cross-section near 3.81 GeV can only be
described with destructive interference between the $\psi(3770)$ state and other resonances or the nonresonant contribution.
The resonance parameters extracted from fitting the cross-section data around $\psi(3770)$ in Ref.~\cite{Julin:2017jcl} were preliminarily determined to be
$M = (3780.8 \pm 0.2 \pm 0.6 \pm 1.3)$ MeV/$c^{2}$, $\Gamma = (24.1 \pm 0.5 \pm 0.6 \pm 1.9)$ MeV, $\Gamma_{e^+e^-} = (216 \pm 9 \pm 11 \pm 17)$ eV and $\phi = (207 \pm 3 \pm 3 \pm 7)^{\circ}$, respectively. Here $\phi$ is the phase between the $\psi(3770)$ state and the nonresonant contribution.  These results were also compared with previous measurements from other experiments. 
Although there are discrepancies with the world averages~\cite{ParticleDataGroup:2024cfk}, the results are consistent with earlier but less precise values obtained by the KEDR experiment~\cite{Anashin:2011kq} and the theoretical prediction of the vector dominance model~\cite{Julin:2017jcl}.
This sheds light on potential differences between different experimental results and their implications for our understanding of charmonium spectroscopy.

\subsubsection{$e^+e^-\to D^{\ast}\bar{D}^{(\ast)}$}
\begin{figure*}[!htbp]
\begin{center}
\includegraphics[width=1.01\textwidth]{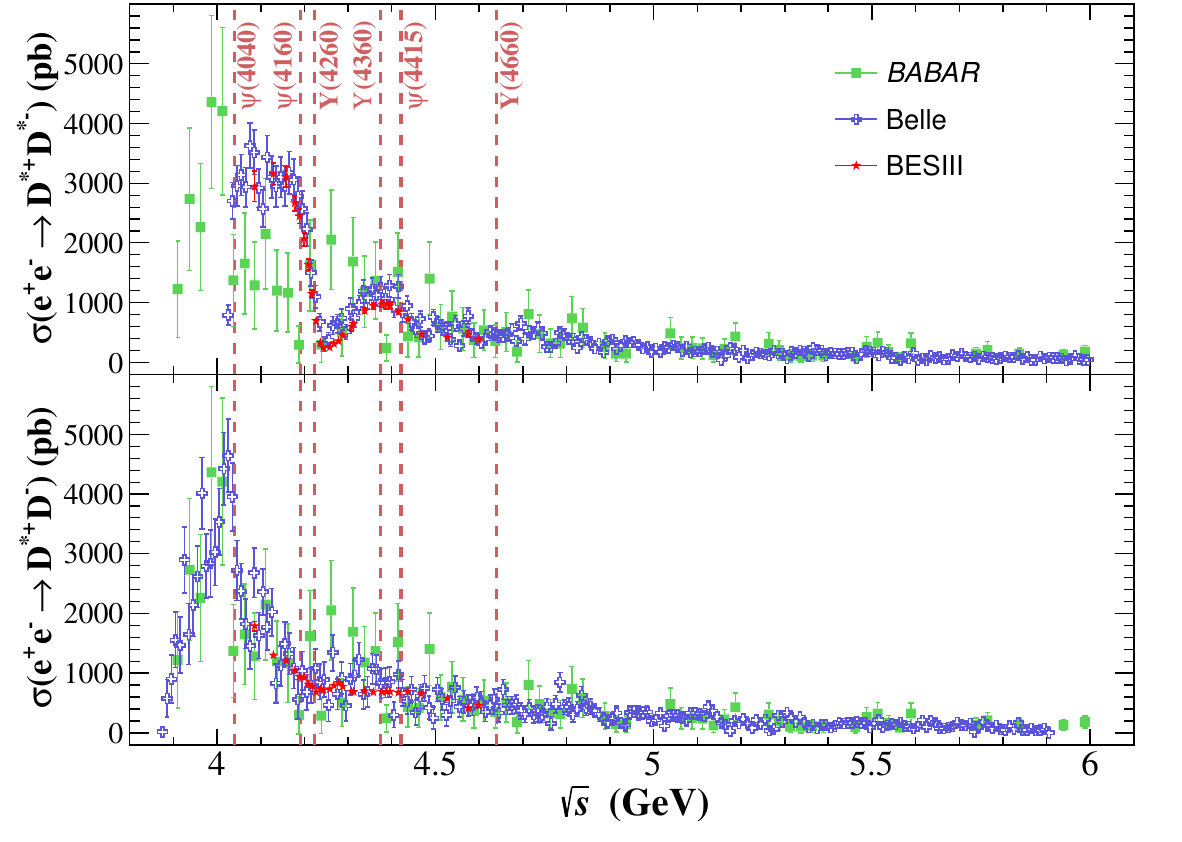}
\caption{ 
Comparisons of cross-sections for the $e^+e^-\to D^{\ast+}D^{\ast-}$ (top panel) and $D^{\ast+}D^-$ reactions (bottom panel) as a function of c.m. energy from 3.9 to 6.2 GeV from different experiments of {\it BABAR}~\cite{BaBar:2009elc}, Belle~\cite{Belle:2006hvs, Belle:2017grj} and BESIII~\cite{BESIII:2021yvc}.
}
\label{BCS:DstarDD}
\end{center}
\end{figure*}
A more in-depth study of charmonium(like) states in open-charm channels would not only enhance our understanding of the characteristics of these states but also contribute valuable insights to various theoretical interpretations. 
The previous measurements of
cross-sections of 
$e^+e^-\to D^{\ast+}D^{\ast-}$
and 
$D^{\ast+}D^{-}$
were performed at c.m. energy of 3.875 to 6 GeV by {\it BABAR} and the Belle experiment~\cite{Belle:2006hvs, Belle:2017grj, BaBar:2009elc} 
based on a partial reconstruction technique  
with the aim of increasing efficiency and suppressing background. Here the mentioned technique was used by requiring full reconstruction of only one of the $D^{\ast+}$ mesons, the $\gamma_{ISR}$, and the slow $\pi^-$ from the other $D^{\ast-}$.
The complex shape of the cross-sections can be explained by the fact that its components can interfere constructively or destructively. The fit of this cross-section is not trivial, because it must take into account the threshold and coupled-channel effects. Recently, using a data sample 
corresponding to the
integrated luminosity of 16 fb$^{-1}$, the BESIII experiment performed a more precise measurement of Born cross-sections of the $e^+e^-\to D^{\ast+}D^{\ast-}$ and $D^{\ast+}D^-$ reactionsat 28 c.m. energies between 4.085 and 4.6 GeV by means of a single-tag
technique~\cite{BESIII:2021yvc} . In the event selection of the BESIII experiment, only the $D^{\ast+}$ meson is reconstructed with the decay chains $D^{\ast+}\to\pi^+D^{0}$ and $D^{0}\to K^-\pi^+$, while the antimeson $D^{\ast-}$ or $D^{-}$ is not reconstructed exclusively but is inferred from energy-momentum conservation.

\begin{figure}[!htbp]
\begin{center}
\includegraphics[width=0.5\textwidth]{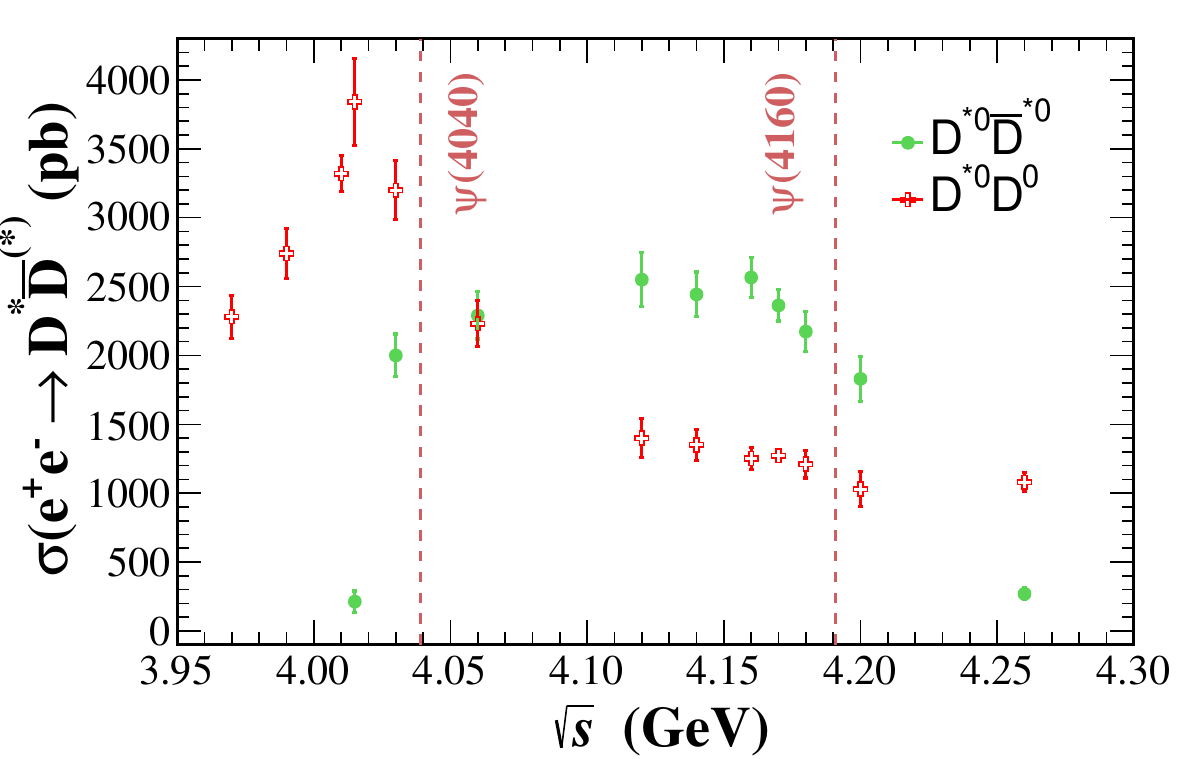}
\caption{cross-section for the reactions of $e^+e^-\to D^{\ast0}\bar{D}^{\ast0}$ and $D^{\ast0}\bar{D}^{0}$ as a function of c.m. energy from the
CLEO-c experiment~\cite{CLEO:2008ojp}.}
\label{BCS:D0sD0s:CLEO}
\end{center}
\end{figure}

Figure~\ref{BCS:DstarDD} shows the comparisons of cross-sections from the different experiments. The results are consistent with each other, and the BESIII experiment presented a more precise measurement than the previous ones by {\it BABAR}, Belle and CLEO-c at c.m. energies between 4.0 and 4.6 GeV.
Figure~\ref{BCS:D0sD0s:CLEO} shows a cross-section of $e^+e^-\to D^{\ast0}\bar{D}^{\ast0}$ and $D^{\ast0}\bar{D}^{0}$
from CLEO-c experiment
only.
The shape of the $e^+e^-\to D^{\ast}\bar{D}^{\ast}$ cross-section is complicated with several local maxima and minima, which is a little different from the line shape of
$e^+e^-\to D\bar{D}$. 
Here a clear plateau can be seen around the mass ranges of $\psi(4040)$ and $\psi(4160)$ between 4.0 and 4.2 GeV for the reaction of $e^+e^-\to D^{\ast+}D^{\ast-}$, a clear peak around the mass range of $\psi(4040)$ for the reaction of $e^+e^-\to D^{\ast+}D^{-}$, and some bumps around the mass ranges of $Y(4230)$, $\psi(4415)$, and $Y(4660)$. 
 The minimum or a clear drop near 4.23 GeV/$c^{2}$ in the
$Y(4230)$ region could be due to 
$D_{s}^{\ast+}D_{s}^{\ast-}, D\bar{D}_{1}(2420)$ or $D\bar{D}_{1}(2430)$
threshold effects described by~\cite{Dubynskiy:2006sg,Rosner:2006vc} or could be due to the destructive interference of this state with other $\psi(nS)$ states. Note that threshold effects is usually explained as a possible 
enhancement in the charm-meson and anticharm-meson system near the mass threshold of $D_{s}^{\ast+}D_{s}^{\ast-}, D\bar{D}_{1}(2420)$ or $D\bar{D}_{1}(2430)$, which exhibits a narrow peak or a steep falloff around the mass threshold, and inspired much speculation  and renewed interest in the threshold bound state.
As for the line shape of the $e^+e^-\to D^{\ast+}D^-$ cross-section, it is relatively featureless except for a prominent excess near $\psi(4040)$.
In  {\it BABAR} experiment~\cite{BaBar:2009elc}, fits to the mass spectra of $D^{\ast+}D^{\ast-}$ and $D^{\ast+}D^-$ 
were performed,
and the amplitudes and relative phases for the charmonium states $\psi(3770)$, $\psi(4040)$, $\psi(4160)$, and $\psi(4415)$,  from which the first measurements of branching fraction ratios are obtained, were measured.

The study of the $Y(4230)$ state continues to be a topic of interest in the field of particle physics. Although initial interpretations suggested that it could be a $1^{--}$ charmonium state with primary decays to the $D^{\ast+}D^{\ast-}$ and $D^{\ast+}D^-$ final states, the current limited data sample size does not provide evidence to support the modes. This has led to alternative hypotheses for the nature of $Y(4230)$, including suggestions that it could be a hybrid, baryonium, molecule, or tetraquark state~\cite{Wang:2019mhs,Wang:2020prx, Llanes-Estrada:2005qvr, Li:2009zu, Cao:2020vab}.
In particular, if $Y(4230)$ were a hybrid state, its decay rates to $D^{\ast+}D^{\ast-}$ and $D^{\ast+}D^-$ would be  small~\cite{Zhu:2005hp, Kou:2005gt, Close:2005iz}. 

\subsubsection{$e^+e^-\to \pi^+D^{(\ast)0}D^{\ast-}$}
\begin{figure*}[!htbp]
\begin{center}
\includegraphics[width=1.01\textwidth]{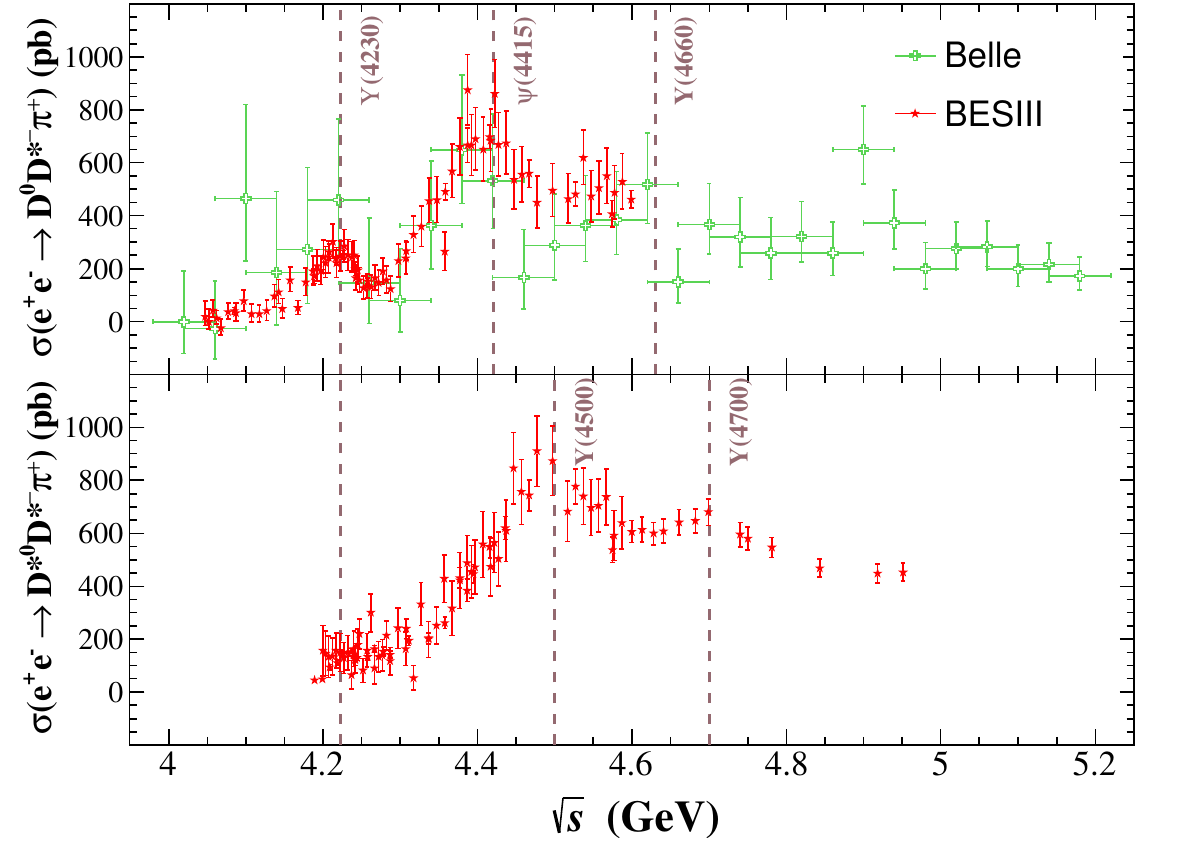}
\caption{ 
Comparisons of cross-sections for the $e^+e^-\to \pi^+D^{0}D^{\ast-}$ (top panel) and $\pi^+D^{\ast0}D^{\ast-}$ reactions  (bottom panel) as a function of c.m. energy from 4 to 5 GeV between Belle~\cite{Belle:2009dus} and BESIII~\cite{BESIII:2018iea}.
}
\label{BCS:piD0Dsm}
\end{center}
\end{figure*}

\begin{figure}[!htbp]
\begin{center}
\includegraphics[width=0.48\textwidth]{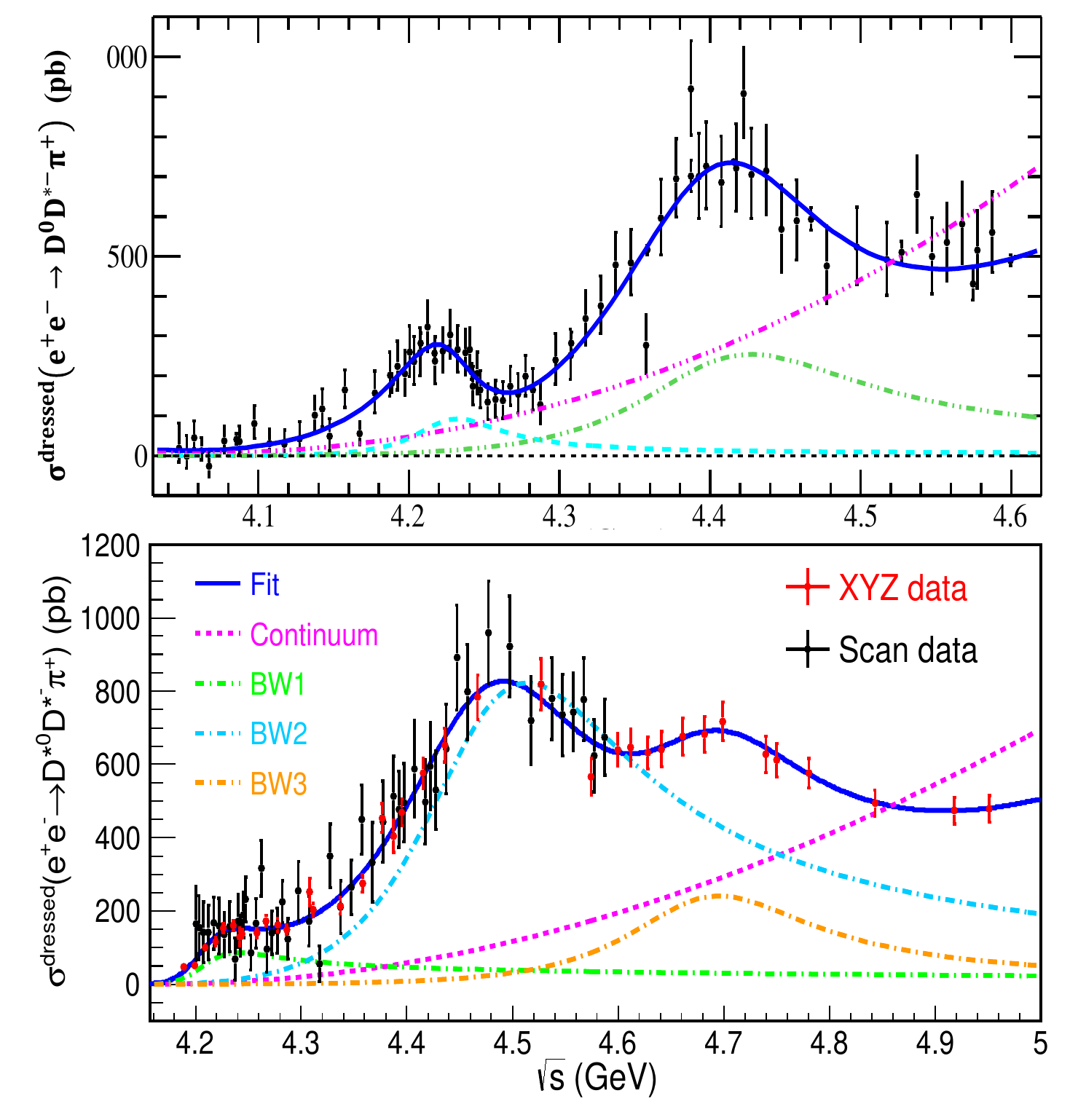}
\caption{Fits to the dressed cross-sections for the reactions of $e^+e^-\to \pi^+D^{0}D^{\ast-}$ (top) and $\pi^+D^{\ast0}D^{\ast-}$ (bottom) from BESIII data~\cite{BESIII:2018iea, BESIII:2023cmv}. Note that the datasets correspond to an integrated luminosity of 17.9 pb$^{-1}$~\cite{BESIII:2017lkp,BESIII:2020eyu,BESIII:2022ulv},where so-called $XYZ$ data include 
49
energy points with integrated luminosities less than 10 pb$^{-1}$ and scan data includes another 
37
energy points with larger integrated luminosities.
}
\label{BCS:piD0Dsp:fit}
\end{center}
\end{figure}

As the first charmoniumlike state with $J^{PC} =1^{--}$, $Y(4230)$ is only about 30 MeV/$c^2$ below the nominal threshold for
$D\bar{D}_{1}(2420)$; its nature remains a mystery.
The mass of the resonance referred to as $Y(4230)$ is consistent
with the prediction of the $D\bar{D}_{1}(2420)$ molecule model~\cite{Wang:2013cya,Cleven:2013mka,Qin:2016spb}. 
The production of $e^+e^-\to\pi^{+}D^{(\ast)0}D^{(\ast)-}$ is expected to be strongly enhanced above the nominal $D\bar{D}_{1}(2420)$ threshold and could be a key to understanding existing puzzles with these $Y$ states.
The cross-section of $e^+e^-\to \pi^+D^{0}D^{\ast-}$ was first measured by the Belle experiment using ISR~\cite{Belle:2009dus} and no evidence for charmonium(like) states was
found within the statistics limitation. A precise measurement of the cross-section of the $e^+e^-\to \pi^+D^{0}D^{\ast-}$ reaction at 84 c.m. energies from 4.05 to 4.60 GeV was presented~\cite{BESIII:2018iea}. Two enhancements are clearly visible in the cross-section around 4.23 and 4.40 GeV, as shown in Fig.~\ref{BCS:piD0Dsm} (top panel).
After the dressed cross-section of $e^+e^-\to \pi^+D^{0}D^{\ast-}$ has been modeled with several fit models, as shown in Fig.~\ref{BCS:piD0Dsp:fit} (top panel), it yields stable parameters for the first enhancement, which has a
mass of $4228.6\pm 4.1\pm 6.3$ MeV/$c^2$ and a width of $77.0 \pm 6.8 \pm 6.3$ MeV.
The resonance parameters obtained from the fit of cross-section of $e^+e^-\to \pi^+D^{0}D^{\ast-}$ are consistent with previous observations of the 
$Y(4230)$
state and the theoretical prediction of a $D\bar{D}_{1}(2420)$ molecule.
This result is the first observation of 
$Y(4230)$
associated with an open-charm final state.
The second enhancement, which has a mass of $4404.7 \pm 7.4$ MeV/$c^2$ and a width of $191.9 \pm 13.0$ MeV, 
is not from a single known resonance based on a current fit as shown in Fig.~\ref{BCS:piD0Dsp:fit}(top panel).
It could contain contributions from $\psi(4415)$ and other resonances, and a detailed amplitude analysis is required to better understand this enhancement.

Meanwhile, the BESIII experiment reported a new charmoniumlike resonance $Y(4500)$ in the reaction of $e^+e^-\to K^+K^-J/\psi$~\cite{BESIII:2022joj}.
The mass and width are determined as $M = (4484.7\pm13.3\pm24.1$) MeV/$c^2$ and $\Gamma = (111.1\pm30.1\pm15.2$) MeV, respectively.
This new resonance, called $Y(4500)$,
was proposed by the prediction of a $c\bar{c}s\bar{s}$ tetraquark state in the lattice QCD calculation~\cite{Chiu:2005ey}, a baryonium state~\cite{Qiao:2007ce},
the $5S$-$4D$ mixing scheme~\cite{Wang:2019mhs,Wang:2022jxj}, a hidden-charm candidate tetraquark in the QCD sum rule~\cite{Wang:2021qus},
and a heavy-antiheavy hadronic molecule state~\cite{Dong:2021juy,Peng:2022nrj}.
Refs.~\cite{Wang:2023zxj,Peng:2024blp} discussed how the unquenched potential model also indicated the good agreement between the experimental observable and the characterized energy level structure.
More explorations are highly desirable in both experimental measurement and theoretical studies to reveal its nature, especially in open-charm final states. Considering that, the BESIII experiment performed a precise measurement of the production cross-section of the $e^+e^-\to \pi^+D^{\ast0}D^{\ast-}$ reaction at 86 c.m. energies between 4.19 and 4.95 GeV including charge-conjugate mode, using a data sample corresponding to a total integrated luminosity of 18 fb$^{-1}$~\cite{BESIII:2023cmv}. 
Here only one $D^{\ast0}(D^{\ast-})$ candidate is reconstructed in the $D^{0}(D^{-})\pi^0$ 
channel
 with the decay of $D^0\to K^-\pi^+$, $K^-\pi^+\pi^0$ and $K^-\pi^+\pi^+\pi^-$, and $D^-\to K^-\pi^+\pi^-$, respectively, while antiparticle is extracted from the recoil side, which has included the charge-conjugate mode.
Three enhancements are visible in the line shape of the cross-section of $e^+e^-\to \pi^+D^{\ast0}D^{\ast-}$ around 4.20, 4.47, and 4.67 GeV, as shown in Fig.~\ref{BCS:piD0Dsm} (bottom panel). 
By performing a fit to the dressed cross-section of $e^+e^-\to \pi^+D^{\ast0}D^{\ast-}$
that takes vacuum polarizations into account~\cite{Jin:2018kjv} as shown in Fig.~\ref{BCS:piD0Dsp:fit} (bottom panel), one can determine the resonance parameters for three enhancements to be 
4209.6 $\pm$ 4.7 $\pm$ 5.9, 4469.1 $\pm$ 26.2 $\pm$ 3.6  and 4675.3 $\pm$ 29.5 $\pm$ 3.5 MeV/$c^{2}$ and widths of 81.6 $\pm$ 17.8 $\pm$ 9.0, 246.3 $\pm$ 36.7 $\pm$ 9.4, and 218.3 $\pm$ 72.9 $\pm$ 9.3 MeV, respectively.
The first and third resonances are consistent with the masses and widths of the $\psi(4230)$ and $\psi(4660)$ states, respectively, while the second one is compatible with the $Y(4500)$ observed in the $e^{+}e^{-}\to K^{+}K^{-}J/\psi$ process. These three charmoniumlike states are observed in the $e^{+}e^{-}\to D^{*0}D^{*-}\pi^{+}$ process for the first time, where the resonance parameters $Y(4500)$ are compatible with those observed in $e^+e^-\to K^+K^-J/\psi$~\cite{BESIII:2022joj}.
While the rate of its decay to $\pi^+D^{\ast0}D^{\ast-}$ is 2 orders of magnitude higher than that to $K\bar{K}J/\psi$, which is inconsistent with the conjectured hidden-strangeness tetraquark nature of the
$Y(4500)$~\cite{Chiu:2005ey,Dong:2021juy, Peng:2022nrj}. 
Further amplitude analyses of different open- and hidden-charm final states are desired to advance our knowledge of the nature of these charmoniumlike states.

\subsubsection{$e^+e^-\to \pi^+\pi^-D^+D^-$ }
\begin{figure}[!htbp]
\begin{center}
\includegraphics[width=0.5\textwidth]{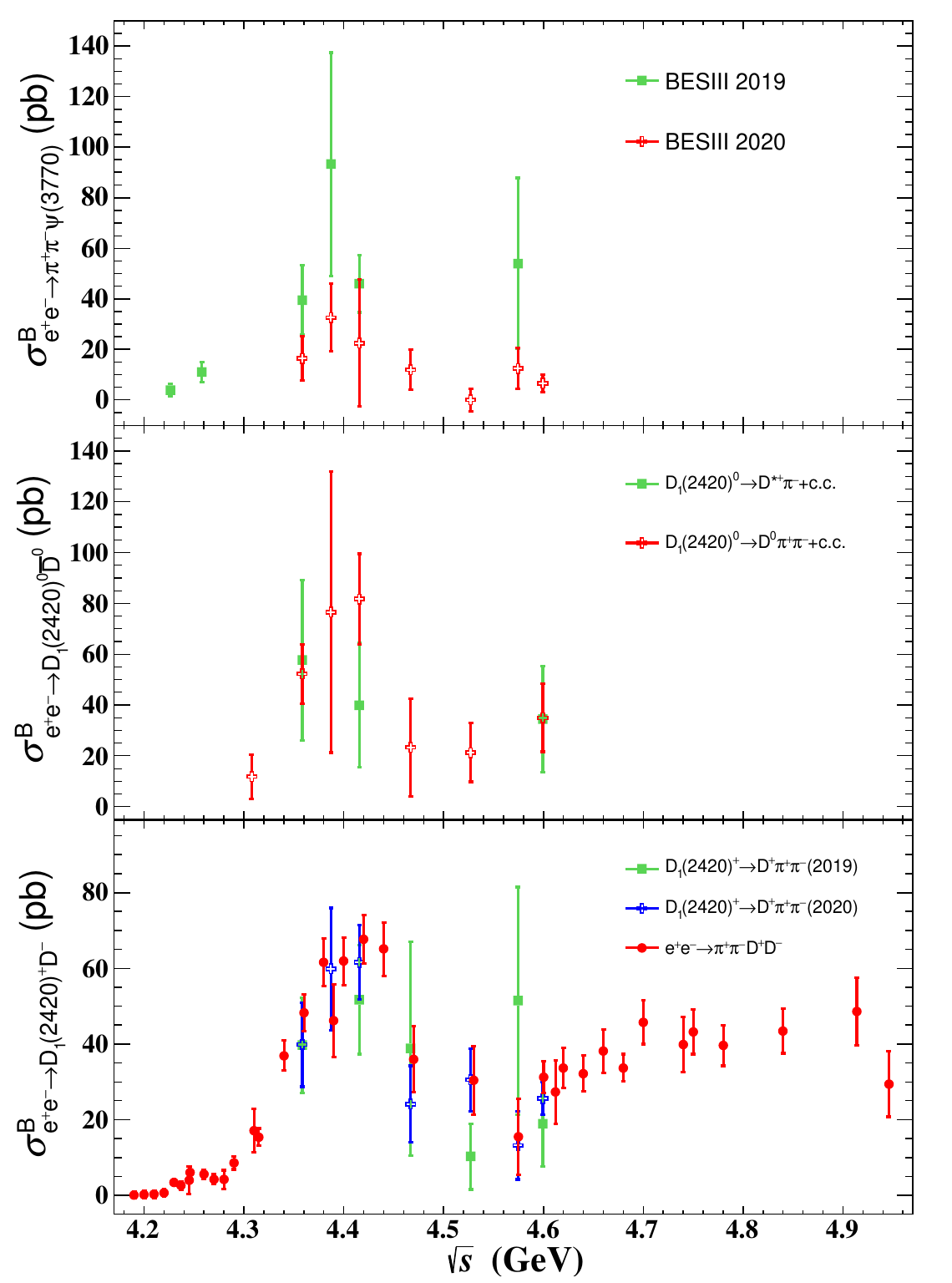}
\caption{Comparisons of Born cross-sections for $e^+e^-\to \pi^+\pi^-\psi(3770)\to\pi^+\pi^-D^{+}D^{-}$,  $e^+e^-\to \pi^+\pi^-D^{+}D^{-}$ and $e^+e^-\to D_{1}(2420)\bar{D}\to \pi^+\pi^-D\bar{D}$ as a function of c.m. energy~\cite{BESIII:2019tdo, BESIII:2019phe}. }
\label{BCS:BCS_pipiDD:ddpi}
\end{center}
\end{figure}

\begin{figure}[!htbp]
\begin{center}
\includegraphics[width=0.5\textwidth]{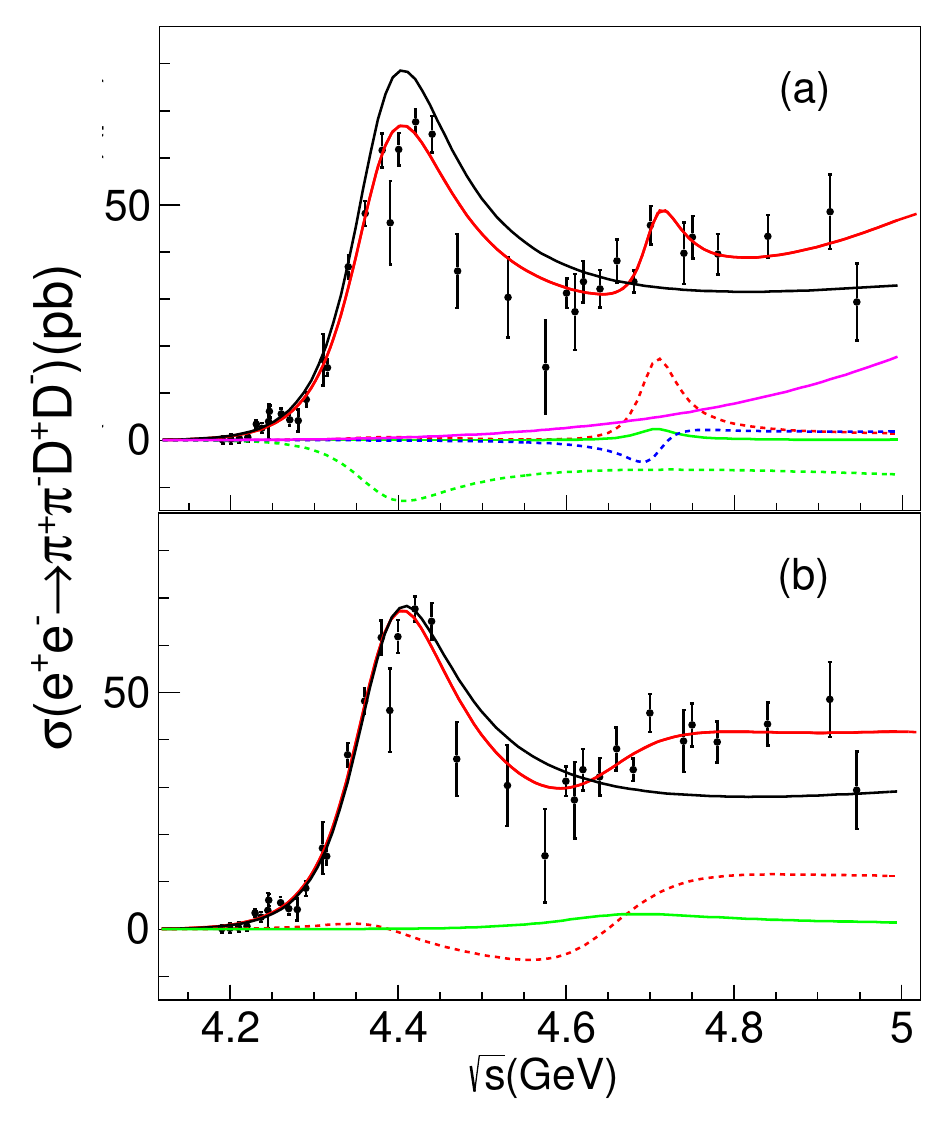}
\caption{
Fits to the dressed cross-section  for the reaction of $e^+e^-\to \pi^+\pi^-D^{+}D^{-}$~\cite{BESIII:2022quc} (a) with the coherent sum of two BW functions and a PHSP term (b) with the coherent sum of two BW functions only.
The dots with error bars are data with
the statistical uncertainties and the red lines show the best fit results. In (a) the black, green, and pink solid lines describe different BWs and PHSP components, respectively,
and the red, green, and blue dashed lines describe interferences between the different resonances and PHSP, respectively.
In (b) the black and green solid lines describe BW components, respectively, and the red dashed line describes the inteference between different BWs.
}
\label{BCS:pipiDmDp:fit}
\end{center}
\end{figure}

Despite the numerous studies that have been conducted to measure the cross-sections for two-body final states with a pair of charmed mesons and three-body final states with a pair of charmed mesons plus a light meson, there is still a lack of experimental information in this area, especially in multibody open-charm final states. This scarcity of data hinders our understanding of the dynamics and interactions involved in these processes. Therefore, further experimental investigations are needed to fill this gap and provide more comprehensive insights into the behavior of charmed mesons in various final-state configurations. Recently, the BESIII experiment reported the first measurement of a Born cross-section for the $e^+e^-\to \pi^+\pi^-D^{+}D^{-}$ reaction at 37 c.m. energies from 4.19 to 4.95 GeV with a partial reconstruction method~\cite{BESIII:2022quc}.  
Figure~\ref{BCS:BCS_pipiDD:ddpi} illustrates the Born cross-sections measured of $e^+e^-\to \pi^+\pi^-\psi(3770)\to\pi^+\pi^-D^{+}D^{-}$ and $e^+e^-\to D_{1}(2420)\bar{D}\to \pi^+\pi^-D\bar{D}$ compared to that cross section of $e^+e^-\to \pi^{+}\pi^{-} D^{+} D^{-}$,
where two clear peaks around the masses of 4.4 and 4.7 GeV
can be seen in the line shape of the Born cross-section of the
reaction $e^+e^-\to \pi^+\pi^-D^{+}D^{-}$.
Performing a fit to the dressed cross-section of $e^+e^-\to \pi^+\pi^-D^{+}D^{-}$ as shown in Fig.~\ref{BCS:pipiDmDp:fit} yields a significant charmoniumlike resonance with
the mass of (4373.1 $\pm$ 4.0 $\pm$ 2.2)~MeV/$c^2$ and the width of (146.5 $\pm$ 7.4 $\pm$ 1.3)~MeV,  which is in agreement with $Y(4390)$.
There is an evidence with a statistical significance of 4.1$\sigma$ for a second resonance, called $R(4700)$, with a mass of (4706 $\pm$ 11 $\pm$ 4) MeV/$c^2$ and
a width of (45 $\pm$ 28 $\pm$ 9) MeV.

\begin{figure}[!htbp]
\begin{center}
\includegraphics[width=0.48\textwidth]{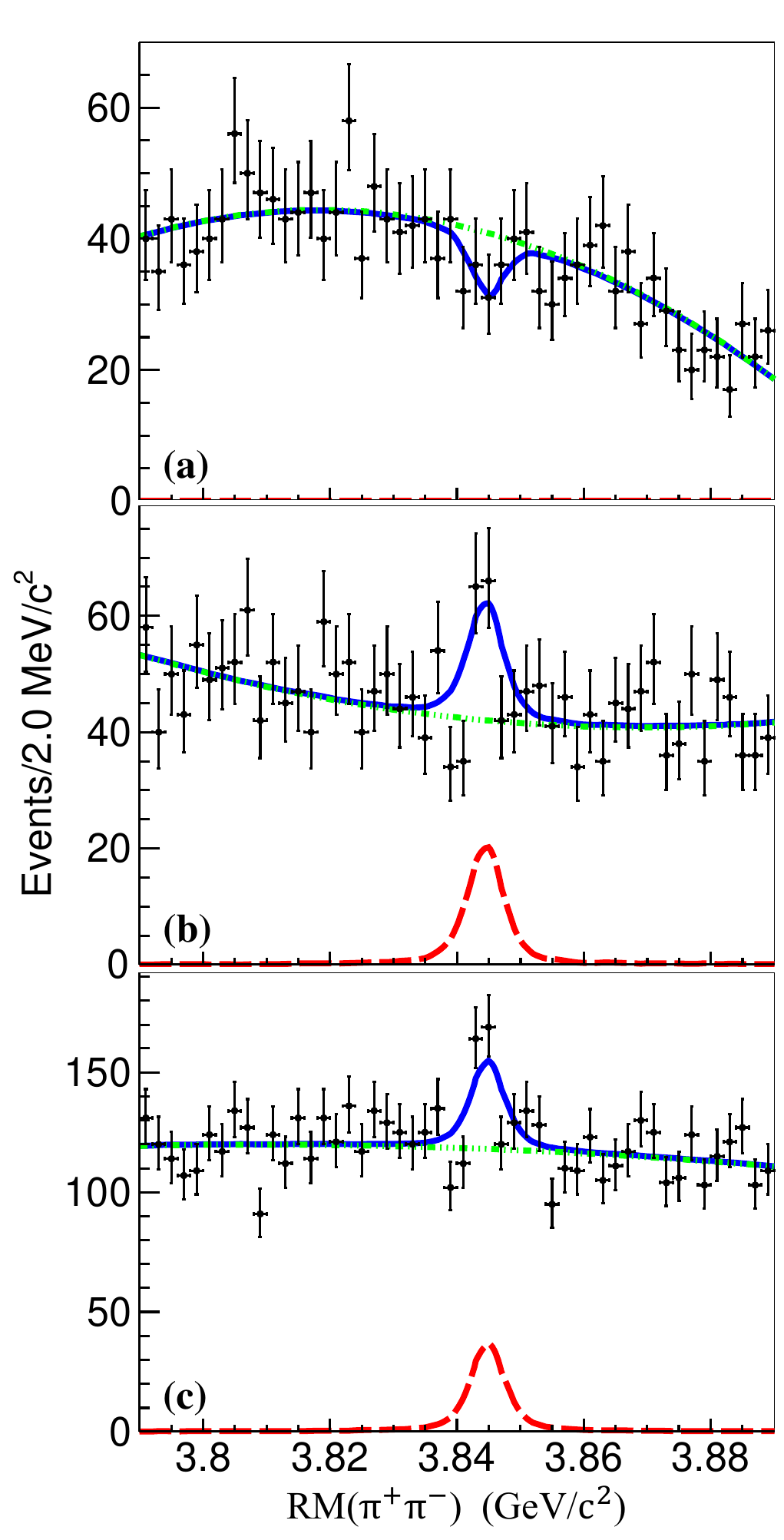}
\caption{Fit to the recoil mass spectra of $\pi^+\pi^-$ system
~\cite{BESIII:2022quc}~(a)
at $\sqrt{s}=$4.420, (b) $\sqrt{s}=$4.680, and~(c) for data samples
with sum of all energy point at $\sqrt{s}= 4.6-4.7$~GeV.
The black dots with error bars are the data sample, and the dashed red, dash-dotted green, and solid blue curves are the
$X(3842)$ shape, background shape, and total fit,
respectively. }
\label{BCS:pipirecoil:fit}
\end{center}
\end{figure}

Furthermore, the search for spin-$3$ $D$-wave charmoniumlike state $X(3842)$ which was first observed in the $D\bar{D}$ mode via the LHCb experiment~\cite{LHCb:2019lnr},
was conducted using the process of $e^+e^-\to \pi^+\pi^-X(3842)\to\pi^+\pi^-D^{+}D^{-}$. Note that unlike all the other spectra, a recoil mass is studied here.
The evidence for this state was found with a significance of $4.2\sigma$ when analyzing all data samples in the c.m. energy range from 4.6 to 4.7 GeV, as depicted in Fig.~\ref{BCS:pipirecoil:fit}. This finding adds valuable information to our understanding of the nature of $X(3842)$ and contributes to ongoing research in related fields.

Meanwhile, $D_{1}(2420)^+$ was also investigated in the mass spectrum of the $D^+\pi^+\pi^-$ system in the $e^+e^-\to \pi^+\pi^-D^{+}D^{-}$ reaction using data collected at $\sqrt{s} = 4.09 -4.60$ GeV~\cite{BESIII:2019tdo, BESIII:2019phe}. The mass and width of $D_{1}(2420)^+$ were determined by fitting the distribution of recoil mass against $D^{+}$, which yields a mass of (2427.2 $\pm$ 1.0 $\pm$ 1.2) MeV/$c^{2}$ and a width of (23.2 $\pm$ 2.3 $\pm$ 2.3) MeV~\cite{BESIII:2019phe}, representing improved precision compared to previous measurements. These results contribute to better constraining uncertainties in theoretical calculations related to molecular explanations for states such as $Y(4230)$ and $Z_c(4430)$, specifically those involving final states of $D_1(2420)\bar{D}^{(\ast)}$~\cite{Meng:2007fu, Wang:2013cya, Ma:2014zua, Chen:2016qju, Liu:2019zoy}. Furthermore, this study marks the first measurement of Born cross-sections for processes such as $e^+e^-\to D_{1}(2420)\bar{D}\to \pi^+\pi^-D\bar{D}$ and $e^+e^- \to \psi(3770) \pi^+ \pi^- \to D^+ D^- \pi^+ \pi^-$. 

\subsection{Charmed-strange meson pair}
\subsubsection{$e^+e^-\to D^{+}_{s}D^{-}_{s}$}
\begin{figure*}[!htbp]
\begin{center}
\includegraphics[width=1.01\textwidth]{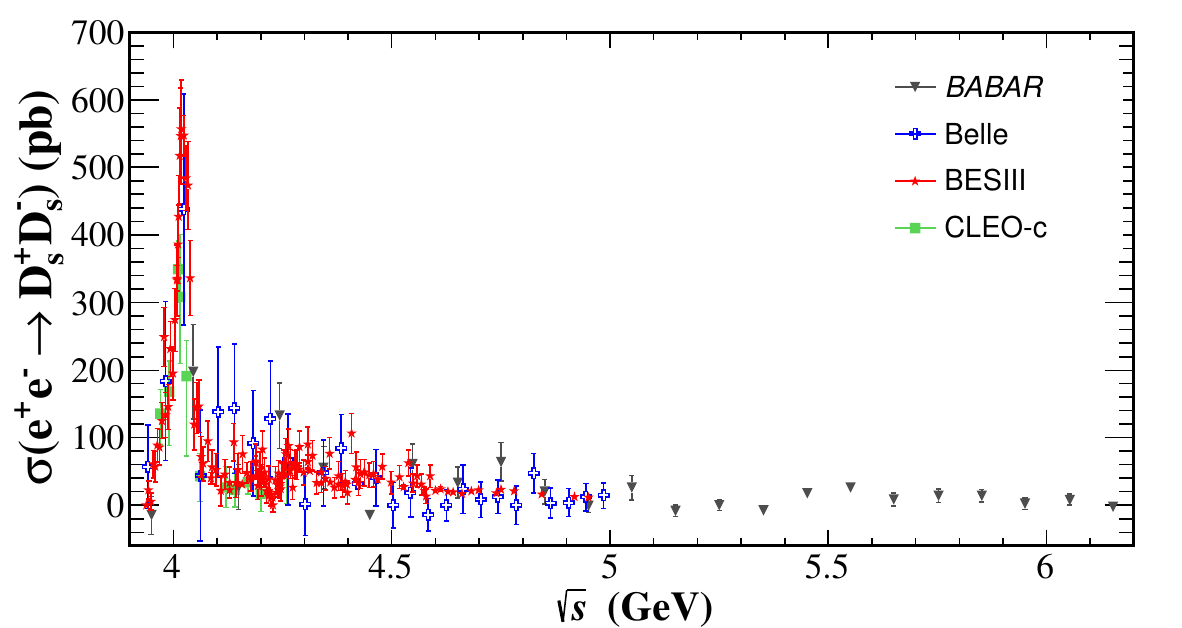}
\put(-340,85){\includegraphics[width=0.6\textwidth]{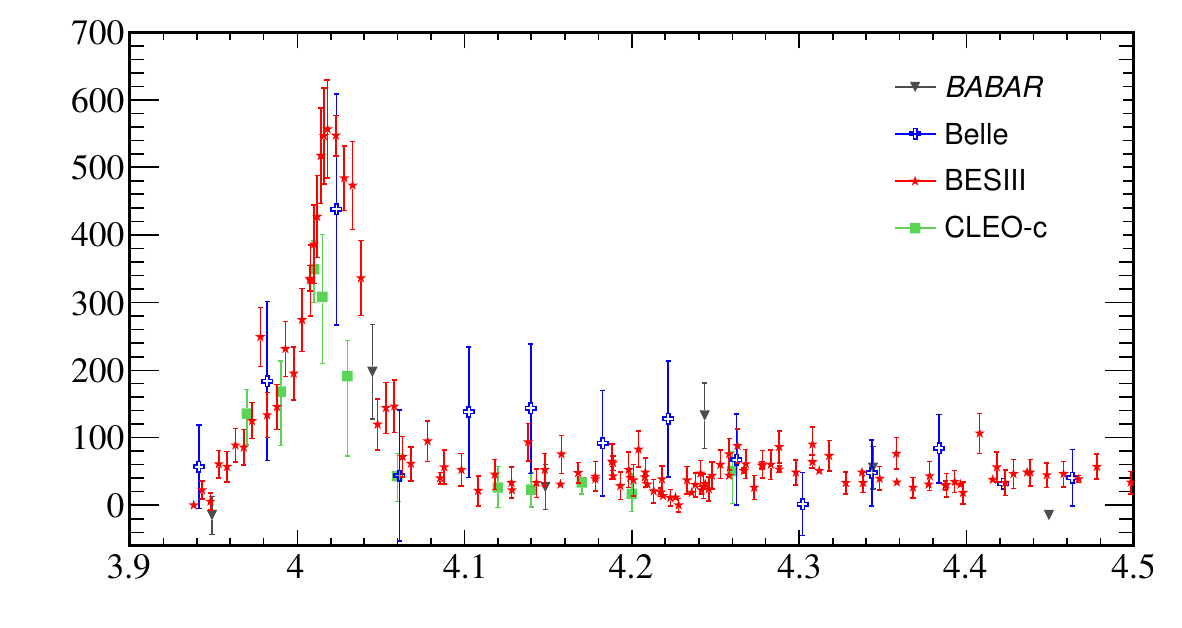}}\\
\caption{ 
Comparison of cross-sections for the $e^+e^-\to D^{+}_{s}D^{-}_{s}$ reaction as a function of c.m. energy from 3.95 to 6.20 GeV from different experiments of {\it BABAR}~\cite{BaBar:2010plp}, Belle~\cite{Belle:2010fwv}, BESIII ~\cite{BESIII:2024zdh}and CLEO-c~\cite{CLEO:2008ojp}.
}
\label{BCS:DDbar02:dsds}
\end{center}
\end{figure*}
The $D_s$ meson is a composite particle consisting of a charm and a strange quark, providing crucial experimental insights into the nature of charmoniumlike states in modern physics through precise measurements of exclusive production cross-sections for the $D^{+}_{s}D^{-}_{s}$ pair.
Although the {\it BABAR} and Belle experiments have conducted measurements of the exclusive cross-sections for $D_s^+D_s^-$ pairs production through ISR processes~\cite{Belle:2010fwv,BaBar:2010plp}, the precision is insufficient to determine the $Y(4230)$ contribution into the process $e^+e^-\to D_s^+D_s^-$. The CLEO-c experiment investigated the process $e^+e^-\to D_s^+D_s^-$ using energy scan data, with a maximum c.m. energy of only 4.26 GeV~\cite{CLEO:2008ojp}.
Recently, the BESIII experiment reported a more precise measurement of Born cross-section of reaction $e^+e^-\to D_s^+D_s^-$ at 138 c.m. energies from 3.94 to 4.95~GeV, corresponding to an integrated luminosity of $23~\textrm{fb}^{-1}$~\cite{BESIII:2024zdh}.
In the event selection of the BESIII experiment, only the $D_{s}$ meson is reconstructed with the 
decays
$D_s^- \rightarrow K^+K^-\pi^-$, 
while the $D_s^+$ is not reconstructed exclusively but is inferred 
from the recoil mass. The charge-conjugated modes are implicit throughout the analysis by default.
Figure~\ref{BCS:DDbar02:dsds} shows the measured Born cross-section of $e^+e^-\to D_s^+D_s^-$ compared with the previous measurements from the {\it BABAR}, Belle, and CLEO-c experiments.
The resulting cross-section reveals several new structures, including a significantly narrower 
structure around the mass range of $\psi(4040)$ and a dip around the $D_s^{*+}D_s^{*-}$ threshold, as well as a broad peak around the mass range of $Y(4230)$, indicating the influence of the open channel effect. Measurements in the reaction of $e^+e^-\to D_s^+D_s^-$ provide a valuable input for coupled-channel analysis and model tests, which are crucial to understanding charmoniumlike states with masses between 4 and 5~GeV. 
\subsubsection{$e^+e^-\to D^{\ast+}_{s}D^{(\ast)-}_{s}$}
\begin{figure*}[!htbp]
\begin{center}
\includegraphics[width=1.01\textwidth]{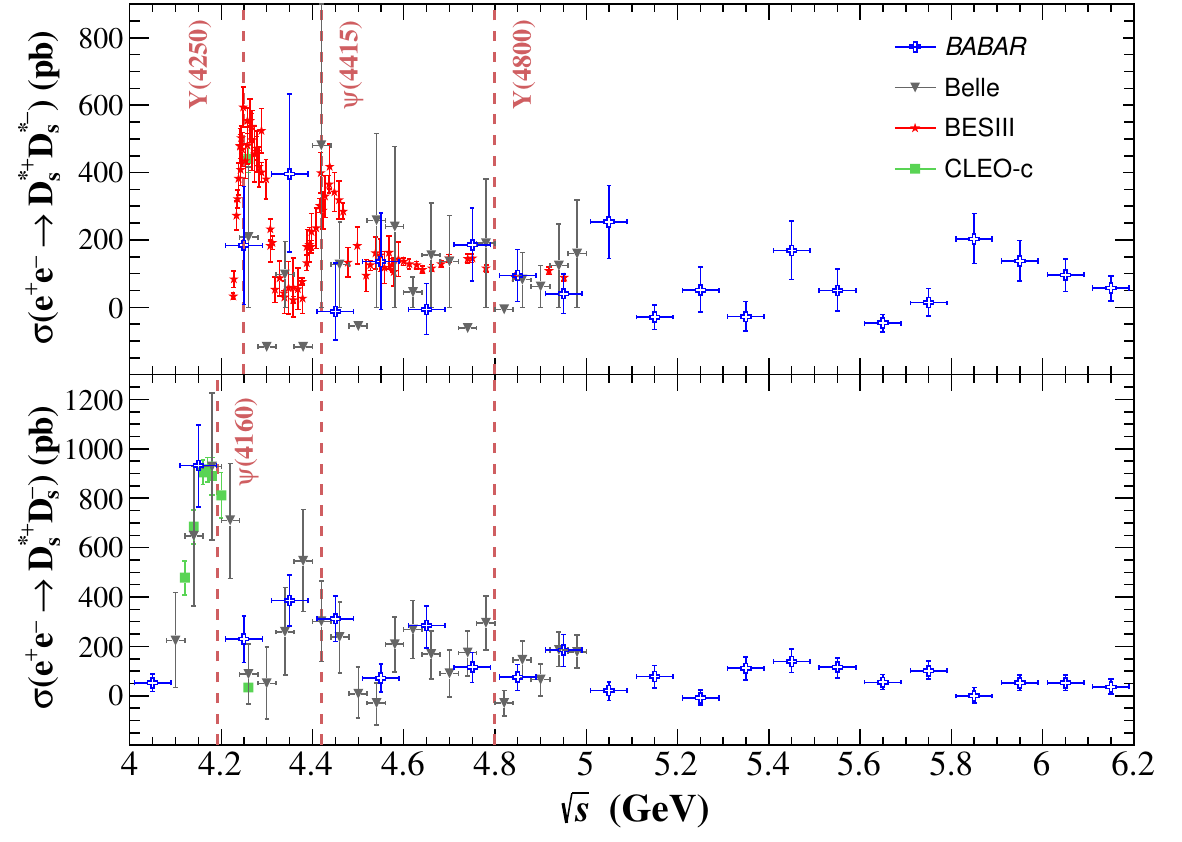}
\caption{ 
Comparisons of cross-sections for the $e^+e^-\to D^{\ast+}_{s}D^{\ast-}_{s}$ (top panel) and $D^{\ast+}_{s}D^{-}_{s}$ reactions (bottom panel) as a function of c.m. energy from 4.0 to 6.2 GeV from different experiments of {\it BABAR}~\cite{BaBar:2010plp}., Belle~\cite{Belle:2010fwv}., BESIII~\cite{BESIII:2023wsc} and CLEO-c~\cite{CLEO:2008ojp}.
}
\label{BCS:DssDssbar}
\end{center}
\end{figure*}
\begin{figure}[!htbp]
\begin{center}
\includegraphics[width=0.48\textwidth]{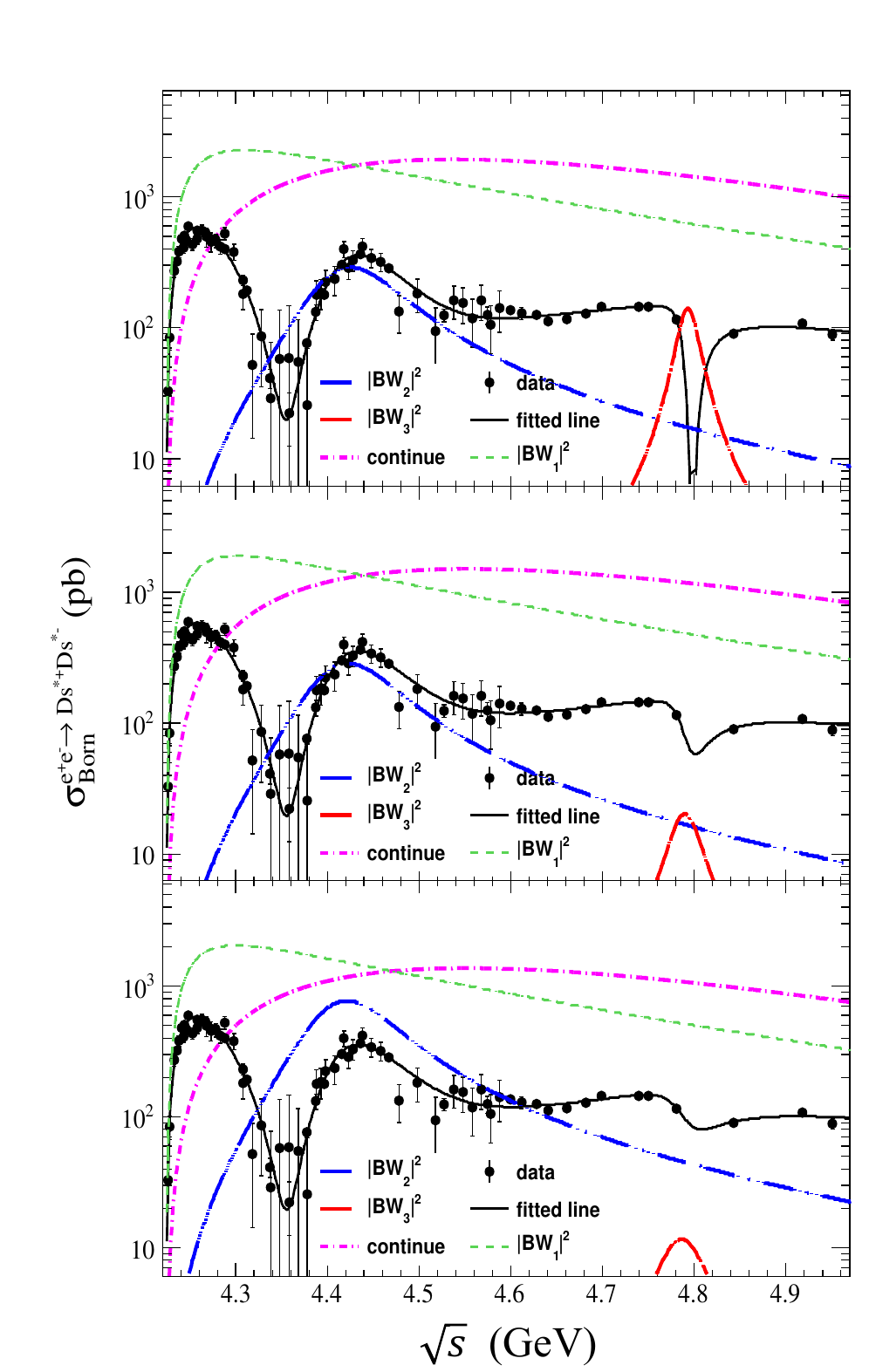}
\caption{Fit to the measured Born cross-sections of $e^{+}e^{-}\rightarrow D_{s}^{\ast+}D_{s}^{\ast-}$ with three best fits
~\cite{BESIII:2023wsc}.
 The dots with error bars are for the measured Born cross-sections.
The black curves represent the fit, the dashed green, two-dashed blue and long-dashed red ones are for the three BW fits, respectively,
 and the dot-dashed pink curves is for the PHSP contributions.
 }
\label{BCS:DPrecoil:BCS}
\end{center}
\end{figure}

The mass of the $Y(4230)$ state is positioned just at the production threshold of the $D_s^{\ast+}D_s^{\ast-}$ pair, indicating a potential correlation between $Y(4230)$ and this open-charm decay mode.
Previously, the {\it BABAR}, Belle and CLEO-c experiments performed measurements of exclusive cross-sections for $e^+e^-\to D^{\ast+}_{s}D^{\ast-}_{s}$ with limited statistics and energy ranges~\cite{Belle:2010fwv,BaBar:2010plp,CLEO:2008ojp}.
To further pin down the nature of charmonium(like) states such as $Y(4230)$,  a more precise measurement of $e^{+}e^{-}\rightarrow D_{s}^{\ast+}D_{s}^{\ast-}$ was performed by BESIII
with a semi-inclusive method using data samples at 76 c.m. energies from threshold to 4.95~GeV corresponding to an integrated luminosity of $16~\textrm{fb}^{-1}$~\cite{BESIII:2023wsc}. Here only the $D_s^{\ast+}$ meson in the $e^+e^-\to D_s^{\ast+}D_s^{\ast-}$ reaction is reconstructed with the decays of $D_s^{\ast+}\to\gamma D_s^{+}$ and $D_s^{+}\to K^+K^-\pi^+$.
Two resonance structures are visible
in the energy-dependent cross-sections around 4.25 and 4.44~GeV as shown in Fig.~\ref{BCS:DssDssbar}. 
When the dressed cross-sections with a coherent sum of three BW amplitudes and one PHSP amplitude as shown in Fig.~\ref{BCS:DPrecoil:BCS}, 
these two significant structures 
have masses measured as (4186.5 $\pm$ 9.0 $\pm$ 30) and (4414.5 $\pm$ 3.2 $\pm$ 6.0)~MeV/$c^{2}$, the widths of (55 $\pm$ 17 $\pm$ 53) and (122.6 $\pm$ 7.0 $\pm$ 8.2)~MeV, where the first errors are statistical and the second ones are systematic~\cite{BESIII:2023wsc}.  
The parameters for the first resonance are consistent with those of $\psi(4160)$ when systematic uncertainties are considered; the state is also
consistent with the $Y(4230)$ observed in the $\pi^+\pi^-J/\psi$ mode. If the contribution is from $Y(4230)$ parameters, it indicates that
$Y(4230)$ is more strongly associated with the $D_s^{\ast+}D_s^{\ast-}$ mode than with the modes with charmonium states since the cross-section of $e^+e^-\to D_s^{\ast+}D_s^{\ast-}$ at 4.23~GeV is roughly 1 order of magnitude higher than that of $e^+e^-\to \pi^+\pi^-J/\psi$. This information is vital to understand the nature of $Y(4230)$.
The mass and width for the second resonance are consistent with the $\psi(4415)$ charmonium state. This could be the first time that we have observed $\psi(4415)$ in the $D_s^{\ast+}D_s^{\ast-}$ mode. Here the parameters for all assumed resonances strongly depend on the chosen fit model and indicating the need for further in-depth research, such as a unitary approach based on the $K$-matrix formalism to fit the cross-section results of various exclusive channels~\cite{Husken:2024hmi}.

In addition, the presence of a third BW amplitude is crucial to accurately characterize the complex structure observed at approximately 4.79 GeV. This additional amplitude provides essential information for understanding the behavior and properties of the system in this energy range. Without it our description of the structure would be incomplete and insufficient to fully capture its dynamics. Therefore, including a third BW amplitude is not just beneficial- it is necessary for a comprehensive analysis of the phenomenon at hand.
Owing to the limited number of data points around 4.79 GeV, the uncertainty in the fitted mass of the third structure ranges from 4786 to 4793 MeV/$c^2$ and the width varies from 27 to 60 MeV. This wide range reflects the challenges of accurately determining these parameters with a small amount of data. It also highlights the need for additional experimental measurements and analysis to further constrain these values and improve our understanding of this energy region in particle physics. 

\begin{figure}[!htbp]
\begin{center}
\includegraphics[width=0.47\textwidth]{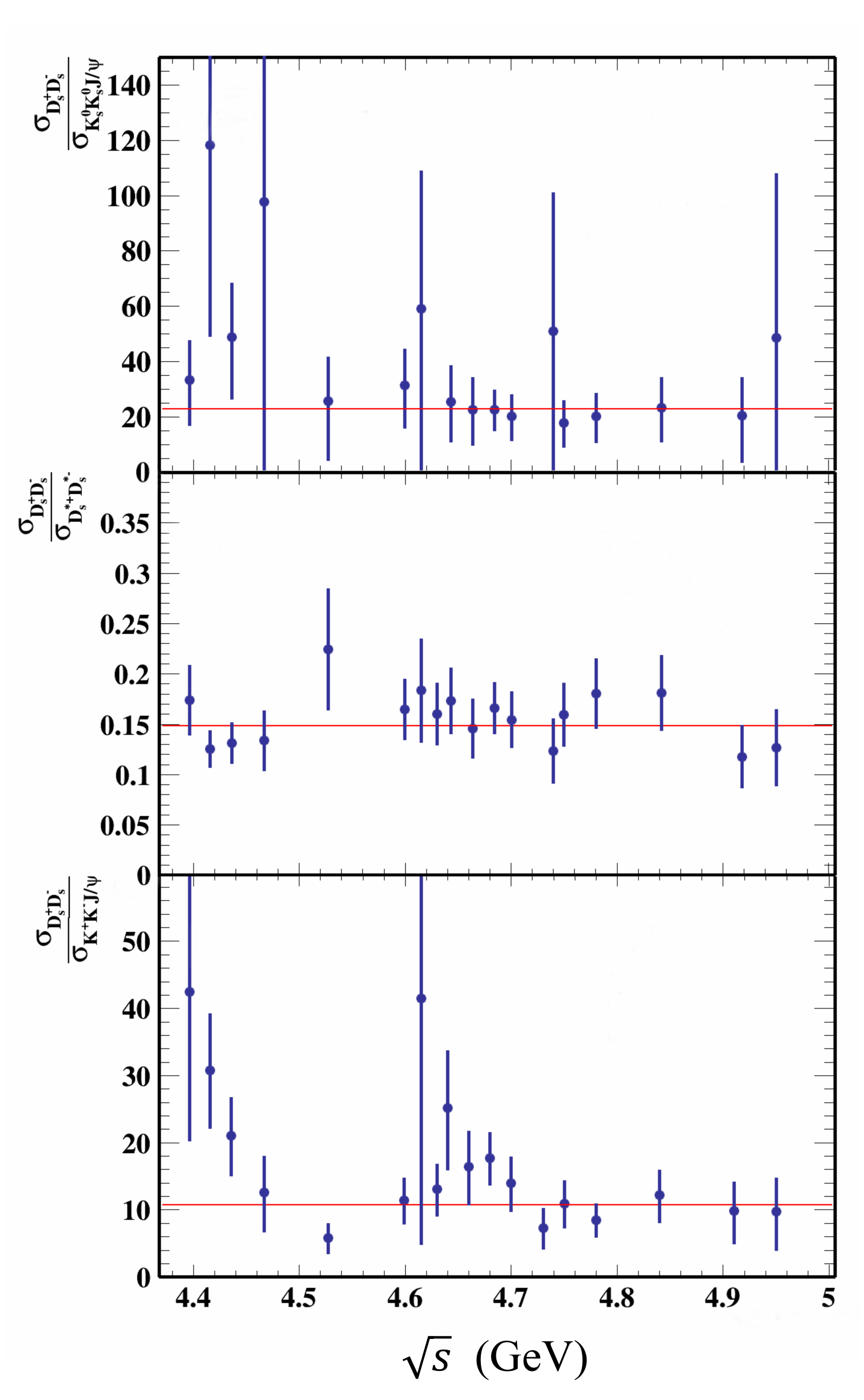}
\caption{Born cross-section ratios of $e^+e^-\to D_s^+D_s^-$ to $e^+e^-\to K^{0}_{S}K^{0}_{S}J/\psi$ (top panel), $D_s^{*+}D_s^{*-}$ (Middle panel) and $ K^+K^-J/\psi$ (bottom panel) as a function of c.m. energy between 4.35 and 5.00 GeV~\cite{BESIII:2024zdh} .}
\label{BCS:DPrecoil:BCS:ratio}
\end{center}
\end{figure}

In addition to determining Born cross-section ratios between $e^+e^-\to D_s^+D_s^-$ and $e^+e^-\to K^{0}_{S}K^{0}_{S}J/\psi$, as well as $e^+e^-\to D_s^{*+}D_s^{*-}$ and $e^+e^-\to K^+K^-J/\psi$ in the energy range of 4.35 to 4.95 GeV as shown in Fig.~\ref{BCS:DPrecoil:BCS:ratio}, further analysis has revealed two distinct structures that bear similarities to those observed in the processes of $e^+e^-\to D_s^{+}D_s^{-}$, $e^+e^-\to K^+K^-J/\psi$, and $e^+e^-\to K_S^{0}K_S^{0}J/\psi$~\cite{BESIII:2024zdh,BESIII:2022joj,BESIII:2022kcv}.

\subsubsection{$e^+e^-\to D^{+}_{s}D_{s1}(2536)^{-}/D^{\ast}_{s2}(2573)^{-}$
}
\begin{figure*}[!htbp]
\begin{center}
\includegraphics[width=1.01\textwidth]{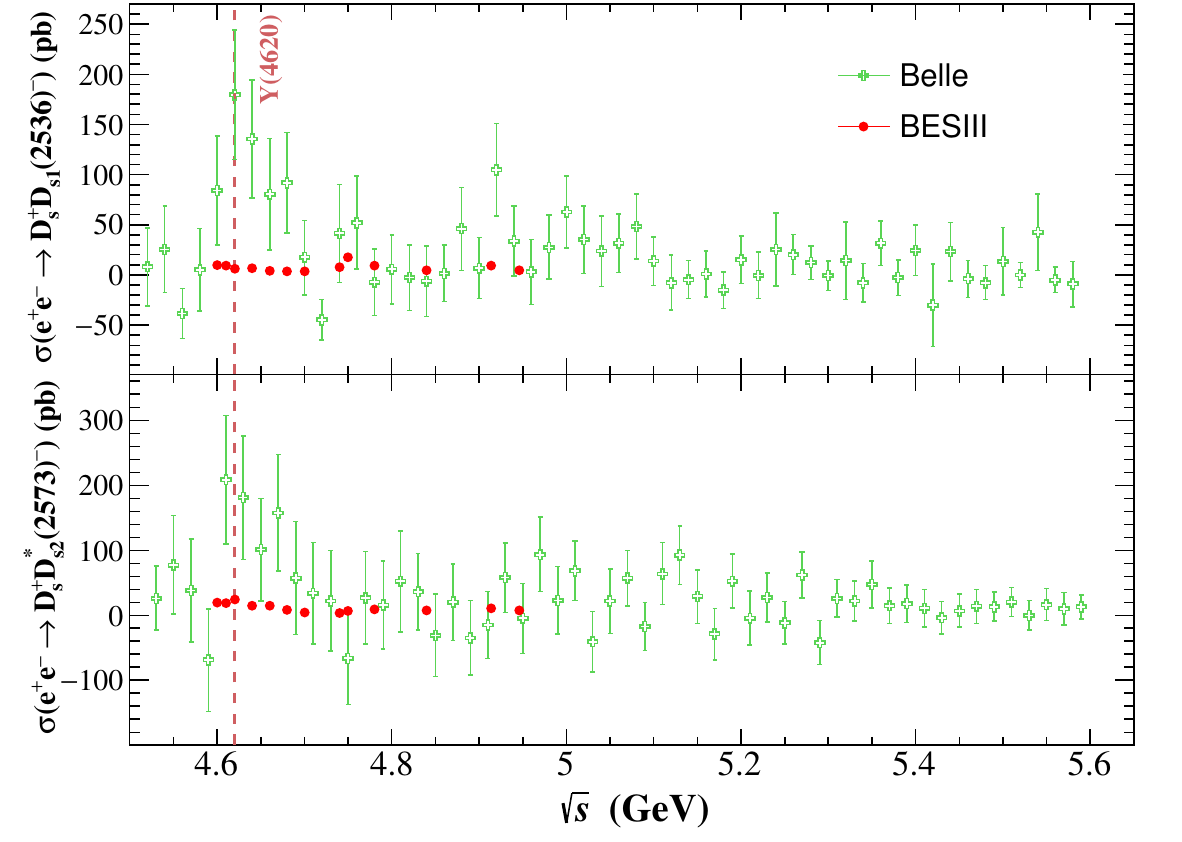}
\caption{ 
Comparisons of cross-sections for the $e^+e^-\to D^{+}_{s}D_{s1}(2536)^{-}$ (top panel) and $D^{+}_{s}D^{\ast}_{s2}(2573)^{-}$ reactions (bottom panel)  as a function of c.m. energy from 4.50 to 5.65 GeV between Belle~\cite{Belle:2019qoi, Belle:2020wtd} and BESIII~\cite{BESIII:2024qfi}.
}
\label{BCS:Ds1Ds2bar:BCS}
\end{center}
\end{figure*}
\begin{figure}[!htbp]
    \centering
    \includegraphics[width=0.49\textwidth]{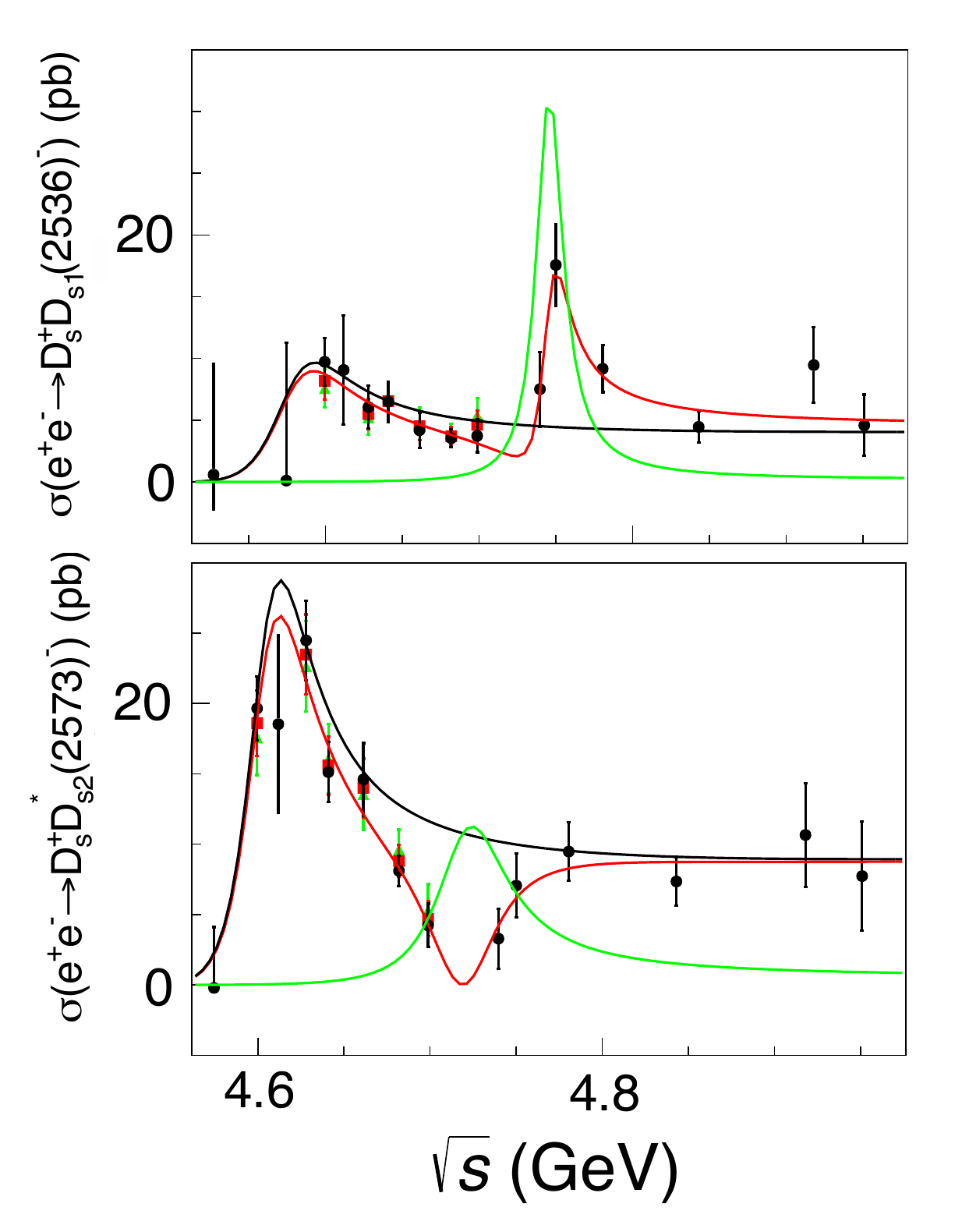}
\caption{Fits to cross-sections of the $e^+e^-\to D^{+}_{s}D_{s1}(2536)^{-}$ and $D^{+}_{s}D_{s2}(2573)^{-}$ reactions~\cite{BESIII:2024qfi}. The black dots, red squares, and green triangles with error bars are the measured cross-sections from the BESIII and Belle experiments. The red, black, and green solid lines are results of total fit, BW shapes. }
    \label{Fit_Ds2536:2573}
\end{figure}

Fairly recently, several high-mass charmonium(like) states were observed near the $D^{+}_{s}D_{s1}(2536)^{-}$ threshold, such as $Y(4500)$~\cite{BESIII:2022joj},  $Y(4790)$~\cite{BESIII:2023wsc} and $Y(4710)$~\cite{BESIII:2023wqy}.
Therefore, it is natural to perform a precise measurement of the Born cross-section of $e^+e^-\to D^{+}_{s}D_{s1}(2536)^{-}$ and $D^{+}_{s}D^{\ast}_{s2}(2573)^{-}$ to further investigate the nature of these charmoniumlike states.
Although this was previously done with the Belle experiment with the ISR process~\cite{Belle:2019qoi, Belle:2020wtd}, the statistics are limited and more precise analyses are needed. 
More recently, the BESIII experiment reported a more precise measurement of $e^+e^-\to D^{+}_{s}D_{s1}(2536)^{-}$ and $D^{+}_{s}D^{\ast}_{s2}(2573)^{-}$
using a data sample corresponding to an integrated luminosity of $6.6~\textrm{fb}^{-1}$ at 15 c.m. energies ranging from $4.53$ to $4.95$~GeV~\cite{BESIII:2024qfi}. 
Here only $D_s^+\to K^-K^+\pi^+$ is reconstructed, while $D_{s1}(2536)^{-}$ and 
$D^{\ast}_{s2}(2573)^{-}$
mesons are inferred in the recoil mass, and thus decay inclusively.
A clear resonance around 4.6 GeV for both reactions, along with a narrow peak around 4.75 GeV for the $e^+e^-\to D^{\ast+}_{s}D_{s1}(2536)^{-}$ process and a clear dip at around 4.71 GeV for $e^+e^-\to D^{+}_{s}D^{\ast}_{s2}(2573)^{-}$ are visible, as shown in Fig.~\ref{BCS:Ds1Ds2bar:BCS}.
when the cross-sections of $e^+e^-\to D^{+}_{s}D_{s1}(2536)^{-}$ and $D^{+}_{s}D_{s2}(2573)^{-}$ were fitted with high-precision measurements from the BESIII experiment as shown 
in Fig.~\ref{Fit_Ds2536:2573}, a resonant structure at around 4.6~GeV with a width of 50 MeV was observed for the first time in $e^+e^-\to D^{+}_{s}D_{s1}(2536)^{-}$, which is consistent
with the evidence reported by the Belle experiment for $Y(4620)$ in the same final state~\cite{Belle:2019qoi, Belle:2020wtd}. 
In addition, two additional resonances are found with widths of 25 MeV at around 4.75 GeV and 50 MeV at around 4.72 GeV in $e^+e^-\to D^{+}_{s}D_{s1}(2536)^{-}$ and $e^+e^-\to D^{+}_{s}D^{\ast}_{s2}(2573)^{-}$, respectively. These findings suggest that a common state at approximately 4.6 GeV decays into both the $D_{s}^{+}D_{s1}(2536)^-$ and $D_{s}^{+}D^{\ast}_{s2}(2573)^-$ final states. Moreover, our observations provide evidence for a structure at 4.75 GeV that may correspond to the $Y(4710)$ or $Y(4790)$ findings previously reported for the BESIII experiment~\cite{BESIII:2023wqy,BESIII:2023wsc}. 
This suggests potential connections between the resonances and opens up new avenues for exploring their properties and interactions within the further research.

\subsubsection{$e^+e^-\to D^{\ast+}_{s}D_{s0}(2317)^{-}/D_{s1}(2460)^{-}$}
\begin{figure}[!htbp]
\begin{center}
\includegraphics[width=0.5\textwidth]{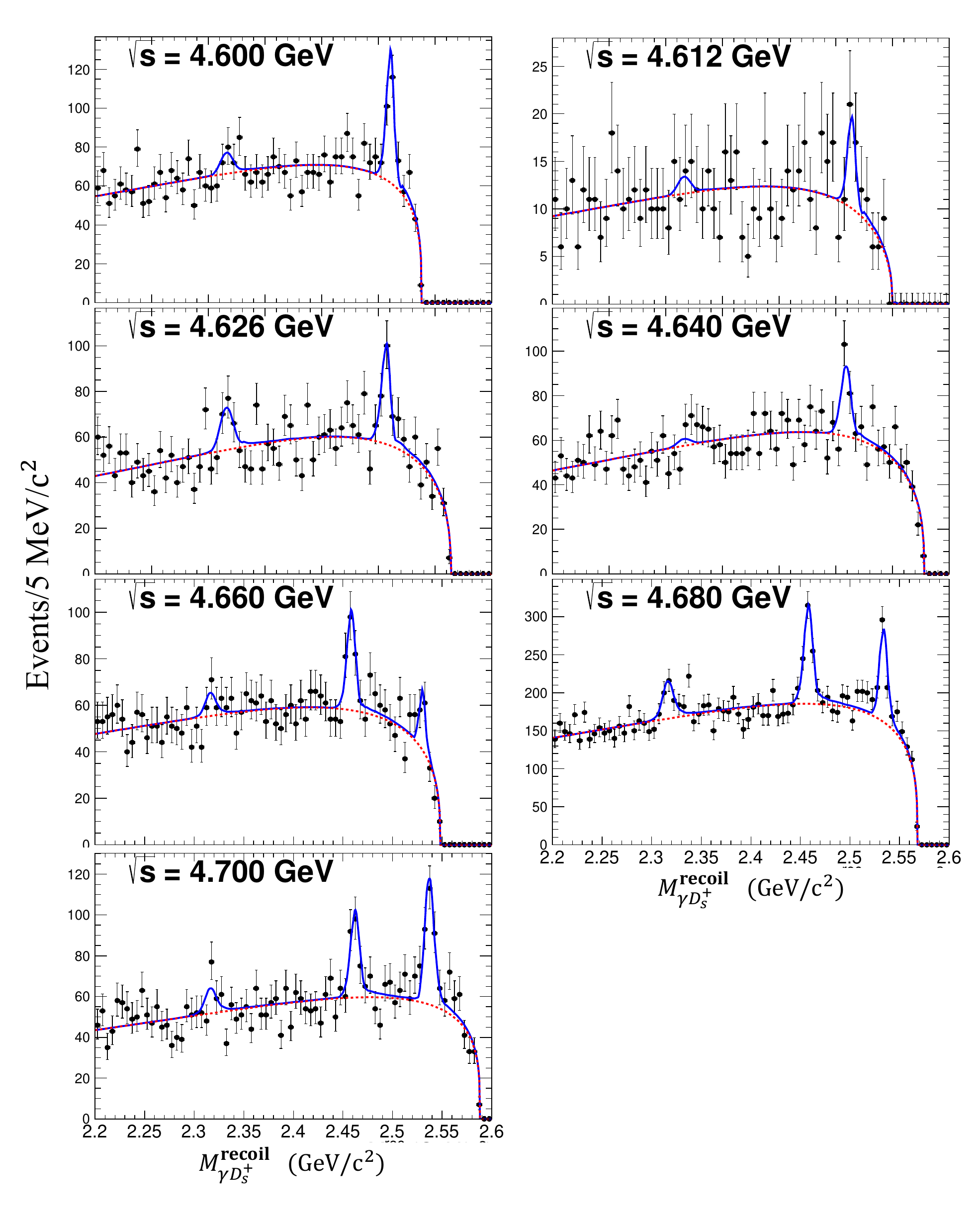}
\caption{Recoil mass spectra of $\gamma D_{s}^{+}$at different c.m. energies~\cite{BESIII:2021xrz} . The dots with error bars are data. Solid lines represent the best fits.
The visible peaks correspond to signals from the $D_{s0}(2317)^{-}$, $D_{s1}(2460)^{-}$ and $D_{s1}(2536)^-$ states.
}
\label{Fits:gammaDsRecoil}
\end{center}
\end{figure}

\begin{figure}[!htbp]
\begin{center}
\includegraphics[width=0.48\textwidth]{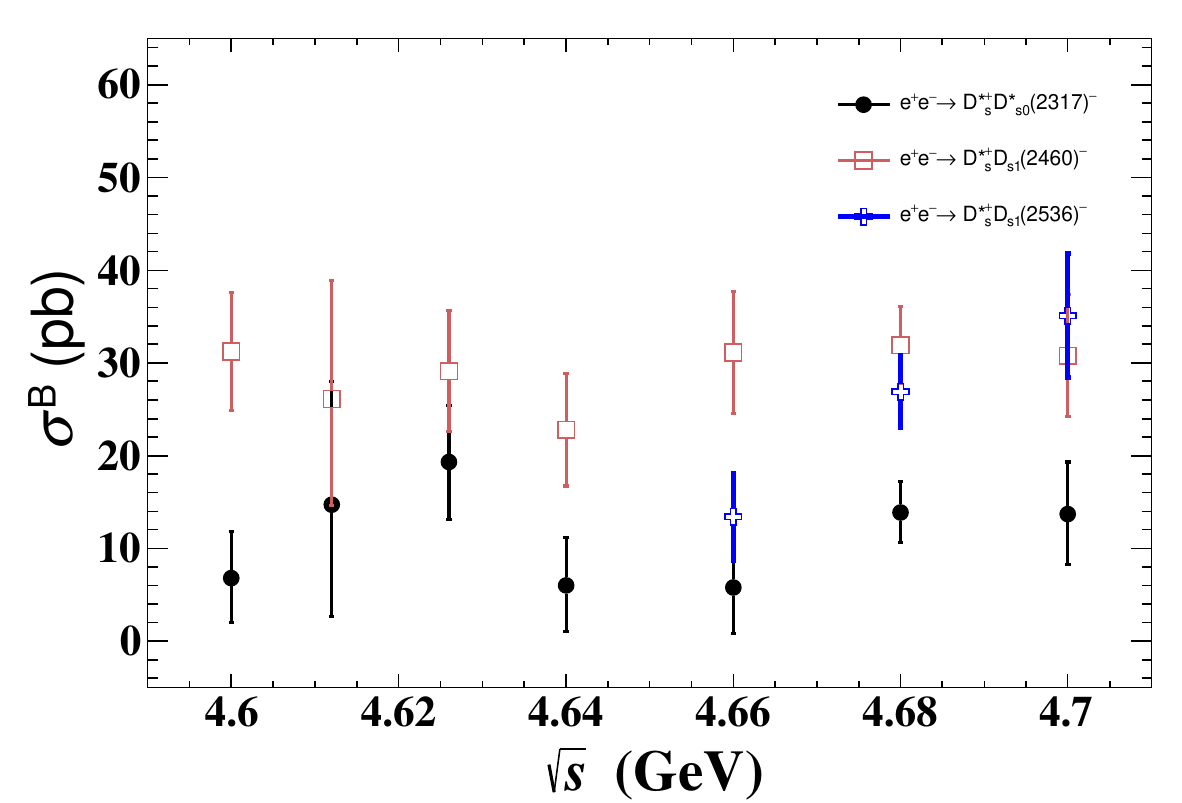}
\caption{Born cross-sections for the $e^+e^-\to D^{\ast+}_{s}D^{\ast}_{s0}(2317)^{-}$, $D^{\ast+}_{s}D_{s1}(2460)^{-}$ and $D^{\ast+}_{s}D_{s1}(2536)^{-}$ reactions as a function of c.m. energy~\cite{BESIII:2021xrz}.
}
\label{BCS:DDbar02:2460}
\end{center}
\end{figure}

Three excited $P$-wave $D_s^+$ states above the $D^{(*)}K$ threshold, namely $D_{s0}(2317)^{-}$, $D_{s1}(2460)^{-}$ and $D^{\ast}_{s2}(2573)^{-}$ mesons, have been observed at the {\it BABAR}, Belle, BESIII and CLEO-c experiments~\cite{CLEO:2003ggt, CLEO:1989qui, BaBar:2003oey, BaBar:2004yux, Bondioli:2004te, BaBar:2011vbs, Belle:2003guh, BESIII:2018fpo}.  
The measured mass and width parameters, as well as the spin parity are consistent with the predictions of the effective theory of heavy quarks~\cite{Dai:2003yg, Zeng:1994vj}. 
The measured masses
of the $D_{s0}(2317)^{-}$ and $D_{s1}(2460)^{-}$ states are notably lower than the theoretical expectations  for the charmed-strange mesons in the $P$-wave doublet~\cite{Chen:2016spr}, leading to various exotic explanations such as the tetraquark states~\cite{Chen:2016spr, Cheng:2003kg, Dmitrasinovic:2012zz,Maiani:2004vq,Wang:2006uba}, $D^{(*)}K$ molecule states~\cite{Barnes:2003dj, Chen:2004dy,Xie:2010zza, Feng:2012zze,Wang:2012bu}, and mixtures of $c\bar{s}$ and $D^{(*)}K$ states~\cite{Browder:2003fk}. 
To further investigate the properties of excited $D_s^+$ states and investigate the nature of charmonium (-like) states above the open-charm threshold, a more precise measurement of Born cross-sections was carried out for the reactions of $e^+e^-\to D^{\ast+}_{s}D_{s0}(2317)^{-}$ and $D^{\ast+}_{s}D_{s1}(2460)^{-}$. 
This measurement involved the use of a semi-inclusive method with data samples at c.m. energies ranging from 4.6 to 4.7~GeV~\cite{BESIII:2021xrz}. The candidates for $e^+e^-\to D^{\ast+}_{s}D_{s0}(2317)^{-}$ and $D^{\ast+}_{s}D_{s1}(2460)^{-}$ were selected through a partial reconstruction method, specifically by reconstructing only the $D^{\ast+}_s$ via its decay into $\gamma D^{+}_s$ final states, where $D^{+}_s$ decays into $\phi\pi^{+}$ and $\bar{K}^{\ast0}K^{+}$ final states with the $\phi\to K^{+}K^{-}$ decay. The candidates for $D_{s0}(2317)^{-}$, and $D_{s1}(2460)^{-}$ could then be searched for on 
the recoil side of the $D^{\ast+}_ s$ candidate. Clear signals for the $D_{s0}(2317)^{-}$, $D_{s1}(2460)^{-}$, and $D_{s1}(2536)^-$ mesons are observed in 
the recoil mass against the $\gamma D_{s}^{+}$ system
as shown in Fig.~\ref{Fits:gammaDsRecoil}.
However, owing to large uncertainties, no significant structures or charmonium(like) states were observed in the measured cross-sections as shown in Fig.~\ref{BCS:DDbar02:2460}. 

\section{Summary and outlook}
Studying hadron spectroscopy has always been a frontier and lively field in particle physics. 
Searching for new potential hadrons is crucial for validating the accuracy of QCD theory and promoting the development of strong interaction theory. A charm quarkonium is located in the transition region between non-pQCD and pQCD. Investigating the decays of charmonium(like) or searching for new candidates of charmonium(like) states or exotic hadron states is a significant way to deepen our understanding of QCD. 
In recent decades a series of charmonoum(-like) states such as $X(3872)$, $Z_{c}(3900)$ and $Y(4230)$, have been observed at $e^+e^-$ colliders via the
{\it BABAR}, Belle, BESIII, and CLEO-c
experiments and have been produced in electron-positron pairs
annihilating into hidden-charm final states. 
Despite various theoretical models having been proposed, including hadron molecular states, hybrid states and  tetraquark states in an attempt to solve the nature of these states, 
there is not yet a
theory that can fully account for experimental measurements.
Among these models, the experimental 
measurements of electron-positron pairs annihilating into open-charm final states provide an important input to probe the nature of vector charmonium(like) states, which have yielded valuable insights into nonstandard hadrons in recent decades. 
This Colloquium examines the contributions from that {\it BABAR}, Belle, BESIII, and CLEO-c experiments have made to such studies through measurements of production cross-section of charmed meson pair and charmed-strange meson pairs. The results for different experiments agree with each other within uncertainties, and the BESIII experiment reported most precision measurements for the production Born cross-sections of $e^+e^-\to D_{(s)}\bar{D}_{(s)}$, $D^{\ast+}\bar{D}^{-}$,  $D^{\ast+}_{(s)}\bar{D}^{\ast-}_{(s)}$,  $\pi^{+}D^{(\ast)0}D^{(\ast)-}$, $\pi^+\pi^-D^+D^-$, $D^{+}_{s}D_{s1}(2536)^{-}$ and $D^{+}_{s}D^{\ast}_{s2}(2573)^{-}$, as well as the first measurement for 
$e^+e^-\to D^{\ast+}_{s}D_{s0}(2317)^{-}$ and $D^{\ast+}_{s}D_{s1}(2460)^{-}$ above the  open-charm threshold to 4.95 GeV.
Many clear peaks in the line shape of the production cross-sections at around the mass ranges of  $G(3900)$, $\psi(4040)$, $\psi(4160)$, $Y(4230)$, $Y(4360)$, $\psi(4415)$, $Y(4660)$ $Y(4660)$, $Y(4700)$, etc., are seen. This implies that there may be some potential contributions from charmoniumlike states. In particular, the possible structure around 3.9 GeV,
which was featured and interpreted as
$G(3900)$ at the $B$ factories~\cite{
BaBar:2006qlj,Belle:2007qxm}, was discussed recently via the theoretical models  as the first $P$-wave $D\bar{D}^{*}$ molecular resonance~\cite{Lin:2024qcq}, threshold enhancement~\cite{Husken:2024hmi} or final-state interaction~\cite{Salnikov:2024wah}.
Thus, more detailed study related to a coupled-channel $K$-matrix analysis is needed to validate this structure.  
Although the BESIII experiment measured the Born cross-section for the production of these open-charm meson pair final states with unprecedented precision and observed several distinct structures, the extraction of resonance parameters remains limited by model dependence.
According to the model calculation by the Cornell group~\cite{Eichten:1979ms}, strong coupled-channel effects need to be considered, which was also proposed in the recent theoretical works~\cite{Lin:2024qcq, Husken:2024hmi, Salnikov:2024wah}.
Indeed, it is out of the scope of the experimental measurements, but a more comprehensive study in theory combined with a $K$-matrix formalism to fit the cross-section results of various exclusive channels 
is expected to discriminate between different
scenarios of charmonium(like) states above the open-charm threshold. 
Returning to the measurements for the production cross-section on electron-positron pairs annihilating into open-charm final states from the {\it BABAR}, Belle, BESIII, and CLEO-c experiments, it is undoubted that these experiments have provided crucial data with unprecedented precision, and valuable insights into the nature of particles in the open-charm region, shedding light on the properties of charmoniumlike states. 
The comprehensive analyses from these experiments have significantly contributed
to our understanding of particle physics and have paved the way for further research in this field, ultimately advancing our knowledge of fundamental interactions in particle physics.

At present, BelleII/SuperKEKB upgraded from the Belle/KEKB facility has started data collection, and in the foreseeable future there are also upgrades planned for the BESIII/BEPCII facility to create two super tau-charm factories in Russia~\cite{Levichev:2018cvd} and China~\cite{Achasov:2023gey}.
The goal of these two facilities is to push the c.m. energies and luminosity to higher order, potentially reaching the c.m. energy of 6 GeV or more and increasing peak luminosities to $10^{35}$ $cm^{-2}s^{-1}$. These advancements would not only represent a significant improvement over BEPCII but also open up new possibilities for groundbreaking research in particle physics.
In addition, an innovative upgrade of the BEPCII facility to the BEPCII-U, during which some small components are being replaced is underway. This upgrade also aims to elevate the c.m. energies and luminosity to a higher level. Specifically, it aims to achieve a c.m. energy of 5.6 GeV or greater and to boost the luminosity by a factor of 3. In addition, the inner drift chamber of the BESIII detector is being replaced by the Cylindrical Gas Electron Multiplier Inner Tracker, which can partially offset the gain loss and address the cathode aging issue. 
With these upgraded and ongoing machines, we can anticipate being able to conduct comprehensive studies on charm particle production with unprecedented precision. Furthermore, we hope to explore the properties of nonstandard hadrons in greater detail. 
In conclusion, the study of charmonium(like) states or exotic hadron states is a challenging but rewarding endeavor that holds the promise of unlocking profound insights into the strong interaction and, ultimately, advancing our understanding of the structure and behavior of matter at the most fundamental level.

\begin{acknowledgments}
We thank doctoral students Ruoyu Zhang, Hao Liu as well as Dr. Anqing Zhang for collecting the data and refining the figures.
This work is supported in part by 
National Natural Science Foundation of China under Contracts No. 12335001 and No. 12247101; 
the Fundamental Research Funds for the Central Universities under Contracts 
No. lzujbky-2025-ytA05, 
No. lzujbky-2025-it06, 
and No. lzujbky-2024-jdzx06;
the Natural Science Foundation of Gansu Province under Contracts 
No. 22JR5RA389 and No. 25JRRA799; 
the 111 Project under Grant No. B20063; the Project for Top-Notch Innovative Talents of Gansu Province; and the Talent Scientific Fund of Lanzhou University.
\end{acknowledgments}

\bibliography{apssamp_deduplicated.bib}

@article{Bjorken:1964gz,
    author = "Bjorken, J. D. and Glashow, S. L.",
    title = "{Elementary Particles and SU(4)}",
    doi = "10.1016/0031-9163(64)90433-0",
    journal = "Phys. Lett.",
    volume = "11",
    pages = "255--257",
    year = "1964"
}

@article{Cabibbo:1963yz,
    author = "Cabibbo, Nicola",
    title = "{Unitary Symmetry and Leptonic Decays}",
    doi = "10.1103/PhysRevLett.10.531",
    journal = "Phys. Rev. Lett.",
    volume = "10",
    pages = "531--533",
    year = "1963"
}

@article{Glashow:1970gm,
    author = "Glashow, S. L. and Iliopoulos, J. and Maiani, L.",
    title = "{Weak Interactions with Lepton-Hadron Symmetry}",
    doi = "10.1103/PhysRevD.2.1285",
    journal = "Phys. Rev. D",
    volume = "2",
    pages = "1285--1292",
    year = "1970"
}

@article{E598:1974sol,
    author = "Aubert, J. J. and others",
    collaboration = "E598",
    title = "{Experimental Observation of a Heavy Particle $J$}",
    reportNumber = "COO-3069-271",
    doi = "10.1103/PhysRevLett.33.1404",
    journal = "Phys. Rev. Lett.",
    volume = "33",
    pages = "1404--1406",
    year = "1974"
}

@article{SLAC-SP-017:1974ind,
    author = "Augustin, J. E. and others",
    collaboration = "SLAC-SP-017",
    title = "{Discovery of a Narrow Resonance in $e^+ e^-$ Annihilation}",
    reportNumber = "SLAC-PUB-1504, LBL-3391",
    doi = "10.1103/PhysRevLett.33.1406",
    journal = "Phys. Rev. Lett.",
    volume = "33",
    pages = "1406--1408",
    year = "1974"
}

@article{Archilli:2017xmu,
    author = "Archilli, F. and Bettler, M. -O. and Owen, P. and Petridis, K. A.",
    title = "{Flavour-changing neutral currents making and breaking the standard model}",
    doi = "10.1038/nature21721",
    journal = "Nature",
    volume = "546",
    number = "7657",
    pages = "221--226",
    year = "2017"
}

@article{Liu:2013waa,
    author = "Liu, Xiang",
    title = "{An overview of $XYZ$ new particles}",
    eprint = "1312.7408",
    archivePrefix = "arXiv",
    primaryClass = "hep-ph",
    doi = "10.1007/s11434-014-0407-2",
    journal = "Chin. Sci. Bull.",
    volume = "59",
    pages = "3815--3830",
    year = "2014"
}

@article{Hosaka:2016pey,
    author = "Hosaka, Atsushi and Iijima, Toru and Miyabayashi, Kenkichi and Sakai, Yoshihide and Yasui, Shigehiro",
    title = "{Exotic hadrons with heavy flavors: X, Y, Z, and related states}",
    eprint = "1603.09229",
    archivePrefix = "arXiv",
    primaryClass = "hep-ph",
    reportNumber = "J-PARC-TH-0046",
    doi = "10.1093/ptep/ptw045",
    journal = "PTEP",
    volume = "2016",
    number = "6",
    pages = "062C01",
    year = "2016"
}

@article{Chen:2016qju,
    author = "Chen, Hua-Xing and Chen, Wei and Liu, Xiang and Zhu, Shi-Lin",
    title = "{The hidden-charm pentaquark and tetraquark states}",
    eprint = "1601.02092",
    archivePrefix = "arXiv",
    primaryClass = "hep-ph",
    doi = "10.1016/j.physrep.2016.05.004",
    journal = "Phys. Rept.",
    volume = "639",
    pages = "1--121",
    year = "2016"
}

@article{Richard:2016eis,
    author = "Richard, Jean-Marc",
    title = "{Exotic hadrons: review and perspectives}",
    eprint = "1606.08593",
    archivePrefix = "arXiv",
    primaryClass = "hep-ph",
    doi = "10.1007/s00601-016-1159-0",
    journal = "Few Body Syst.",
    volume = "57",
    number = "12",
    pages = "1185--1212",
    year = "2016"
}

@article{Esposito:2016noz,
    author = "Esposito, A. and Pilloni, A. and Polosa, A. D.",
    title = "{Multiquark Resonances}",
    eprint = "1611.07920",
    archivePrefix = "arXiv",
    primaryClass = "hep-ph",
    reportNumber = "JLAB-THY-16-2301",
    doi = "10.1016/j.physrep.2016.11.002",
    journal = "Phys. Rept.",
    volume = "668",
    pages = "1--97",
    year = "2017"
}

@article{Ali:2017jda,
    author = {Ali, Ahmed and Lange, Jens S\"oren and Stone, Sheldon},
    title = "{Exotics: Heavy Pentaquarks and Tetraquarks}",
    eprint = "1706.00610",
    archivePrefix = "arXiv",
    primaryClass = "hep-ph",
    reportNumber = "DESY-17-071",
    doi = "10.1016/j.ppnp.2017.08.003",
    journal = "Prog. Part. Nucl. Phys.",
    volume = "97",
    pages = "123--198",
    year = "2017"
}

@article{Olsen:2017bmm,
    author = "Olsen, Stephen Lars and Skwarnicki, Tomasz and Zieminska, Daria",
    title = "{Nonstandard heavy mesons and baryons: Experimental evidence}",
    eprint = "1708.04012",
    archivePrefix = "arXiv",
    primaryClass = "hep-ph",
    doi = "10.1103/RevModPhys.90.015003",
    journal = "Rev. Mod. Phys.",
    volume = "90",
    number = "1",
    pages = "015003",
    year = "2018"
}

@article{Guo:2017jvc,
    author = "Guo, Feng-Kun and Hanhart, Christoph and Mei\ss{}ner, Ulf-G. and Wang, Qian and Zhao, Qiang and Zou, Bing-Song",
    title = "{Hadronic molecules}",
    eprint = "1705.00141",
    archivePrefix = "arXiv",
    primaryClass = "hep-ph",
    doi = "10.1103/RevModPhys.90.015004",
    journal = "Rev. Mod. Phys.",
    volume = "90",
    number = "1",
    pages = "015004",
    year = "2018",
    note = "[Erratum: Rev.Mod.Phys. 94, 029901 (2022)]"
}

@article{Liu:2019zoy,
    author = "Liu, Yan-Rui and Chen, Hua-Xing and Chen, Wei and Liu, Xiang and Zhu, Shi-Lin",
    title = "{Pentaquark and Tetraquark states}",
    eprint = "1903.11976",
    archivePrefix = "arXiv",
    primaryClass = "hep-ph",
    doi = "10.1016/j.ppnp.2019.04.003",
    journal = "Prog. Part. Nucl. Phys.",
    volume = "107",
    pages = "237--320",
    year = "2019"
}

@article{Yuan:2019zfo,
    author = "Yuan, Chang-Zheng and Olsen, Stephen Lars",
    title = "{The BESIII physics programme}",
    eprint = "2001.01164",
    archivePrefix = "arXiv",
    primaryClass = "hep-ex",
    doi = "10.1038/s42254-019-0082-y",
    journal = "Nature Rev. Phys.",
    volume = "1",
    number = "8",
    pages = "480--494",
    year = "2019"
}

@article{Brambilla:2019esw,
    author = "Brambilla, Nora and Eidelman, Simon and Hanhart, Christoph and Nefediev, Alexey and Shen, Cheng-Ping and Thomas, Christopher E. and Vairo, Antonio and Yuan, Chang-Zheng",
    title = "{The $XYZ$ states: experimental and theoretical status and perspectives}",
    eprint = "1907.07583",
    archivePrefix = "arXiv",
    primaryClass = "hep-ex",
    reportNumber = "TUM-EFT 125/19",
    doi = "10.1016/j.physrep.2020.05.001",
    journal = "Phys. Rept.",
    volume = "873",
    pages = "1--154",
    year = "2020"
}

@article{Chen:2022asf,
    author = "Chen, Hua-Xing and Chen, Wei and Liu, Xiang and Liu, Yan-Rui and Zhu, Shi-Lin",
    title = "{An updated review of the new hadron states}",
    eprint = "2204.02649",
    archivePrefix = "arXiv",
    primaryClass = "hep-ph",
    doi = "10.1088/1361-6633/aca3b6",
    journal = "Rept. Prog. Phys.",
    volume = "86",
    number = "2",
    pages = "026201",
    year = "2023"
}

@article{Meng:2022ozq,
    author = "Meng, Lu and Wang, Bo and Wang, Guang-Juan and Zhu, Shi-Lin",
    title = "{Chiral perturbation theory for heavy hadrons and chiral effective field theory for heavy hadronic molecules}",
    eprint = "2204.08716",
    archivePrefix = "arXiv",
    primaryClass = "hep-ph",
    doi = "10.1016/j.physrep.2023.04.003",
    journal = "Phys. Rept.",
    volume = "1019",
    pages = "1--149",
    year = "2023"
}

@article{ParticleDataGroup:2024cfk,
    author = "Navas, S. and others",
    collaboration = "Particle Data Group",
    title = "{Review of particle physics}",
    doi = "10.1103/PhysRevD.110.030001",
    journal = "Phys. Rev. D",
    volume = "110",
    number = "3",
    pages = "030001",
    year = "2024"
}

@article{BESIII:2020nme,
    author = "Ablikim, M. and others",
    collaboration = "BESIII",
    title = "{Future Physics Programme of BESIII}",
    eprint = "1912.05983",
    archivePrefix = "arXiv",
    primaryClass = "hep-ex",
    reportNumber = "HEP-Physics-Report-BESIII-2019-12-13",
    doi = "10.1088/1674-1137/44/4/040001",
    journal = "Chin. Phys. C",
    volume = "44",
    number = "4",
    pages = "040001",
    year = "2020"
}

@article{Kozanecki:2000cm,
    author = "Kozanecki, W.",
    editor = "Erhan, S. and Krizan, P. and Lohse, T.",
    title = "{The PEP-II B factory: Status and prospects}",
    doi = "10.1016/S0168-9002(00)00022-X",
    journal = "Nucl. Instrum. Meth. A",
    volume = "446",
    pages = "59--64",
    year = "2000"
}

@article{BaBar:2001yhh,
    author = "Aubert, Bernard and others",
    collaboration = "BaBar",
    title = "{The BaBar detector}",
    eprint = "hep-ex/0105044",
    archivePrefix = "arXiv",
    reportNumber = "SLAC-PUB-8569, BABAR-PUB-01-08",
    doi = "10.1016/S0168-9002(01)02012-5",
    journal = "Nucl. Instrum. Meth. A",
    volume = "479",
    pages = "1--116",
    year = "2002"
}

@article{BaBar:2005hhc,
    author = "Aubert, Bernard and others",
    collaboration = "BaBar",
    title = "{Observation of a broad structure in the $\pi^+ \pi^- J/\psi$ mass spectrum around 4.26-GeV/c$^2$}",
    eprint = "hep-ex/0506081",
    archivePrefix = "arXiv",
    reportNumber = "BABAR-PUB-05-29, SLAC-PUB-11320, BABAR-PUB-05-029",
    doi = "10.1103/PhysRevLett.95.142001",
    journal = "Phys. Rev. Lett.",
    volume = "95",
    pages = "142001",
    year = "2005"
}

@article{Belle:2000cnh,
    author = "Abashian, A. and others",
    collaboration = "Belle",
    title = "{The Belle Detector}",
    reportNumber = "KEK-PROGRESS-REPORT-2000-4",
    doi = "10.1016/S0168-9002(01)02013-7",
    journal = "Nucl. Instrum. Meth. A",
    volume = "479",
    pages = "117--232",
    year = "2002"
}

@inproceedings{Yu:2016cof,
    author = "Yu, Chenghui and others",
    title = "{BEPCII Performance and Beam Dynamics Studies on Luminosity}",
    booktitle = "{7th International Particle Accelerator Conference}",
    doi = "10.18429/JACoW-IPAC2016-TUYA01",
    pages = "TUYA01",
    year = "2016"
}

@article{Li:2021zrf,
    author = "Li, Gang and Ye, Rui and Sang, Mingjing and Li, Shaopeng and Zhang, Zhuo and Zhang, Jiehao and Han, Ruixiong and Lei, Ge and Ge, Rui",
    title = "{The cryogenic control system of BEPCII}",
    doi = "10.1140/epjti/s40485-021-00064-9",
    journal = "EPJ Tech. Instrum.",
    volume = "8",
    number = "1",
    pages = "7",
    year = "2021"
}

@article{BESIII:2013ris,
    author = "Ablikim, M. and others",
    collaboration = "BESIII",
    title = "{Observation of a Charged Charmoniumlike Structure in $e^+e^- \to \pi^+\pi^- J/\psi$ at $\sqrt{s}$ =4.26 GeV}",
    eprint = "1303.5949",
    archivePrefix = "arXiv",
    primaryClass = "hep-ex",
    doi = "10.1103/PhysRevLett.110.252001",
    journal = "Phys. Rev. Lett.",
    volume = "110",
    pages = "252001",
    year = "2013"
}

@article{Fast:1999rf,
    author = "Fast, J and others",
    editor = "Loukas, D. and Markou, C.",
    title = "{Design, performance and status of the CLEO III silicon detector}",
    doi = "10.1016/S0168-9002(99)00405-2",
    journal = "Nucl. Instrum. Meth. A",
    volume = "435",
    pages = "9--15",
    year = "1999"
}

@article{Richichi:2003md,
    author = "Richichi, S. J.",
    editor = "Adeva, Bernardo and Erhan, Samim",
    title = "{CLEO-c and CESR-c: A new frontier in strong and weak interactions}",
    doi = "10.1016/S0920-5632(03)01878-4",
    journal = "Nucl. Phys. B Proc. Suppl.",
    volume = "120",
    pages = "27--34",
    year = "2003"
}

@article{BESIII:2009fln,
    author = "Ablikim, M. and others",
    collaboration = "BESIII",
    title = "{Design and Construction of the BESIII Detector}",
    eprint = "0911.4960",
    archivePrefix = "arXiv",
    primaryClass = "physics.ins-det",
    doi = "10.1016/j.nima.2009.12.050",
    journal = "Nucl. Instrum. Meth. A",
    volume = "614",
    pages = "345--399",
    year = "2010"
}

@article{Ablikim:2013ntc,
    author = "Ablikim, M.",
    collaboration = "BESIII",
    title = "{Measurement of the integrated luminosities of the data taken by BESIII at $\sqrt{s}=$3.650 and 3.773 GeV}",
    eprint = "1307.2022",
    archivePrefix = "arXiv",
    primaryClass = "hep-ex",
    doi = "10.1088/1674-1137/37/12/123001",
    journal = "Chin. Phys. C",
    volume = "37",
    pages = "123001",
    year = "2013"
}

@article{BESIII:2015qfd,
    author = "Ablikim, M. and others",
    collaboration = "BESIII",
    title = "{Precision measurement of the integrated luminosity of the data taken by BESIII at center of mass energies between 3.810 GeV and 4.600 GeV}",
    eprint = "1503.03408",
    archivePrefix = "arXiv",
    primaryClass = "hep-ex",
    doi = "10.1088/1674-1137/39/9/093001",
    journal = "Chin. Phys. C",
    volume = "39",
    number = "9",
    pages = "093001",
    year = "2015"
}

@article{BESIII:2017lkp,
    author = "Ablikim, M. and others",
    collaboration = "BESIII",
    title = "{Luminosity measurements for the R scan experiment at BESIII}",
    eprint = "1702.04977",
    archivePrefix = "arXiv",
    primaryClass = "hep-ex",
    doi = "10.1088/1674-1137/41/6/063001",
    journal = "Chin. Phys. C",
    volume = "41",
    number = "6",
    pages = "063001",
    year = "2017"
}

@article{BESIII:2022dxl,
    author = "Ablikim, M. and others",
    collaboration = "BESIII",
    title = "{Measurement of integrated luminosities at BESIII for data samples at center-of-mass energies between 4.0 and 4.6 GeV}",
    eprint = "2203.03133",
    archivePrefix = "arXiv",
    primaryClass = "hep-ex",
    doi = "10.1088/1674-1137/ac80b4",
    journal = "Chin. Phys. C",
    volume = "46",
    number = "11",
    pages = "113002",
    year = "2022"
}

@article{BESIII:2022ulv,
    author = "Ablikim, M. and others",
    collaboration = "BESIII",
    title = "{Luminosities and energies of e $^{+}$ e $^{−}$ collision data taken between =4.61 GeV and 4.95 GeV at BESIII*}",
    eprint = "2205.04809",
    archivePrefix = "arXiv",
    primaryClass = "hep-ex",
    doi = "10.1088/1674-1137/ac84cc",
    journal = "Chin. Phys. C",
    volume = "46",
    number = "11",
    pages = "113003",
    year = "2022"
}

@article{BESIII:2024lbn,
    author = "Ablikim, Medina and others",
    collaboration = "BESIII",
    title = "{Measurement of integrated luminosity of data collected at 3.773 GeV by BESIII from 2021 to 2024*}",
    eprint = "2406.05827",
    archivePrefix = "arXiv",
    primaryClass = "hep-ex",
    doi = "10.1088/1674-1137/ad70a0",
    journal = "Chin. Phys. C",
    volume = "48",
    number = "12",
    pages = "123001",
    year = "2024"
}

@article{Belle:2003nnu,
    author = "Choi, S. K. and others",
    collaboration = "Belle",
    title = "{Observation of a narrow charmonium-like state in exclusive $B^\pm \to K^\pm \pi^+ \pi^- J/\psi$ decays}",
    eprint = "hep-ex/0309032",
    archivePrefix = "arXiv",
    doi = "10.1103/PhysRevLett.91.262001",
    journal = "Phys. Rev. Lett.",
    volume = "91",
    pages = "262001",
    year = "2003"
}

@article{Eichten:1974af,
    author = "Eichten, E. and Gottfried, K. and Kinoshita, T. and Kogut, John B. and Lane, K. D. and Yan, Tung-Mow",
    title = "{The Spectrum of Charmonium}",
    reportNumber = "PRINT-74-1715 (CORNELL)",
    doi = "10.1103/PhysRevLett.34.369",
    journal = "Phys. Rev. Lett.",
    volume = "34",
    pages = "369--372",
    year = "1975",
    note = "[Erratum: Phys.Rev.Lett. 36, 1276 (1976)]"
}

@article{Eichten:1978tg,
    author = "Eichten, E. and Gottfried, K. and Kinoshita, T. and Lane, K. D. and Yan, Tung-Mow",
    title = "{Charmonium: The Model}",
    reportNumber = "CLNS-375",
    doi = "10.1103/PhysRevD.17.3090",
    journal = "Phys. Rev. D",
    volume = "17",
    pages = "3090",
    year = "1978",
    note = "[Erratum: Phys.Rev.D 21, 313 (1980)]"
}

@article{Eichten:1979ms,
    author = "Eichten, E. and Gottfried, K. and Kinoshita, T. and Lane, K. D. and Yan, Tung-Mow",
    title = "{Charmonium: Comparison with Experiment}",
    reportNumber = "CLNS-425",
    doi = "10.1103/PhysRevD.21.203",
    journal = "Phys. Rev. D",
    volume = "21",
    pages = "203",
    year = "1980"
}

@article{Barnes:2005pb,
    author = "Barnes, T. and Godfrey, S. and Swanson, E. S.",
    title = "{Higher charmonia}",
    eprint = "hep-ph/0505002",
    archivePrefix = "arXiv",
    doi = "10.1103/PhysRevD.72.054026",
    journal = "Phys. Rev. D",
    volume = "72",
    pages = "054026",
    year = "2005"
}

@article{Radford:2007vd,
    author = "Radford, Stanley F. and Repko, Wayne W.",
    title = "{Potential model calculations and predictions for heavy quarkonium}",
    eprint = "hep-ph/0701117",
    archivePrefix = "arXiv",
    doi = "10.1103/PhysRevD.75.074031",
    journal = "Phys. Rev. D",
    volume = "75",
    pages = "074031",
    year = "2007"
}

@article{Belle:2008xmh,
    author = "Pakhlova, G. and others",
    collaboration = "Belle",
    title = "{Observation of a near-threshold enhancement in the e+e- ---\ensuremath{>} Lambda+(c) Lambda-(c) cross section using initial-state radiation}",
    eprint = "0807.4458",
    archivePrefix = "arXiv",
    primaryClass = "hep-ex",
    doi = "10.1103/PhysRevLett.101.172001",
    journal = "Phys. Rev. Lett.",
    volume = "101",
    pages = "172001",
    year = "2008"
}

@article{Belle:2007umv,
    author = "Wang, X. L. and others",
    collaboration = "Belle",
    title = "{Observation of Two Resonant Structures in e+e- to pi+ pi- psi(2S) via Initial State Radiation at Belle}",
    eprint = "0707.3699",
    archivePrefix = "arXiv",
    primaryClass = "hep-ex",
    reportNumber = "BELLE-2007-33, KEK-2007-27",
    doi = "10.1103/PhysRevLett.99.142002",
    journal = "Phys. Rev. Lett.",
    volume = "99",
    pages = "142002",
    year = "2007"
}

@article{Belle-II:2018jsg,
    author = "Altmannshofer, W. and others",
    editor = "Kou, E. and Urquijo, P.",
    collaboration = "Belle-II",
    title = "{The Belle II Physics Book}",
    eprint = "1808.10567",
    archivePrefix = "arXiv",
    primaryClass = "hep-ex",
    reportNumber = "KEK Preprint 2018-27, BELLE2-PUB-PH-2018-001, FERMILAB-PUB-18-398-T, JLAB-THY-18-2780, INT-PUB-18-047, UWThPh 2018-26",
    doi = "10.1093/ptep/ptz106",
    journal = "PTEP",
    volume = "2019",
    number = "12",
    pages = "123C01",
    year = "2019",
    note = "[Erratum: PTEP 2020, 029201 (2020)]"
}

@article{Belle:2007qxm,
    author = "Pakhlova, G. and others",
    editor = "Barlow, Roger",
    collaboration = "Belle",
    title = "{Measurement of the near-threshold e+ e- ---\ensuremath{>} D anti-D cross section using initial-state radiation}",
    eprint = "0708.0082",
    archivePrefix = "arXiv",
    primaryClass = "hep-ex",
    reportNumber = "BELLE-CONF-0704",
    doi = "10.1103/PhysRevD.77.011103",
    journal = "Phys. Rev. D",
    volume = "77",
    pages = "011103",
    year = "2008"
}

@article{BaBar:2006qlj,
    author = "Aubert, Bernard and others",
    editor = "Sissakian, Alexey and Kozlov, Gennady and Kolganova, Elena",
    collaboration = "BaBar",
    title = "{Study of the Exclusive Initial-State Radiation Production of the D anti-D System}",
    eprint = "hep-ex/0607083",
    archivePrefix = "arXiv",
    reportNumber = "SLAC-PUB-11983, BABAR-CONF-06-33, BABAR-CONF-06-033",
    doi = "10.1103/PhysRevD.76.111105",
    journal = "Phys. Rev. D",
    volume = "76",
    pages = "111105",
    year = "2007"
}

@article{CLEO:2008ojp,
    author = "Cronin-Hennessy, D. and others",
    collaboration = "CLEO",
    title = "{Measurement of Charm Production Cross Sections in e+e- Annihilation at Energies between 3.97 and 4.26-GeV}",
    eprint = "0801.3418",
    archivePrefix = "arXiv",
    primaryClass = "hep-ex",
    reportNumber = "CLNS07-2015, CLEO-07-19",
    doi = "10.1103/PhysRevD.80.072001",
    journal = "Phys. Rev. D",
    volume = "80",
    pages = "072001",
    year = "2009"
}

@article{BESIII:2024ths,
    author = "Ablikim, Medina and others",
    collaboration = "BESIII",
    title = "{Precise Measurement of Born Cross Sections for e+e-\textrightarrow{}DD\textasciimacron{} at s=3.80-4.95\,\,GeV}",
    eprint = "2402.03829",
    archivePrefix = "arXiv",
    primaryClass = "hep-ex",
    doi = "10.1103/PhysRevLett.133.081901",
    journal = "Phys. Rev. Lett.",
    volume = "133",
    number = "8",
    pages = "081901",
    year = "2024"
}

@article{WorkingGrouponRadiativeCorrections:2010bjp,
    author = "Actis, S. and others",
    collaboration = "Working Group on Radiative Corrections, Monte Carlo Generators for Low Energies",
    title = "{Quest for precision in hadronic cross sections at low energy: Monte Carlo tools vs. experimental data}",
    eprint = "0912.0749",
    archivePrefix = "arXiv",
    primaryClass = "hep-ph",
    reportNumber = "BIHEP-TH-2009-005, BU-HEPP-09-08, CERN-PH-TH-2009-201, FNT-T-2009-03, FREIBURG-PHENO-09-07, HEPTOOLS-09-018, IEKP-KA-2009-33, LNF-09-14-P, LPSC-09-157, LPT-ORSAY-09-95, LTH-851, MZ-TH-09-38, PITHA-09-14",
    doi = "10.1140/epjc/s10052-010-1251-4",
    journal = "Eur. Phys. J. C",
    volume = "66",
    pages = "585--686",
    year = "2010"
}

@article{Sun:2020ehv,
    author = "Sun, Wenyu and Liu, Tong and Jing, Maoqiang and Wang, Liangliang and Zhong, Bin and Song, Weimin",
    title = "{An iterative weighting method to apply ISR correction to e$^{+}$e$^{−}$ hadronic cross-section measurements}",
    eprint = "2011.07889",
    archivePrefix = "arXiv",
    primaryClass = "hep-ex",
    doi = "10.1007/s11467-021-1085-6",
    journal = "Front. Phys. (Beijing)",
    volume = "16",
    number = "6",
    pages = "64501",
    year = "2021"
}

@article{Jegerlehner:2011ti,
    author = "Jegerlehner, Fred and Szafron, Robert",
    title = "{$\rho^0 - \gamma$ mixing in the neutral channel pion form factor $F_{\pi}^{e}$ and its role in comparing $e^+ e^-$ with $\tau$ spectral functions}",
    eprint = "1101.2872",
    archivePrefix = "arXiv",
    primaryClass = "hep-ph",
    reportNumber = "DESY-11-008, HU-EP-11-04",
    doi = "10.1140/epjc/s10052-011-1632-3",
    journal = "Eur. Phys. J. C",
    volume = "71",
    pages = "1632",
    year = "2011"
}

@phdthesis{Julin:2017jcl,
    author = "Julin, Andy Jarod",
    title = "{Measurement of $D\overline{D}$ Decays from the psi(3770) Resonance}",
    school = "Minnesota U.",
    year = "2017"
}

@article{Husken:2024hmi,
    author = {H\"usken, Nils and Lebed, Richard F. and Mitchell, Ryan E. and Swanson, Eric S. and Wang, Ya-Qian and Yuan, Chang-Zheng},
    title = "{Poles and poltergeists in e+e-\textrightarrow{}DD\textasciimacron{} data}",
    eprint = "2404.03896",
    archivePrefix = "arXiv",
    primaryClass = "hep-ph",
    doi = "10.1103/PhysRevD.109.114010",
    journal = "Phys. Rev. D",
    volume = "109",
    number = "11",
    pages = "114010",
    year = "2024"
}

@article{Anashin:2011kq,
    author = "Anashin, V. V. and others",
    title = "{Measurement of \textbackslash{}psi(3770) parameters}",
    eprint = "1109.4205",
    archivePrefix = "arXiv",
    primaryClass = "hep-ex",
    doi = "10.1016/j.physletb.2012.04.019",
    journal = "Phys. Lett. B",
    volume = "711",
    pages = "292--300",
    year = "2012"
}

@article{Belle:2006hvs,
    author = "Abe, Kazuo and others",
    collaboration = "Belle",
    title = "{Measurement of the near-threshold e+ e- ---\ensuremath{>} D(*)+- D(*)-+ cross section using initial-state radiation}",
    eprint = "hep-ex/0608018",
    archivePrefix = "arXiv",
    doi = "10.1103/PhysRevLett.98.092001",
    journal = "Phys. Rev. Lett.",
    volume = "98",
    pages = "092001",
    year = "2007"
}

@article{Belle:2017grj,
    author = "Zhukova, V. and others",
    collaboration = "Belle",
    title = "{Angular analysis of the $e^+ e^- \to D^{(*) \pm} D^{* \mp}$ process near the open charm threshold using initial-state radiation}",
    eprint = "1707.09167",
    archivePrefix = "arXiv",
    primaryClass = "hep-ex",
    reportNumber = "BELLE-PREPRINT-2017-15, KEK-PREPRINT-2017-16",
    doi = "10.1103/PhysRevD.97.012002",
    journal = "Phys. Rev. D",
    volume = "97",
    number = "1",
    pages = "012002",
    year = "2018"
}

@article{BaBar:2009elc,
    author = "Aubert, Bernard and others",
    collaboration = "BaBar",
    title = "{Exclusive Initial-State-Radiation Production of the D anti-D, D* anti-D*, and D* anti-D* Systems}",
    eprint = "0903.1597",
    archivePrefix = "arXiv",
    primaryClass = "hep-ex",
    reportNumber = "SLAC-PUB-13560, BABAR-PUB-08-057",
    doi = "10.1103/PhysRevD.79.092001",
    journal = "Phys. Rev. D",
    volume = "79",
    pages = "092001",
    year = "2009"
}

@article{BESIII:2021yvc,
    author = "Ablikim, Medina and others",
    collaboration = "BESIII",
    title = "{Cross section measurements of the $e^+e^-\to D^{*+}D^{*-}$ and $e^+e^-\to D^{*+}D^{-}$ processes at center-of-mass energies from 4.085 to 4.600 GeV}",
    eprint = "2112.06477",
    archivePrefix = "arXiv",
    primaryClass = "hep-ex",
    doi = "10.1007/JHEP05(2022)155",
    journal = "JHEP",
    volume = "05",
    pages = "155",
    year = "2022"
}

@article{Dubynskiy:2006sg,
    author = "Dubynskiy, S. and Voloshin, M. B.",
    title = "{Possible new resonance at the D* anti-D* threshold in e+ e- annihilation}",
    eprint = "hep-ph/0608179",
    archivePrefix = "arXiv",
    reportNumber = "TPI-MINN-06-28-T, UMN-TH-2514-06",
    doi = "10.1142/S0217732306022195",
    journal = "Mod. Phys. Lett. A",
    volume = "21",
    pages = "2779--2788",
    year = "2006"
}

@article{Rosner:2006vc,
    author = "Rosner, Jonathan L.",
    title = "{Effects of S-wave thresholds}",
    eprint = "hep-ph/0608102",
    archivePrefix = "arXiv",
    reportNumber = "EFI-06-14",
    doi = "10.1103/PhysRevD.74.076006",
    journal = "Phys. Rev. D",
    volume = "74",
    pages = "076006",
    year = "2006"
}

@article{Wang:2019mhs,
    author = "Wang, Jun-Zhang and Chen, Dian-Yong and Liu, Xiang and Matsuki, Takayuki",
    title = "{Constructing $J/\psi$ family with updated data of charmoniumlike $Y$ states}",
    eprint = "1903.07115",
    archivePrefix = "arXiv",
    primaryClass = "hep-ph",
    doi = "10.1103/PhysRevD.99.114003",
    journal = "Phys. Rev. D",
    volume = "99",
    number = "11",
    pages = "114003",
    year = "2019"
}

@article{Wang:2020prx,
    author = "Wang, Jun-Zhang and Qian, Ri-Qing and Liu, Xiang and Matsuki, Takayuki",
    title = "{Are the $Y$ states around 4.6 GeV from $e^+e^-$ annihilation higher charmonia?}",
    eprint = "2001.00175",
    archivePrefix = "arXiv",
    primaryClass = "hep-ph",
    doi = "10.1103/PhysRevD.101.034001",
    journal = "Phys. Rev. D",
    volume = "101",
    number = "3",
    pages = "034001",
    year = "2020"
}

@article{Llanes-Estrada:2005qvr,
    author = "Llanes-Estrada, Felipe J.",
    title = "{Y(4260) and possible charmonium assignment}",
    eprint = "hep-ph/0507035",
    archivePrefix = "arXiv",
    reportNumber = "SLAC-PUB-11338",
    doi = "10.1103/PhysRevD.72.031503",
    journal = "Phys. Rev. D",
    volume = "72",
    pages = "031503",
    year = "2005"
}

@article{Li:2009zu,
    author = "Li, Bai-Qing and Chao, Kuang-Ta",
    title = "{Higher Charmonia and X,Y,Z states with Screened Potential}",
    eprint = "0903.5506",
    archivePrefix = "arXiv",
    primaryClass = "hep-ph",
    doi = "10.1103/PhysRevD.79.094004",
    journal = "Phys. Rev. D",
    volume = "79",
    pages = "094004",
    year = "2009"
}

@article{Cao:2020vab,
    author = "Cao, Qin-Fang and Qi, Hong-Rong and Tang, Guang-Yi and Xue, Yun-Feng and Zheng, Han-Qing",
    title = "{On leptonic width of $X(4260)$}",
    eprint = "2002.05641",
    archivePrefix = "arXiv",
    primaryClass = "hep-ph",
    doi = "10.1140/epjc/s10052-020-08813-y",
    journal = "Eur. Phys. J. C",
    volume = "81",
    number = "1",
    pages = "83",
    year = "2021"
}

@article{Zhu:2005hp,
    author = "Zhu, Shi-Lin",
    title = "{The Possible interpretations of Y(4260)}",
    eprint = "hep-ph/0507025",
    archivePrefix = "arXiv",
    doi = "10.1016/j.physletb.2005.08.068",
    journal = "Phys. Lett. B",
    volume = "625",
    pages = "212",
    year = "2005"
}

@article{Kou:2005gt,
    author = "Kou, E. and Pene, O.",
    title = "{Suppressed decay into open charm for the Y(4260) being an hybrid}",
    eprint = "hep-ph/0507119",
    archivePrefix = "arXiv",
    reportNumber = "UCL-IPT-05-07",
    doi = "10.1016/j.physletb.2005.09.013",
    journal = "Phys. Lett. B",
    volume = "631",
    pages = "164--169",
    year = "2005"
}

@article{Close:2005iz,
    author = "Close, Frank E. and Page, Philip R.",
    title = "{Gluonic charmonium resonances at BaBar and BELLE?}",
    eprint = "hep-ph/0507199",
    archivePrefix = "arXiv",
    doi = "10.1016/j.physletb.2005.09.016",
    journal = "Phys. Lett. B",
    volume = "628",
    pages = "215--222",
    year = "2005"
}

@article{BESIII:2016bnd,
    author = "Ablikim, Medina and others",
    collaboration = "BESIII",
    title = "{Precise measurement of the $e^+e^-\to \pi^+\pi^-J/\psi$ cross section at center-of-mass energies from 3.77 to 4.60 GeV}",
    eprint = "1611.01317",
    archivePrefix = "arXiv",
    primaryClass = "hep-ex",
    doi = "10.1103/PhysRevLett.118.092001",
    journal = "Phys. Rev. Lett.",
    volume = "118",
    number = "9",
    pages = "092001",
    year = "2017"
}

@article{Wang:2013cya,
    author = "Wang, Qian and Hanhart, Christoph and Zhao, Qiang",
    title = "{Decoding the riddle of $Y(4260)$ and $Z_c(3900)$}",
    eprint = "1303.6355",
    archivePrefix = "arXiv",
    primaryClass = "hep-ph",
    doi = "10.1103/PhysRevLett.111.132003",
    journal = "Phys. Rev. Lett.",
    volume = "111",
    number = "13",
    pages = "132003",
    year = "2013"
}

@article{Cleven:2013mka,
    author = "Cleven, Martin and Wang, Qian and Guo, Feng-Kun and Hanhart, Christoph and Mei\ss{}ner, Ulf-G. and Zhao, Qiang",
    title = "{$Y(4260)$ as the first $S$-wave open charm vector molecular state?}",
    eprint = "1310.2190",
    archivePrefix = "arXiv",
    primaryClass = "hep-ph",
    doi = "10.1103/PhysRevD.90.074039",
    journal = "Phys. Rev. D",
    volume = "90",
    number = "7",
    pages = "074039",
    year = "2014"
}

@article{Qin:2016spb,
    author = "Qin, Wen and Xue, Si-Run and Zhao, Qiang",
    title = "{Production of $Y(4260)$ as a hadronic molecule state of $\bar{D}D_1 +c.c.$ in $e^+e^-$ annihilations}",
    eprint = "1605.02407",
    archivePrefix = "arXiv",
    primaryClass = "hep-ph",
    doi = "10.1103/PhysRevD.94.054035",
    journal = "Phys. Rev. D",
    volume = "94",
    number = "5",
    pages = "054035",
    year = "2016"
}

@article{Belle:2009dus,
    author = "Pakhlova, G. and others",
    editor = "Behnke, Ties and Mnich, Joachim",
    collaboration = "Belle",
    title = "{Measurement of the e+ e- ---\ensuremath{>} D0 D*- pi+ cross section using initial-state radiation}",
    eprint = "0908.0231",
    archivePrefix = "arXiv",
    primaryClass = "hep-ex",
    doi = "10.1103/PhysRevD.80.091101",
    journal = "Phys. Rev. D",
    volume = "80",
    pages = "091101",
    year = "2009"
}

@article{BESIII:2018iea,
    author = "Ablikim, Medina and others",
    collaboration = "BESIII",
    title = "{Evidence of a resonant structure in the $e^+e^-\to \pi^+D^0D^{*-}$ cross section between 4.05 and 4.60 GeV}",
    eprint = "1808.02847",
    archivePrefix = "arXiv",
    primaryClass = "hep-ex",
    doi = "10.1103/PhysRevLett.122.102002",
    journal = "Phys. Rev. Lett.",
    volume = "122",
    number = "10",
    pages = "102002",
    year = "2019"
}

@article{BESIII:2023cmv,
    author = "Ablikim, M. and others",
    collaboration = "BESIII",
    title = "{Observation of Three Charmoniumlike States with JPC=1-- in e+e-\textrightarrow{}D*0D*-\ensuremath{\pi}+}",
    eprint = "2301.07321",
    archivePrefix = "arXiv",
    primaryClass = "hep-ex",
    doi = "10.1103/PhysRevLett.130.121901",
    journal = "Phys. Rev. Lett.",
    volume = "130",
    number = "12",
    pages = "121901",
    year = "2023"
}

@article{BESIII:2022joj,
    author = "Ablikim, Medina and others",
    collaboration = "BESIII",
    title = "{Observation of the Y(4230) and a new structure in *}",
    eprint = "2204.07800",
    archivePrefix = "arXiv",
    primaryClass = "hep-ex",
    doi = "10.1088/1674-1137/ac945c",
    journal = "Chin. Phys. C",
    volume = "46",
    number = "11",
    pages = "111002",
    year = "2022"
}

@article{Chiu:2005ey,
    author = "Chiu, Ting-Wai and Hsieh, Tung-Han",
    collaboration = "TWQCD",
    title = "{Y(4260) on the lattice}",
    eprint = "hep-lat/0512029",
    archivePrefix = "arXiv",
    reportNumber = "NTUTH-05-505F",
    doi = "10.1103/PhysRevD.73.094510",
    journal = "Phys. Rev. D",
    volume = "73",
    pages = "094510",
    year = "2006"
}

@article{Qiao:2007ce,
    author = "Qiao, Cong-Feng",
    title = "{A Uniform description of the states recently observed at B-factories}",
    eprint = "0709.4066",
    archivePrefix = "arXiv",
    primaryClass = "hep-ph",
    reportNumber = "GUCAS-CPS-07-006",
    doi = "10.1088/0954-3899/35/7/075008",
    journal = "J. Phys. G",
    volume = "35",
    pages = "075008",
    year = "2008"
}

@article{Wang:2022jxj,
    author = "Wang, Jun-Zhang and Liu, Xiang",
    title = "{Confirming the existence of a new higher charmonium \ensuremath{\psi}(4500) by the newly released data of e+e-\textrightarrow{}K+K-J/\ensuremath{\psi}}",
    eprint = "2212.13512",
    archivePrefix = "arXiv",
    primaryClass = "hep-ph",
    doi = "10.1103/PhysRevD.107.054016",
    journal = "Phys. Rev. D",
    volume = "107",
    number = "5",
    pages = "054016",
    year = "2023"
}

@article{Wang:2021qus,
    author = "Wang, Zhi-Gang",
    title = "{Analysis of the vector hidden-charm tetraquark states without explicit P-waves via the QCD sum rules}",
    eprint = "2108.05759",
    archivePrefix = "arXiv",
    primaryClass = "hep-ph",
    doi = "10.1016/j.nuclphysb.2021.115592",
    journal = "Nucl. Phys. B",
    volume = "973",
    pages = "115592",
    year = "2021"
}

@article{Dong:2021juy,
    author = "Dong, Xiang-Kun and Guo, Feng-Kun and Zou, Bing-Song",
    title = "{A survey of heavy-antiheavy hadronic molecules}",
    eprint = "2101.01021",
    archivePrefix = "arXiv",
    primaryClass = "hep-ph",
    doi = "10.13725/j.cnki.pip.2021.02.001",
    journal = "Progr. Phys.",
    volume = "41",
    pages = "65--93",
    year = "2021"
}

@article{Peng:2022nrj,
    author = "Peng, Fang-Zheng and Yan, Mao-Jun and S\'anchez S\'anchez, Mario and Pavon Valderrama, Manuel",
    title = "{Light- and heavy-quark symmetries and the Y(4230), Y(4360), Y(4500), Y(4620), and X(4630) resonances}",
    eprint = "2205.13590",
    archivePrefix = "arXiv",
    primaryClass = "hep-ph",
    doi = "10.1103/PhysRevD.107.016001",
    journal = "Phys. Rev. D",
    volume = "107",
    number = "1",
    pages = "016001",
    year = "2023"
}

@article{Jin:2018kjv,
    author = "Jin, Hong-Dou and Zhou, Li-Peng and Zhang, Bing-Xin and Hu, Hai-Ming",
    title = "{A discussion on vacuum polarization correction to the cross-section of $e^+e^-\to\gamma^\ast/\psi\to\mu^+\mu^-$}",
    eprint = "1805.03803",
    archivePrefix = "arXiv",
    primaryClass = "hep-ph",
    doi = "10.1088/1674-1137/43/1/013104",
    journal = "Chin. Phys. C",
    volume = "43",
    number = "1",
    pages = "013104",
    year = "2019"
}

@article{BESIII:2022quc,
    author = "Ablikim, M. and others",
    collaboration = "BESIII",
    title = "{Measurement of $e^{+}e^{-}\rightarrow\pi^{+}\pi^{-}D^{+}D^{-}$ cross sections at center-of-mass energies from 4.190 to 4.946 GeV}",
    eprint = "2208.00099",
    archivePrefix = "arXiv",
    primaryClass = "hep-ex",
    doi = "10.1103/PhysRevD.106.052012",
    journal = "Phys. Rev. D",
    volume = "106",
    number = "5",
    pages = "052012",
    year = "2022"
}

@article{BESIII:2019tdo,
    author = "Ablikim, M. and others",
    collaboration = "BESIII",
    title = "{Observation of $e^{+}e^{-}\rightarrow \pi^{+}\pi^{-}\psi(3770)$ and $D_{1}(2420)^{0}\bar{D}^{0} + c.c.$}",
    eprint = "1903.08126",
    archivePrefix = "arXiv",
    primaryClass = "hep-ex",
    doi = "10.1103/PhysRevD.100.032005",
    journal = "Phys. Rev. D",
    volume = "100",
    number = "3",
    pages = "032005",
    year = "2019"
}

@article{BESIII:2019phe,
    author = "Ablikim, Medina and others",
    collaboration = "BESIII",
    title = "{Study of $e^{+}e^{-} \to D^{+} D^{-} \pi^{+} \pi^{-} $ at center-of-mass energies from 4.36 to 4.60 GeV}",
    eprint = "1909.12478",
    archivePrefix = "arXiv",
    primaryClass = "hep-ex",
    doi = "10.1016/j.physletb.2020.135395",
    journal = "Phys. Lett. B",
    volume = "804",
    pages = "135395",
    year = "2020"
}

@article{Meng:2007fu,
    author = "Meng, Ce and Chao, Kuang-Ta",
    title = "{Z+(4430) as a resonance in the D(1)(D(1)-prime)D* channel}",
    eprint = "0708.4222",
    archivePrefix = "arXiv",
    primaryClass = "hep-ph",
    month = "8",
    year = "2007"
}

@article{Ma:2014zua,
    author = "Ma, Li and Liu, Xiao-Hai and Liu, Xiang and Zhu, Shi-Lin",
    title = "{Exotic Four Quark Matter: $Z_1(4475)$}",
    eprint = "1404.3450",
    archivePrefix = "arXiv",
    primaryClass = "hep-ph",
    doi = "10.1103/PhysRevD.90.037502",
    journal = "Phys. Rev. D",
    volume = "90",
    number = "3",
    pages = "037502",
    year = "2014"
}

@article{Belle:2010fwv,
    author = "Pakhlova, G. and others",
    collaboration = "Belle",
    title = "{Measurement of $e^+e^-\to D_s^{(*)+} D_s^{(*)-}$ cross sections near threshold using initial-state radiation}",
    eprint = "1011.4397",
    archivePrefix = "arXiv",
    primaryClass = "hep-ex",
    doi = "10.1103/PhysRevD.83.011101",
    journal = "Phys. Rev. D",
    volume = "83",
    pages = "011101",
    year = "2011"
}

@article{BaBar:2010plp,
    author = "del Amo Sanchez, P. and others",
    collaboration = "BaBar",
    title = "{Exclusive Production of $D^+_s D^-_s$, $D^{*+}_s D^-_s$, and $D^{*+}_s D^{*-}_s$ via $e^+ e^-$ Annihilation with Initial-State-Radiation}",
    eprint = "1008.0338",
    archivePrefix = "arXiv",
    primaryClass = "hep-ex",
    reportNumber = "SLAC-PUB-14209, BABAR-PUB-10-015",
    doi = "10.1103/PhysRevD.82.052004",
    journal = "Phys. Rev. D",
    volume = "82",
    pages = "052004",
    year = "2010"
}

@article{BESIII:2024zdh,
    author = "Ablikim, Medina and others",
    collaboration = "BESIII",
    title = "{Precise Measurement of the e+e-\textrightarrow{}Ds+Ds- Cross Section at Center-of-Mass Energies from Threshold to 4.95~GeV}",
    eprint = "2403.14998",
    archivePrefix = "arXiv",
    primaryClass = "hep-ex",
    doi = "10.1103/PhysRevLett.133.261902",
    journal = "Phys. Rev. Lett.",
    volume = "133",
    number = "26",
    pages = "261902",
    year = "2024"
}

@article{BESIII:2023wsc,
    author = "Ablikim, M. and others",
    collaboration = "BESIII",
    title = "{Precise Measurement of the e+e-\textrightarrow{}Ds*+Ds*- Cross Sections at Center-of-Mass Energies from Threshold to 4.95~GeV}",
    eprint = "2305.10789",
    archivePrefix = "arXiv",
    primaryClass = "hep-ex",
    doi = "10.1103/PhysRevLett.131.151903",
    journal = "Phys. Rev. Lett.",
    volume = "131",
    number = "15",
    pages = "151903",
    year = "2023"
}

@article{BESIII:2023wqy,
    author = "Ablikim, M. and others",
    collaboration = "BESIII",
    title = "{Observation of a Vector Charmoniumlike State at 4.7\,\,GeV/c2 and Search for Zcs in e+e-\textrightarrow{}K+K-J/\ensuremath{\psi}}",
    eprint = "2308.15362",
    archivePrefix = "arXiv",
    primaryClass = "hep-ex",
    doi = "10.1103/PhysRevLett.131.211902",
    journal = "Phys. Rev. Lett.",
    volume = "131",
    number = "21",
    pages = "211902",
    year = "2023"
}

@article{Belle:2019qoi,
    author = "Jia, S. and others",
    collaboration = "Belle",
    title = "{Observation of a vector charmoniumlike state in $e^+e^- \to D^+_sD_{s1}(2536)^-+c.c.$}",
    eprint = "1911.00671",
    archivePrefix = "arXiv",
    primaryClass = "hep-ex",
    reportNumber = "Belle Preprint \# 2019-20, and KEK Preprint \#: 2019-42",
    doi = "10.1103/PhysRevD.100.111103",
    journal = "Phys. Rev. D",
    volume = "100",
    number = "11",
    pages = "111103",
    year = "2019"
}

@article{Belle:2020wtd,
    author = "Jia, S. and others",
    collaboration = "Belle",
    title = "{Evidence for a vector charmoniumlike state in $e^+e^- \to D^+_sD^*_{s2}(2573)^-+c.c.$}",
    eprint = "2004.02404",
    archivePrefix = "arXiv",
    primaryClass = "hep-ex",
    doi = "10.1103/PhysRevD.101.091101",
    journal = "Phys. Rev. D",
    volume = "101",
    number = "9",
    pages = "091101",
    year = "2020"
}

@article{BESIII:2024qfi,
    author = "Ablikim, M. and others",
    collaboration = "BESIII",
    title = "{Study of the Decay and Production Properties of Ds1(2536) and Ds2*(2573)}",
    eprint = "2407.07651",
    archivePrefix = "arXiv",
    primaryClass = "hep-ex",
    doi = "10.1103/PhysRevLett.133.171903",
    journal = "Phys. Rev. Lett.",
    volume = "133",
    number = "17",
    pages = "171903",
    year = "2024"
}

@article{CLEO:2003ggt,
    author = "Besson, D. and others",
    collaboration = "CLEO",
    title = "{Observation of a narrow resonance of mass 2.46-GeV/c**2 decaying to D*+(s) pi0 and confirmation of the D*(sJ)(2317) state}",
    eprint = "hep-ex/0305100",
    archivePrefix = "arXiv",
    reportNumber = "CLNS-03-1826, CLEO-03-09, CLNS03-1826",
    doi = "10.1103/PhysRevD.68.032002",
    journal = "Phys. Rev. D",
    volume = "68",
    pages = "032002",
    year = "2003",
    note = "[Erratum: Phys.Rev.D 75, 119908 (2007)]"
}

@article{CLEO:1989qui,
    author = "Avery, P. and others",
    collaboration = "CLEO",
    title = "{$P$ Wave Charmed Mesons in $e^+ e^-$ Annihilation}",
    reportNumber = "CLNS-89/939, CLEO-89-11",
    doi = "10.1103/PhysRevD.41.774",
    journal = "Phys. Rev. D",
    volume = "41",
    pages = "774",
    year = "1990"
}

@article{BaBar:2003oey,
    author = "Aubert, B. and others",
    collaboration = "BaBar",
    title = "{Observation of a narrow meson decaying to $D_s^+ \pi^0$ at a mass of 2.32-GeV/c$^2$}",
    eprint = "hep-ex/0304021",
    archivePrefix = "arXiv",
    reportNumber = "SLAC-PUB-9711, BABAR-PUB-03-011",
    doi = "10.1103/PhysRevLett.90.242001",
    journal = "Phys. Rev. Lett.",
    volume = "90",
    pages = "242001",
    year = "2003"
}

@article{BaBar:2004yux,
    author = "Aubert, Bernard and others",
    collaboration = "BaBar",
    title = "{Study of $B \to D_{sJ}^{(*)+} \bar{D}^{(*)}$ decays}",
    eprint = "hep-ex/0408041",
    archivePrefix = "arXiv",
    reportNumber = "SLAC-PUB-10627, BABAR-PUB-04-24, BABAR-PUB-04-024",
    doi = "10.1103/PhysRevLett.93.181801",
    journal = "Phys. Rev. Lett.",
    volume = "93",
    pages = "181801",
    year = "2004"
}

@article{Bondioli:2004te,
    author = "Bondioli, M.",
    editor = "Narison, S.",
    collaboration = "BaBar",
    title = "{Observation in the BABAR, experiment of new narrow states at 2.317-GeV/c and 2.458- GeV/c(2) in the D(s) system}",
    reportNumber = "SLAC-REPRINT-2004-206",
    doi = "10.1016/j.nuclphysbps.2004.04.155",
    journal = "Nucl. Phys. B Proc. Suppl.",
    volume = "133",
    pages = "158--161",
    year = "2004"
}

@article{BaBar:2011vbs,
    author = "Lees, J. P. and others",
    collaboration = "BaBar",
    title = "{Measurement of the mass and width of the D(s1)(2536)+ meson}",
    eprint = "1103.2675",
    archivePrefix = "arXiv",
    primaryClass = "hep-ex",
    reportNumber = "BABAR-PUB-10-031, SLAC-PUB-14400",
    doi = "10.1103/PhysRevD.83.072003",
    journal = "Phys. Rev. D",
    volume = "83",
    pages = "072003",
    year = "2011"
}

@article{Belle:2003guh,
    author = "Krokovny, P. and others",
    collaboration = "Belle",
    title = "{Observation of the D(sJ)(2317) and D(sJ)(2457) in B decays}",
    eprint = "hep-ex/0308019",
    archivePrefix = "arXiv",
    doi = "10.1103/PhysRevLett.91.262002",
    journal = "Phys. Rev. Lett.",
    volume = "91",
    pages = "262002",
    year = "2003"
}

@article{BESIII:2018fpo,
    author = "Ablikim, Medina and others",
    collaboration = "BESIII",
    title = "{Observation of $e^+e^- \rightarrow D_s^+ \overline{D}{}^{(*)0} K^-$ and study of the $P$-wave $D_s$ mesons}",
    eprint = "1812.09800",
    archivePrefix = "arXiv",
    primaryClass = "hep-ex",
    doi = "10.1088/1674-1137/43/3/031001",
    journal = "Chin. Phys. C",
    volume = "43",
    number = "3",
    pages = "031001",
    year = "2019",
    note = "[Erratum: Chin.Phys.C 43, 129102 (2019)]"
}

@article{Zeng:1994vj,
    author = "Zeng, J. and Van Orden, J. W. and Roberts, W.",
    title = "{Heavy mesons in a relativistic model}",
    eprint = "hep-ph/9412269",
    archivePrefix = "arXiv",
    reportNumber = "CEBAF-TH-94-08",
    doi = "10.1103/PhysRevD.52.5229",
    journal = "Phys. Rev. D",
    volume = "52",
    pages = "5229--5241",
    year = "1995"
}

@article{Dai:2003yg,
    author = "Dai, Yuan-Ben and Huang, Chao-Shang and Liu, Chun and Zhu, Shi-Lin",
    title = "{Understanding the D+(sJ)(2317) and D+(sJ)(2460) with sum rules in HQET}",
    eprint = "hep-ph/0306274",
    archivePrefix = "arXiv",
    doi = "10.1103/PhysRevD.68.114011",
    journal = "Phys. Rev. D",
    volume = "68",
    pages = "114011",
    year = "2003"
}

@article{Chen:2016spr,
    author = "Chen, Hua-Xing and Chen, Wei and Liu, Xiang and Liu, Yan-Rui and Zhu, Shi-Lin",
    title = "{A review of the open charm and open bottom systems}",
    eprint = "1609.08928",
    archivePrefix = "arXiv",
    primaryClass = "hep-ph",
    doi = "10.1088/1361-6633/aa6420",
    journal = "Rept. Prog. Phys.",
    volume = "80",
    number = "7",
    pages = "076201",
    year = "2017"
}

@article{Cheng:2003kg,
    author = "Cheng, Hai-Yang and Hou, Wei-Shu",
    title = "{B decays as spectroscope for charmed four quark states}",
    eprint = "hep-ph/0305038",
    archivePrefix = "arXiv",
    doi = "10.1016/S0370-2693(03)00834-7",
    journal = "Phys. Lett. B",
    volume = "566",
    pages = "193--200",
    year = "2003"
}

@article{Dmitrasinovic:2012zz,
    author = "Dmitrasinovic, V.",
    title = "{Chiral symmetry of heavy-light scalar mesons with $U_A(1)$ symmetry breaking}",
    doi = "10.1103/PhysRevD.86.016006",
    journal = "Phys. Rev. D",
    volume = "86",
    pages = "016006",
    year = "2012"
}

@article{Maiani:2004vq,
    author = "Maiani, L. and Piccinini, F. and Polosa, A. D. and Riquer, V.",
    title = "{Diquark-antidiquarks with hidden or open charm and the nature of X(3872)}",
    eprint = "hep-ph/0412098",
    archivePrefix = "arXiv",
    reportNumber = "ROMA1-1396-2004, FNT-T-2004-20, BA-TH-502-04, CERN-PH-TH-2004-239",
    doi = "10.1103/PhysRevD.71.014028",
    journal = "Phys. Rev. D",
    volume = "71",
    pages = "014028",
    year = "2005"
}

@article{Wang:2006uba,
    author = "Wang, Zhi-Gang and Wan, Shao-Long",
    title = "{D(s)(2317) as a tetraquark state with QCD sum rules in heavy quark limit}",
    eprint = "hep-ph/0602080",
    archivePrefix = "arXiv",
    doi = "10.1016/j.nuclphysa.2006.07.041",
    journal = "Nucl. Phys. A",
    volume = "778",
    pages = "22--29",
    year = "2006"
}

@article{Barnes:2003dj,
    author = "Barnes, T. and Close, F. E. and Lipkin, H. J.",
    title = "{Implications of a DK molecule at 2.32-GeV}",
    eprint = "hep-ph/0305025",
    archivePrefix = "arXiv",
    doi = "10.1103/PhysRevD.68.054006",
    journal = "Phys. Rev. D",
    volume = "68",
    pages = "054006",
    year = "2003"
}

@article{Chen:2004dy,
    author = "Chen, Yu-Qi and Li, Xue-Qian",
    title = "{A Comprehensive four-quark interpretation of D(s)(2317), D(s)(2457) and D(s)(2632)}",
    eprint = "hep-ph/0407062",
    archivePrefix = "arXiv",
    doi = "10.1103/PhysRevLett.93.232001",
    journal = "Phys. Rev. Lett.",
    volume = "93",
    pages = "232001",
    year = "2004"
}

@article{Xie:2010zza,
    author = "Xie, Zhen-Xing and Feng, Guan-Qiu and Guo, Xin-Heng",
    title = "{Analyzing Ds0*(2317)(+) in the DK molecule picture in the Beth-Salpeter approach}",
    doi = "10.1103/PhysRevD.81.036014",
    journal = "Phys. Rev. D",
    volume = "81",
    pages = "036014",
    year = "2010"
}

@article{Feng:2012zze,
    author = "Feng, G. Q. and Guo, X. H. and Zhang, Z. H.",
    title = "{Studying the D* K molecular structure of D/s(2460) in the Bethe-Salpeter approach}",
    doi = "10.1140/epjc/s10052-012-2033-y",
    journal = "Eur. Phys. J. C",
    volume = "72",
    pages = "2033",
    year = "2012"
}

@article{Wang:2012bu,
    author = "Wang, P. and Wang, X. G.",
    title = "{Study on $0^+$ states with open charm in unitarized heavy meson chiral approach}",
    eprint = "1204.5553",
    archivePrefix = "arXiv",
    primaryClass = "hep-ph",
    doi = "10.1103/PhysRevD.86.014030",
    journal = "Phys. Rev. D",
    volume = "86",
    pages = "014030",
    year = "2012"
}

@article{Browder:2003fk,
    author = "Browder, Thomas E. and Pakvasa, Sandip and Petrov, Alexey A.",
    title = "{Comment on the new D(s)(*)+ pi0 resonances}",
    eprint = "hep-ph/0307054",
    archivePrefix = "arXiv",
    reportNumber = "WSU-HEP-0306",
    doi = "10.1016/j.physletb.2003.10.067",
    journal = "Phys. Lett. B",
    volume = "578",
    pages = "365--368",
    year = "2004"
}

@article{BESIII:2021xrz,
    author = "Ablikim, Medina and others",
    collaboration = "BESIII",
    title = "{Measurements of Born Cross Sections of $e^+e^-\to D_s^{*+} D_{sJ}^{-} +c.c.$}",
    eprint = "2106.02298",
    archivePrefix = "arXiv",
    primaryClass = "hep-ex",
    doi = "10.1103/PhysRevD.104.032012",
    journal = "Phys. Rev. D",
    volume = "104",
    pages = "032012",
    year = "2021"
}

@article{Lin:2024qcq,
    author = "Lin, Zi-Yang and Wang, Jun-Zhang and Cheng, Jian-Bo and Meng, Lu and Zhu, Shi-Lin",
    title = "{Identification of the $G(3900)$ as the P-wave $D\bar{D}^*/\bar{D}D^*$ resonance}",
    eprint = "2403.01727",
    archivePrefix = "arXiv",
    primaryClass = "hep-ph",
    doi = "10.1103/PhysRevLett.133.241903",
    journal = "Phys. Rev. Lett.",
    volume = "133",
    number = "24",
    pages = "241903",
    year = "2024"
}

@article{Zhang:2009gy,
    author = "Zhang, Yuan-Jiang and Zhao, Qiang",
    title = "{The Lineshape of psi(3770) and low-lying vector charmonium resonance parameters in e+ e- ---\ensuremath{>} D anti-D}",
    eprint = "0911.5651",
    archivePrefix = "arXiv",
    primaryClass = "hep-ph",
    doi = "10.1103/PhysRevD.81.034011",
    journal = "Phys. Rev. D",
    volume = "81",
    pages = "034011",
    year = "2010"
}

@article{Du:2016qcr,
    author = "Du, Meng-Lin and Mei\ss{}ner, Ulf-G. and Wang, Qian",
    title = "{$P$-wave coupled channel effects in electron-positron annihilation}",
    eprint = "1608.02537",
    archivePrefix = "arXiv",
    primaryClass = "hep-ph",
    doi = "10.1103/PhysRevD.94.096006",
    journal = "Phys. Rev. D",
    volume = "94",
    number = "9",
    pages = "096006",
    year = "2016"
}

@article{Salnikov:2024wah,
    author = "Salnikov, S. G. and Milstein, A. I.",
    title = "{Production of D(*)D\textasciimacron{}(*) near the thresholds in e+e- annihilation}",
    eprint = "2404.06160",
    archivePrefix = "arXiv",
    primaryClass = "hep-ph",
    doi = "10.1103/PhysRevD.109.114015",
    journal = "Phys. Rev. D",
    volume = "109",
    number = "11",
    pages = "114015",
    year = "2024"
}

@article{Levichev:2018cvd,
    author = "Levichev, E. B. and Skrinsky, A. N. and Tumaikin, G. M. and Shatunov, Yu M.",
    title = "{Electron\textendash{}positron beam collision studies at the Budker Institute of Nuclear Physics}",
    doi = "10.3367/UFNe.2018.01.038300",
    journal = "Phys. Usp.",
    volume = "61",
    number = "5",
    pages = "405--423",
    year = "2018"
}

@article{Achasov:2023gey,
    author = "Achasov, M. and others",
    title = "{STCF conceptual design report (Volume 1): Physics \& detector}",
    eprint = "2303.15790",
    archivePrefix = "arXiv",
    primaryClass = "hep-ex",
    doi = "10.1007/s11467-023-1333-z",
    journal = "Front. Phys. (Beijing)",
    volume = "19",
    number = "1",
    pages = "14701",
    year = "2024"
}

@article{Okubo:1963fa,
    author = "Okubo, S.",
    title = "{Phi meson and unitary symmetry model}",
    doi = "10.1016/S0375-9601(63)92548-9",
    journal = "Phys. Lett.",
    volume = "5",
    pages = "165--168",
    year = "1963"
}

@inbook{Zweig:1964jf,
    author = "Zweig, G.",
    editor = "Lichtenberg, D. B. and Rosen, Simon Peter",
    title = "{An SU(3) model for strong interaction symmetry and its breaking. Version 2}",
    booktitle = "{DEVELOPMENTS IN THE QUARK THEORY OF HADRONS. VOL. 1. 1964 - 1978}",
    reportNumber = "CERN-TH-412, NP-14146, PRINT-64-170",
    doi = "10.17181/CERN-TH-412",
    pages = "22--101",
    month = "2",
    year = "1964"
}

@article{Iizuka:1966fk,
    author = "Iizuka, Jugoro",
    title = "{Systematics and phenomenology of meson family}",
    doi = "10.1143/PTPS.37.21",
    journal = "Prog. Theor. Phys. Suppl.",
    volume = "37",
    pages = "21--34",
    year = "1966"
}

@article{BESIII:2020eyu, 
author = "Ablikim, Medina and others", 
collaboration = "BESIII", 
title = "{Measurements of the center-of-mass energies of collisions at BESIII}", eprint = "2012.14750", archivePrefix = "arXiv", primaryClass = "hep-ex", doi = "10.1088/1674-1137/ac1575", journal = "Chin. Phys. C", volume = "45", number = "10", pages = "103001", year = "2021" }

@article{Chung:1995dx,
    author = "Chung, S. U. and Brose, J. and Hackmann, R. and Klempt, E. and Spanier, S. and Strassburger, C.",
    title = "{Partial wave analysis in K matrix formalism}",
    doi = "10.1002/andp.19955070504",
    journal = "Annalen Phys.",
    volume = "4",
    pages = "404--430",
    year = "1995"
}

@article{Oller:2025leg,
    author = "Oller, Jos{\'e} Antonio",
    title = "{Coupled-channel formalism}",
    eprint = "2501.10000",
    archivePrefix = "arXiv",
    primaryClass = "hep-ph",
    month = "1",
    year = "2025"
}

@article{Henner:2022ksn,
    author = "Henner, Victor and Belozerova, Tatyana",
    title = "{Analysis of Overlapping Resonances with Unitary Breit{\textendash}Wigner and K-Matrix Approaches}",
    doi = "10.3390/particles5040035",
    journal = "Particles",
    volume = "5",
    number = "4",
    pages = "451--487",
    year = "2022"
}

@article{Druzhinin:2011qd,
    author = "Druzhinin, V. P. and Eidelman, S. I. and Serednyakov, S. I. and Solodov, E. P.",
    title = "{Hadron Production via e+e- Collisions with Initial State Radiation}",
    eprint = "1105.4975",
    archivePrefix = "arXiv",
    primaryClass = "hep-ex",
    doi = "10.1103/RevModPhys.83.1545",
    journal = "Rev. Mod. Phys.",
    volume = "83",
    pages = "1545",
    year = "2011"
}

@article{Wang:2023zxj,
    author = "Wang, Jun-Zhang and Liu, Xiang",
    title = "{Identifying a characterized energy level structure of higher charmonium well matched to the peak structures in e+e{\ensuremath{-}} {\textrightarrow} {\ensuremath{\pi}}+D0D{\textasteriskcentered}{\ensuremath{-}}}",
    eprint = "2306.14695",
    archivePrefix = "arXiv",
    primaryClass = "hep-ph",
    doi = "10.1016/j.physletb.2024.138456",
    journal = "Phys. Lett. B",
    volume = "849",
    pages = "138456",
    year = "2024"
}

@article{Peng:2024blp,
    author = "Peng, Tian-Cai and Bai, Zi-Yue and Wang, Jun-Zhang and Liu, Xiang",
    title = "{Reevaluating the {\ensuremath{\psi}}(4160) resonance parameter using B+{\textrightarrow}K+{\ensuremath{\mu}}+{\ensuremath{\mu}}- data in the context of unquenched charmonium spectroscopy}",
    eprint = "2412.11096",
    archivePrefix = "arXiv",
    primaryClass = "hep-ph",
    doi = "10.1103/PhysRevD.111.054023",
    journal = "Phys. Rev. D",
    volume = "111",
    number = "5",
    pages = "054023",
    year = "2025"
}

@article{LHCb:2019lnr,
    author = "Aaij, Roel and others",
    collaboration = "LHCb",
    title = "{Near-threshold $D\bar{D}$ spectroscopy and observation of a new charmonium state}",
    eprint = "1903.12240",
    archivePrefix = "arXiv",
    primaryClass = "hep-ex",
    reportNumber = "CERN-EP-2019-047, LHCb-PAPER-2019-005",
    doi = "10.1007/JHEP07(2019)035",
    journal = "JHEP",
    volume = "07",
    pages = "035",
    year = "2019"
}

@article{BESIII:2022kcv,
    author = "Ablikim, M. and others",
    collaboration = "BESIII",
    title = "{Observation of the Y(4230) and evidence for a new vector charmoniumlike state Y(4710) in e+e-{\textrightarrow}KS0KS0J/{\ensuremath{\psi}}}",
    eprint = "2211.08561",
    archivePrefix = "arXiv",
    primaryClass = "hep-ex",
    doi = "10.1103/PhysRevD.107.092005",
    journal = "Phys. Rev. D",
    volume = "107",
    number = "9",
    pages = "092005",
    year = "2023"
}

\end{document}